\documentclass[11pt,a4paper]{article}
\usepackage[utf8]{inputenc}
\usepackage[english]{babel}
\usepackage{amsmath}
\usepackage{amsfonts}
\usepackage{amssymb}
\usepackage{makeidx}
\usepackage{graphicx}
\usepackage{lmodern}
\usepackage{kpfonts}
\usepackage{bm}
\usepackage{multirow}
\usepackage{supertabular}
\usepackage{dcolumn}
\usepackage{pdflscape}
\usepackage{pgfplots}
\usepackage{accents}
\usepackage{setspace}
\usepackage{caption}
\usepackage[round]{natbib}
\usepackage{lscape}
\usepackage[left=2cm,right=2cm,top=2cm,bottom=2cm]{geometry}

\begin{document}

{\center {\Large A Machine Learning Approach for Flagging Incomplete Bid-rigging Cartels}{\large
\vspace{0.1cm}}\smallskip\\

{\large Hannes Wallimann*, David Imhof** and Martin Huber***}\smallskip\\
{\small {* University of Applied Sciences and Arts Lucerne, Competence Center for Mobility,  University of Fribourg, Dept.\ of Economics}}\\[0pt]
{\small {** Corresponding Author, Swiss Competition Commission, University of Fribourg, Dept.\ of Economics and Unidistance (Switzerland)}}\\[0pt]
{\small {*** University of Fribourg, Dept.\ of Economics}}\\[0pt]}\smallskip

\vspace{2cm} \noindent \textbf{Abstract:} {\small \textit{We propose a new method for flagging bid rigging, which is particularly useful for detecting incomplete bid-rigging cartels. Our approach combines screens, i.e. statistics derived from the distribution of bids in a tender, with machine learning to predict the probability of collusion. As a methodological innovation, we calculate such screens for all possible subgroups of three or four bids within a tender and use summary statistics like the mean, median, maximum, and minimum of each screen as predictors in the machine learning algorithm. This approach tackles the issue that competitive bids in incomplete cartels distort the statistical signals produced by bid rigging. We demonstrate that our algorithm outperforms previously suggested methods in applications to incomplete cartels based on empirical data from Switzerland.}}
\vspace{0cm}\smallskip\\

\vspace{2pt}

\vspace{2pt}

{\footnotesize \noindent \textbf{Address for correspondence:} David Imhof, Hallwylstrasse 4, 3003 Bern, Switzerland; david.imhof@weko.admin.ch.}\vspace{2pt}

\begin{spacing}{1}
{\footnotesize \noindent \textbf{Acknowledgement} The authors would like to thank Christoph Brunner, Philippe Sulger and Niklaus Wallimann for support and helpful comments.}

{\footnotesize \noindent \textbf{Disclaimer:} All views contained in this paper are solely those of the authors and cannot be attributed to the Swiss Competition Commission, its Secretariat, the University of Fribourg, University of Applied Sciences and Arts Lucerne, or Unidistance (Switzerland).}
\end{spacing}

\thispagestyle{empty}\pagebreak  {\small
\renewcommand{\thefootnote}{\arabic{footnote}}
\setcounter{footnote}{0} \pagebreak \setcounter{footnote}{0}
\pagebreak \setcounter{page}{1} }

\newpage

\section{Introduction}
Bid rigging is a pervasive problem and has been considered in a range of studies. A few examples include bid rigging in a Swedish asphalt cartel \citep[][]{Bergman2019}, in public procurement auctions for construction works in Japan \citep[][]{Ishii2014}, of stamp dealers in New York \citep[][]{Asker2010}, among seafood processors in the US \citep[][]{Abrantes2006}, of construction firms in Korea \citep[][]{Lee2002}, for school milk contracts in Ohio \citep[][]{Porter1999}, Florida and Texas \citep[][]{Pesendorfer2000} and related to US highway construction contracts \citep[][]{Porter1993, Feinstein1985}. When firms deviate from competitive bidding and form a bid-rigging cartel, they secretly conspire to raise prices or lower the quality of goods or services. As such, bid-rigging conspiracies directly harm taxpayers, buyers or sellers. In order to enhance the fight against bid rigging, the OECD recommends competition agencies to promote pro-active methods for uncovering bid-rigging cartels \citep[][]{OECD2014}. Answering the need of competition agencies for pro-active statistical methods, \citet{Imhof2017a}, \citet{Huberimhof2019} and \citet{Imhof2020} have proposed different methods for uncovering bid-rigging cartels with screens. However, the three papers deal solely with one form of bid rigging: Complete bid-rigging cartels. The reality is more complex, and competitive bidders often participate in tenders in which a cartel is active. Competitive bids distort the statistical signals produced by cartels in the distribution of the bids so that conventional detection methods fail to recognize the track of bid rigging. With competitive bids, the correct classification rate of collusive and competitive tenders based on the methods suggested in the three previously mentioned papers shrinks. This study improves on this issue by constructing a more robust detection method obtaining a decent correct classification rate similar to \citet{Huberimhof2019} even in the presence of competitive bids.

As methodological contribution, our approach considers a larger set of screens than \citet{Huberimhof2019}, likely more powerful for detecting incomplete cartels. Screens are statistics derived from the distribution of bids in a tender aiming to capture the distributional changes produced by bid rigging. However, since competitive bids distort the statistical signals produced by bid rigging, screens based on the total of bids in some tender are less effective in detecting incomplete bid-rigging cartels. Our approach overcomes this issue by calculating screens for subgroups of bids in a tender. For this purpose, we form all possible subgroups of three and four bids per tender and calculate screens for each subgroup. We then calculate the mean, median, minimum and maximum of each screen for the different subgroups within some tender. This permits restoring the original statistical signals produced by bid rigging in presence of competitive bids.
We use data from Switzerland in which the incidence of collusive and competitive tenders is known in order to apply the screens as predictors for collusion and assess the correct classification rate of collusive and competitive tenders. As in \citet{Huberimhof2019}, we use machine learning for the optimal prediction (which does not require the construction of explicit structural or causal models for collusion) and focus on the random forest, see \citet{Breiman2001}, a widely used algorithm for statistical learning \citep[see][]{Athey2019}.

We first apply our method to the Ticino bid-rigging cartel (hereafter: the Ticino cartel), which was a complete cartel involving all firms active in road construction in the canton of Ticino. The cartel members rigged all tenders from January 1999 until the end of March 2005 \citep[see][for details]{Imhof2020}. Since we also have the data from the post-cartel period, we use them to simulate competitive bids that we add to the collusive tenders. We create five new datasets with the collusive tenders: The first dataset contains one simulated competitive bid in each collusive tender. The second one includes two simulated competitive bids in each collusive tender, and so on. This stepwise approach permits investigating how competitive bids distort the statistical patterns of bid rigging in the distribution of the bids. We also note that the simulation process produces competitive bids exhibiting the same features as the bids in the post-cartel period. We compare the correct classification rate of three methods. The first method of \citet{Imhof2017a} uses two screens applied with benchmarks. The second method of \citet{Huberimhof2019} uses machine learning, namely lasso and an ensemble method (averaging over several machine learners, namely bagged trees, random forests and neural networks) to a range of screens in order to predict collusion. In this paper, we implement their suggested method using random forests as machine learner. The third method, which is an original contribution of this paper, includes summary statistics of the screens (\textit{median, mean, maximum and minimum}) calculated in all possible subgroups of bids in a tender as predictors in the random forest.

We find that the method of \citet{Imhof2017a} performs poorly when the number of competitive bids per tender increases in the data of the Ticino cartel. With three simulated competitive bids, its correct classification rate does not perform better than flipping a coin. It is particularly unsatisfactory in the case of collusive tenders, with a correct classification rate of only 15\%. The method of \citet{Huberimhof2019} entails a correct classification rate between 72\% and 84\%, again decreasing in the number of competitive bids added to the collusive tenders. Including summary statistics of screens calculated for all possible subgroups of bids in a tender, as suggested in this paper, exhibits a higher correct classification rate of 77\% to 86\%. We also observe that the correct classification rate does not decrease in the number of competitive bids added to the collusive tenders. Therefore, the relative performance of the method suggested in this paper when compared to \citet{Huberimhof2019} increases in the number of competitive bids. With five simulated competitive bids, the difference between both methods accounts for over 10 percentage points. If we consider the error rate, defined as one minus the correct classification rate, it decreases by 43\% with our approach. Cutting the error rate almost by half is substantial with regard to the heavy legal consequences of flagging firms as bid-rigging cartels. The detection method constructed in our paper can therefore flag both complete and incomplete bid-rigging cartels with a decent correct classification rate.

However, the simulation exercise based on the Ticino case is limited to the purpose of examining the correct classification rate of different methods. It does not account for the reaction of competitive and collusive bidders when they are aware of their reciprocal existence. In fact, competitive bidders might try to benefit from the umbrella effect of the cartel by bidding higher than they would have in a competitive situation \citep[][]{Bos2010}. Some may also try to join temporarily a cartel as observed in some cases. Conversely, collusive bidders react to the presence of competitive bidders. They might adopt a more competitive behavior by bidding more aggressively as attested in the Swiss case of See-Gaster in which cartel participants decided together on which tender they would form a cartel and on which they would compete. We therefore need to apply our methods to fully empirical data to verify the performance of our simulation exercise based on the Ticino cartel.

As an important empirical contribution, we therefore consider rather unique data from two investigations of the Swiss competition commission (hereafter: COMCO): See-Gaster and Strassenbau Graubünden. Both cases were characterized by well-organized bid-rigging cartels that were for a large share of tenders complete. However, from time to time, they faced competition from outsiders. We mainly focus on the collusive tenders with competitive bids, and again evaluate the performance of the three previously mentioned methods, namely the approach of \citet{Imhof2017a}, the method of \citet{Huberimhof2019} using a random forest, and the method suggested in this paper that includes summary statistics calculated for subgroups of bids in a tender.

We find that in See-Gaster and Graubünden data, the approach of \citet{Imhof2017a} again exhibits a low correct classification rate which roughly amounts to 50\% for incomplete cartels (collusive tenders with competitive bids). Considering tenders with incomplete cartels only, the correct classification rate is even much lower and varies between 8\% and 14\%. Even for complete cartels, the correct classification rate of \citet{Imhof2017a} barely improves. This result might suggest that the benchmarks applied in \citet{Imhof2017a} for detecting collusive tenders in Swiss data are generally ill-posed, due to the substantial amount of false negative predictions (incorrectly proposing the absence of a cartel).
However, benchmarks require few information to be implemented and perform relatively well in the data of the Ticino cartel (when complete), and they therefore remain a first step in developing screening methods for flagging cartels. Moreover, we also illustrate that one can use the decision tree to recalibrate the benchmarks used by \citet{Imhof2017a} and increase the correct prediction rate.

Considering machine learning methods, we note that for complete cartels, there are very little differences across approaches in terms of the correct classification rate, which amounts to 81-83\%. When competitive bidders are present, however, the correct classification rate is higher (amounting to 67\% to 84\% depending on the sample) when including summary screens for subgroups of bids in a tender than when using the screens of \citet{Huberimhof2019} alone (61\% to 77\%). We note that the performance of all machine learning approaches decreases in the proportion of competitive bids. From less to more competitive bids, the correct classification rate shrinks by 16 to 17 percentage points for the various methods. Contrary to the simulation based on the Ticino case, this result suggests that cartel participants partially endogenise the presence of competitive bidders by adopting in some cases a more competitive behavior.

Our paper is related to other studies using screens for uncovering bid-rigging cartels. \citet{Huberimhof2019} and \citet{Imhof2020} construct detection methods based on screens using data from Swiss bid-rigging cases, in which they can distinguish between collusive and competitive periods. Such analyses are \textit{ex-post} and allow assessing the performance of the screens and detection methods. Our paper is also an \textit{ex-post} analysis and contrasts with \citet{Imhof2017a}, who also develop a method based on screens, however without information on the existence of bid-rigging cartels in the data. By constructing different self-reinforcing tests, \citet{Imhof2017a} uncovered a bid-rigging cartel in the region of See-Gaster in the canton of St Gallen. COMCO opened an investigation in 2013 essentially based on such a statistical analysis and sanctioned the firms involved in the cartel in 2016.\footnote{See press release: https://www.weko.admin.ch/weko/fr/home/actualites/communiques-de-presse/nsb-news.msg-id-64011.html.} In a broader context, our paper is also related to contributions using screens for uncovering price-fixing conspiracies \citep[see][]{Abrantes2006, Esposito2006, Hueschelrath2011, Jimenez2012, Abrantes2012, Crede2019}.

Our study is also related to further papers on detecting bid-rigging cartels, however, not relying on screens. One seminal paper is \citet{Bajari2003}, who propose two econometric tests for classifying pairs of firms as collusive. The idea of the first test consists of verifying whether bids are independent conditional on the costs of each firm by estimating a bidding function and testing if residuals are correlated between firms. Correlation between firms would indicate that bids are not independent and thus point to potential collusion. The second test examines if firms react in the same way as a function of their own costs. The test verifies if estimated coefficients on cost variables are identical for each firm. Divergent coefficients between firms could indicate potential collusive issues. Subsequent papers apply and refine the econometric tests suggested by Bajari and Ye (2003) \citep[see][]{Jakobsson2007, Aryal2013, Chotibhongs2012a, Chotibhongs2012b, Imhof2017b, Bergman2019}. \citet{Imhof2017b} questions the performance of the econometric tests proposed by \citet{Bajari2003} for detecting the Ticino cartel because the econometric tests produce too many false negative results by failing to classify pairs of firms as collusive. Moreover, econometric tests require data on the firm level, which are difficult to access especially if a competition agency wishes to apply such tests ex ante. In contrast, the test by \citet{Bergman2019} only uses data from the bid summaries\footnote{Bid summaries are the official records of a bid opening. At a fixed date, the procurement agency open all received bids and writes a record of all the submitted bids for some specific contract. Then, the procurement agency examines all the submitted bids in detail.} combined with a spatial model for uncovering bid-rigging cartels in Sweden. Our research is also related to papers analyzing the effect of bid rigging \citep[][]{Pesendorfer2000, Ishii2009} and to papers investigating the change in bidding patterns when bid rigging occurs \citep[][]{Porter1993, Porter1999}. As a further relevant approach, \citet{Chassang2020} investigate the occurrence of bid values with zero densities in the distribution of bids. Finding such gaps in the observed values of bids, especially between the first and the second lowest bid, might point to collusive behavior as bids with zero densities (between the two lowest bids) should not systematically occur in competitive markets.

The remainder of this study is organized as follows. Section 2 presents the bid-rigging cartels uncovered in Switzerland from which our data are drawn. Section 3 outlines the detection methods for flagging both complete and incomplete bid-rigging cartels. Section 4 applies our methods to a simulation of incomplete cartels based on data from the Ticino bid-rigging cartel and as well as to empirical data from the cases of See-Gaster and Strassenbau Graubünden. Section 5 concludes.

\section{Bid-Rigging Cartels and Data}

The Swiss Parliament revised the federal Cartel Act and introduced a sanction regime in April 2004, with an adaptation period of one year, alongside with a compliance program. This legislative change helped in initiating a change in the praxis towards economic harmful bid-rigging cartels. The same year, a strategic body of the federal administration for procurement of goods and services prepared a questionnaire for both procurement agencies and bidding firms. One result was that all surveyed persons declared to know the existence of bid rigging and 50\% to have a concrete experience in bid rigging.\footnote{We suspect this rate to be underestimated since the bidding firms participating in a bid-rigging conspiracy might have an incentive for misreporting.} At the end of 2004, COMCO began to investigate the Ticino cartel and rendered its decision in 2007. The Ticino cartel dissolved without sanctions since it stopped its illegal conduct precisely before April 2005, after having consumed the whole adaptation period. It however stressed the damages and mischiefs of a bid-rigging cartel with a price increase over 30\% \citep[see][]{Imhof2020}. In 2008, COMCO decided to put priority to fight bid rigging with a strategy involving three pillars: prosecution, prevention, and detection.

In the pillar prosecution, COMCO treated many bid-rigging cases subsequent to the Ticino cartel. Table \ref{decisions} lists the most important decisions of COMCO in bid-rigging cases and the sanctions for each case. Initially, COMCO rendered an important decision against bid rigging every other year. From 2015 on, however, COMCO rendered a decision each year and increased sanctioning, emphasizing the determination to prosecute bid-rigging conspiracies. This appeared necessary for disciplining firms and deterring the formation of bid-rigging cartels.

{\renewcommand{\arraystretch}{1.1}
\begin{table} [!htp]
\caption{Decisions of COMCO in bid-rigging cases} \label{decisions}
\begin{center}
\begin{tabular}{lccc}\hline\hline
Decisions of COMCO (excerpt)&Year&Sanctions in CHF mio.&Number of firms\\\hline
Road asphalting in Ticino&2007&--&17\\
Electric Installations Bern&2009&1,2&7\\
Road Construction and Civil Engineering Aargau&2011&7&18\\
Road Construction and Civil Engineering Zurich&2013&0,5&12\\
Tunnel Cleaning&2015&0,16&3\\
Road Construction and Civil Engineering See-Gaster&2016&5&8\\
Construction in Val Mustair&2017&--&5\\
Six short Decisions in Engadine&2017&1&12\\
Construction in Lower Engadine&2018&7,5&7\\
One short Decision in Engadine&2019&0,5&3\\
Road construction Graubünden&2019&11&12\\\hline\hline
\end{tabular}
\end{center}
\end{table}}

The sanctions presented in Table \ref{decisions} depend on the proven facts of misconduct in the various cases. In Switzerland, the sanctions depend on the definition of the relevant market, which is derived from the proven facts. Whenever COMCO only dismantles the existence of single agreements between firms, it defines the relevant market as being restricted to the tenders in which firms have settled the single agreements. However, whenever COMCO can prove a system agreement, which goes beyond the scope of a single tender implying contract allocation over time, then the relevant market comprises all market activities affected by bid rigging. For example, the Ticino cartel was a system agreement since all firms, active on the road construction and asphalting market, rigged all public and private contracts for more than five years. In such a case, COMCO considers all the revenues in the relevant market for sanctioning, not only in tenders for which COMCO has proved an agreement. The Ticino cartel went unsanctioned since it stopped its illegal conduct before April 2005, but if it had been sanctioned, the total fine would have amounted up to 30 million CHF. In the cases of road construction in Graubünden, building and civil engineering in Lower Engadine as well as road construction and civil engineering in See-Gaster, COMCO proved system agreements. Unsurprisingly, the sanctions are consistently higher than in the case of single agreements.

In parallel with the prosecution of bid rigging, COMCO repeatedly organized information sessions for procurement agencies in Switzerland to increase their awareness of bid rigging. Teaching procurement agencies how to recognize bid rigging is important since they are key players in organizing public procurement markets and may design procurement in a way that fosters competition and decreases the risk of bid rigging. If procurement agencies demonstrate that they care about competition and clearly stand against bid rigging, they send a clear deterrent signal toward bidders tempted to collude. Finally, procurement agencies are a source of information for COMCO in prosecuting bid-rigging cartels. The information and evidence gathered by procurement agencies can lead to the opening of an investigation. It is therefore advantageous for COMCO to make procurement agencies aware of relevant information of bid rigging so that they can transfer them to COMCO for assessment.

The last pillar concerns the detection of bid rigging. COMCO opens an investigation if there are reasonable grounds to assume the existence of a bid-rigging cartel. Compliance programs, whistleblowers and procurement agencies can provide insightful information leading to the opening of an investigation. However, COMCO decided to mitigate its dependency on those sources and started to develop statistical methods for detecting bid rigging based on screens \citep[see][]{Imhof2017a}. Based on the latter method, COMCO opened an investigation of bid rigging in the region of See-Gaster in 2013.

\citet{Imhof2017a} indicate that their screening method can detect partial cartels, as it succeeded in dismantling the quasi-complete bid-rigging cartel in the region of See-Gaster. However, as all firms in the road construction sector in that region were involved in the case, the term “\textit{partial cartel}” in their study does not imply incomplete bid-rigging cartels as in our paper. The cartel only concerned one region out of eight regions in the canton of St. Gallen (and not the whole canton). In our simulation that combines competitive bids with the Ticino cartel, we show that the method of \citet{Imhof2017a} is not suitable for flagging incomplete bid-rigging cartels. Considering the evolution of laws and practice of COMCO toward bid-rigging cartels, incomplete bid-rigging cartels occur more often than well-organized and complete cartels. Therefore, if COMCO desires to reinforce the detection pillar, it must continue to improve detection methods. The broader approach for flagging both incomplete and complete bid-rigging cartels proposed in this paper responds to that need and is likely of interest for competition agencies around the world.

In the empirical analyses, we use data from the three most important cases in Switzerland: the Ticino cartel, the cartel of See-Gaster and the asphalt cartel of Graubünden. After discussing procurement in Switzerland, we synthetize the key aspects of procurement data in Switzerland and each case below.

\subsection{Procurement Data}
Procurement agencies of cantons and cities in Switzerland follow the Agreement on Public Markets between cantons and their own cantonal laws for public procurement. A procurement agency can choose between four procedures: the open procedure, the procedure on invitation, the selection procedure and the discretionary procedure.\footnote{The selection procedure allows procurement agency to select and qualify a set of bidders for participating in a tender. This procedure is useful when bidders are too numerous as for example in architecture designing where hundreds of architects are interested in submitting to the project. However, such a high number of bidders are rarely a problem in the construction sector.} In the construction sector, a procurement agency generally uses either the open procedure or the procedure on invitation. The selection procedure is rare in the construction sector and when a procurement agency tenders a contract with a discretionary procedure, its application is restrictive since the laws authorize it only under specific conditions. The open procedure does not restrict the participation of submitting firms. In contrast, the procedure on invitation restricts participation as the procurement agency invites only a small number of firms to submit a bid, in general three to five firms. This changes the nature of competition, as the firms are aware of the restricted number of potential competitors.

A procurement agency announces future contracts along with the deadline for submitting bids (varying according to the procedure) in an official journal. If a firm is interested in submitting, the procurement agency provides the firm with all the relevant documents or information for the contract. Between the time of the announcement and the deadline, firms prepare their bids for submission. If they occur, then collusive agreements between firms are typically made during this period.

At a pre-announced date, the procurement agency gathers the incoming bids for the contract and opens them. It officially records all the bids received on time in a bid summary or so-called official record of the bid opening. The procurement agency registers the names of the firms, their addresses and their bids. After having written the official record of the bid opening, the procurement agency proceeds with the detailed examination of the bids. In the awarding of the contract, the agency does not only consider the price of the bids, but also other criteria as the quality, references, environmental or social aspects might play a role. However, as contracts are relatively homogeneous in the construction sector, especially in the road construction and the related civil engineering, the price remains the most important criterion for awarding the contract in practice. Furthermore, the differences in firms’ criteria other than the price are typically small. We therefore consider the procurement process as an almost first-price sealed-bid auction.

\subsection{The Ticino Cartel}
The Ticino cartel started in January 1999 and dissolved itself at the end of March 2005, precisely when the adaptation year to the new cartel Act, entered in force in April 2004, terminated. The cartel was well-organized with a convention of two pages explaining the rules to follow \citep[see][]{Imhof2020}. All firms active in the road construction sector participated in the cartel and rigged all public tenders and all private contracts above 20’000 CHF.\footnote{Approximatively 23'500 USA dollars given an exchange rate of 0.85 (indirect quotation) in March 2005.} The convention allocated contracts among cartel participants according to different criteria. The first criterion was revenue, putting cartel participants with many contracts recently attributed at the bottom of a priority list for allocating new contracts that was updated each week, and those with few contracts at the top. The geographic distance between the firm and the location of the contract was the second most important criterion and played an important role in the decision of allocating small contracts. Ties with private clients was an important criterion in the attribution of private contracts. In particular, cartel participants that had already produced a quote for some private client were privileged. After allocating contracts, cartel participants decided on the price of the bid that the designated winner by the cartel should submit. COMCO stated in its decision that the Ticino cartel roused prices by 30\% for contracts in the road construction and asphalting market.\footnote{For the decision of COMCO, see \textit{Strassenbeläge Tessin} (LPC 2008-1, pp. 85-112).}

We consider data from the cartel and the post-cartel periods. Table \ref{OvervTI} summarizes key information about contracts with four or more bids in our sample. We observe 149 tenders in the collusive period, whose value amounts up to 160 million CHF. In total, we recorded 974 bids for the collusive period, henceforth referred to as collusive bids. For the post-cartel period, we observe only 33 tenders, accounting for a value of 23 million CHF, in which firms submitted 222 competitive bids.

{\renewcommand{\arraystretch}{1.1}
\begin{table} [!htp]
\caption{Overview sample Ticino cartel} \label{OvervTI}
\begin{center}
\begin{tabular}{lc}\hline\hline
Tenders in the cartel period&149\\
Volume of the collusive tenders in mio CHF&160.7\\
Collusive bids&974\\
Tenders in the post-cartel period&33\\
Volume of the competitive tenders in mio CHF&22.79\\
Competitive bids&222\\\hline\hline
\end{tabular}
\end{center}
\end{table}}

\begin{figure}[!htp] \begin{center}
\includegraphics[height=8cm, width=14cm]{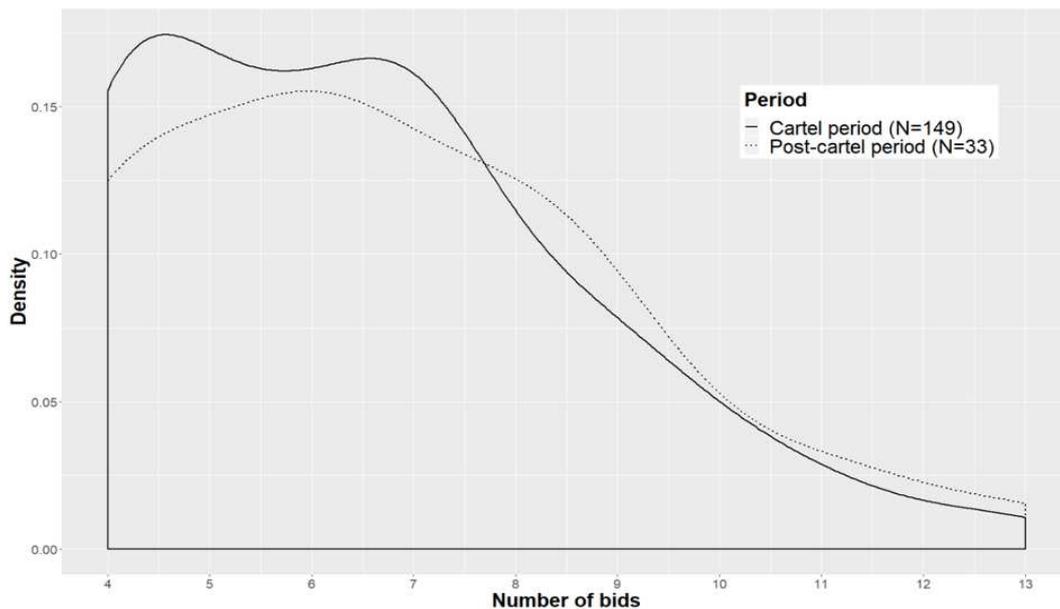}
\caption{Distribution of tenders for a predetermined number of bids for the Ticino data\label{densityTI}}
\end{center}
\end{figure}

Figure \ref{densityTI} visualizes the distribution of tenders for a predetermined number of bids. In either periods, most of the tenders consist of 4 to 8 bids (see also Table \ref{nbrbidTI} in Appendix). Table \ref{empiridistTI} provides the empirical distribution of the bids for each period. Both periods contain contracts of different values varying from several tens of CHF to up 3 to 5 million CHF. The mean and the median of the cartel period are superior to those of the post-cartel period. In either periods, the contract values exhibit higher means than medians, indicating a right-skewed distribution with outliers of comparably high contract values.

{\renewcommand{\arraystretch}{1.1}
\begin{table} [!htp]
\caption{Empirical distributions of collusive and competitive bids in the Ticino data} \label{empiridistTI}
\begin{center}
\begin{tabular}{lcc}\hline\hline
&Cartel period&Post-cartel Period\\\hline
Mean&1.08&0.69\\
Std&1.01&0.75\\
Min&0.02&0.04\\
Lower Q.&0.36&0.25\\
Median&0.78&0.44\\
Upper Q.&1.47&0.68\\
Max&4.85&2.95\\
N&149&33\\\hline
\hline
\end{tabular}
\end{center}
{\footnotesize Note: “Mean”, “Std”, “Min”, “Lower Q.”, “Median”, “Upper Q.”, “Max”, and “N” denote the mean, standard deviation, minimum, lower quartile, median, upper quartile, maximum, and number of observations, respectively.}
\end{table}}

\newpage

\subsection{The Cartel in See-Gaster}
COMCO opened its investigation against six firms in the region of See-Gaster mainly because of a statistical investigation based on the procurement data from 2004 to 2010, which were provided by the canton of St. Gallen \citep[see][]{Imhof2017a}.\footnote{Report release: See the decision \textit{Bauleistung See-Gaster: Verfügung vom 8. Juli 2016}, available on the following internet page: https://www.weko.admin.ch/weko/fr/home/actualites/dernieres-decisions.html.} Six months later, COMCO extended the investigation to two additional firms, undetected by \citet{Imhof2017a}. One firm did not bid in tenders of the canton of St. Gallen and had therefore not been detected previously. The second firm was small such that its participation in conspicuous contracts was insufficient for being flagged as potential cartel candidate.

In total, eight firms participated in bid-rigging conspiracies in the region of See-Gaster, including the district of See-Gaster in the canton of St. Gallen and the districts of March and Höfe in the canton of Schwyz.\footnote{See the decision \textit{Bauleistung See-Gaster: Verfügung vom 8. Juli 2016}, available on the following internet page: https://www.weko.admin.ch/weko/fr/home/actualites/dernieres-decisions.html.} Cartel participants regularly met once or twice per month. In their meetings, they mainly discussed about future contracts tendered, including road construction, asphalting and civil engineering, and exchanged their interest for them. Before each meeting, one cartel participant sent an actualized table to all others, listing all future contracts in the region of See-Gaster. Each cartel participant had a column and could put a star for each contract if interested in obtaining the contract or two stars in the case of very high interest.\footnote{See the decision \textit{Bauleistung See-Gaster}, R. 809 ff.} When the tender procedure for a contract started, the cartel typically designated the cartel participant who should win the contract. The allocation mechanism was based on the interests announced and fairness in terms of allocation among participants for insuring cartel stability.\footnote{See the decision \textit{Bauleistung See-Gaster}, R 587, R 608 and R 623.} Besides, when two cartel participants had put two stars for a specific contract, they possibly formed a consortium for sharing the contract and the other participants covered the consortium.\footnote{See the decision \textit{Bauleistung See-Gaster}, R 620 ff. and R. 645.}

The cartel took the decision of contract allocation during the meetings in which they discussed the list, but organized separate meetings for discussing the price of the bids.\footnote{See the decision \textit{Bauleistung See-Gaster}, R. 649 ff.} One reason for separate meetings is that not all cartel participants were interested in fixing the price since some did not necessarily participate in the tender. Second, discussions about price could have taken too much time such that the cartel preferred the designated winner to invite the other bidders to a separate meeting for discussing the price. COMCO found some evidence that the cartel from time to time used the mechanism of the mean for determining the bid of the designated winner,\footnote{See the decision \textit{Bauleistung See-Gaster}, R. 714 ff.} which implies that the latter had to submit either his own bid or the mean of all exchanged bids in the separate meetings. With such a mechanism, the designated winner had some incentive to provide a relatively high bid to influence the calculated mean in the separate meeting. All other cartel participants whose announced bids were below the mean or below the bid of the winner increased their bids to cover the designated winner. They generally ensured a minimal price difference of 2-3\% between the bid of the designated winner and their bids.\footnote{See the decision \textit{Bauleistung See-Gaster}, R. 714 ff. and R 718.}

Finally, the cartel also decided about contracts free for competitive bidding.\footnote{See the decision \textit{Bauleistung See-Gaster}, R 681 ff. and R. 815 ff.} This decision was also determined by the presence of external bidders. The more external bidders, the smaller was the incentive to collude because the chances of success were lower. This was the case for some contracts of high value, for which more non-cartel firms were interested in bidding. Sometimes, the cartel also tried including such external firms in the agreement.

In June 2009, the cartel stopped its illegal conduct after COMCO launched house searches in the canton of Aargau, which to a certain extent explained the breakdown of the cartel. In its decision, COMCO attested that the cartel had discussed more than 400 contracts in the region of See-Gaster from 2004 to 2009 with a value of 198 million CHF. COMCO also proved that they attempted to rig at least 200 contracts for a value of 67.5 million CHF.\footnote{See the decision \textit{Bauleistung See-Gaster}, R. 797 ff. and table 15.} In the rendering of its decision, COMCO sanctioned the involved firms for bid-rigging conspiracies with more than 5 million CHF. Two firms applied to the leniency program and two other firms settled an agreement to close the case. Four firms appealed against the decision.

\subsection{The Strassenbau Cartel in Graubünden}
The cartel in the canton of Graubünden was organized by the members of the local trade association for road construction. The members were firms active in road construction and asphalting. COMCO proved in its decision that the cartel participants met regularly in the investigated period from 2004 to the end of May 2010. The meetings called “allocation meeting” or “calculation meeting” were mainly held in the beginning of the year since the canton and the local municipalities tendered most of the contracts in spring.\footnote{See the decision \textit{Strassenbau Graubünden}, R. 139.} They discussed contracts for road construction and asphalting tendered by the canton of Graubünden and the local municipalities. Since the geography of Graubünden is profoundly marked by mountains and valleys, the cartel was divided into firms operating in the North and South, respectively.

In the North of Graubünden, the cartel mostly organized the meetings in the office of the most important mixing plant in the canton, and to a lesser extent in the offices of cartel participants. The meetings included either all of the 12 to 13 cartel participants\footnote{See the decision \textit{Strassenbau Graubünden}, R 247 ff.} or two different subgroups.\footnote{See the decision \textit{Strassenbau Graubünden}, R 195 ff.} In the South, the all in all six cartel participants\footnote{See the decision \textit{Strassenbau Graubünden}, R 248.} also organized such meetings, however, with alternating locations.

COMCO stated in its press release that the cartel decided upon the allocation of contracts based on a contingent determined for all the cartel participants in the canton of Graubünden.\footnote{See press release: https://www.newsd.admin.ch/newsd/message/attachments/58229.pdf.} The cartel allocated contracts according to the interests of each firm and fixed the price of the designated winner following a specific calculation method.\footnote{The online published decision \textit{Strassenbau Graubünden.} and the press release give for now no precision about the calculation method.} The price of the designated winner was in general above the minimal announced bid in the respective meeting. The calculation method therefore contributed to raise the price.

During the investigated period from 2004 to the end of May 2010, the cartel allocated 70 to 80\% of the total value of the cantonal and communal road construction contracts in North and South Graubünden among its participants. The cartel approximatively rigged 650 road construction contracts concerning in total 190 million of CHF of market volume.\footnote{See press release: https://www.newsd.admin.ch/newsd/message/attachments/58229.pdf.} The cartel ceased its illegal conduct in summer 2010 in both the North and South, since some firms decided to stop, mainly because arising concerns regarding the Cartel Act.\footnote{See the decision \textit{Strassenbau Graubünden}, R 197.}

\subsection{Data from the Cases See-Gaster and Graubünden}
We requested data on all bid summaries from the investigations of See-Gaster and Graubünden based on the Federal Act on Freedom of Information in the Administration (Freedom of information Act, FoIA).\footnote{https://www.admin.ch/opc/en/classified-compilation/20022540/index.html.} COMCO approved the request and transmitted us the data, referred to as Swiss data hereafter. They contain the bids, a running number for each contract, a dummy variable for each of the anonymized cartel participants and a dummy variable indicating whether the tender took place in the cartel period (taking the value 1 for cartel and 0 otherwise), a categorical variable for the contract type (taking the value 1 for contracts in road construction and asphalting, 2 for mixed contracts including road construction and civil engineering and 3 for civil engineering contracts), as well as an anonymized date and year. The first year in our sample begins with the value of 1, the last year ends with the value of 14. The first anonymized date equals 42 and the last 4886. To ensure the anonymization of the bids, COMCO multiplied them with a factor between 1 and 1.2. Such a transformation does not affect the calculation of the screens.

{\renewcommand{\arraystretch}{1.1}
\begin{table} [!htp]
\caption{Overview for the Swiss data} \label{OvervSWI}
\begin{center}
\begin{tabular}{lc}\hline\hline
Tenders with complete cartels&310\\
Volume of tenders with complete cartels in million CHF&111.74\\
Collusive bids in tenders with complete cartels&2031\\
Tenders with incomplete cartels&287\\
Volume of tenders with incomplete cartels in million CHF&114.73\\
Competitive bids in tenders with incomplete cartels&650\\
Collusive bids in tenders with incomplete cartels&1414\\
Competitive tenders&2398\\
Volume of competitive tenders&1735.91\\
Competitive bids in competitive tenders&13925\\
Tenders for road construction and asphalting&1389\\
Tenders for civil engineering&1286\\
Tenders for mixed contracts&273\\
\hline\hline
\end{tabular}
\end{center}
\end{table}}

Table \ref{OvervSWI} provides key information on the Swiss data. As for the data from Ticino, we only consider tenders with four bids or more. In total, there are 310 tenders with complete cartels with a total value of more than 110 million CHF and 2'031 bids submitted by the cartel participants. Furthermore, there are 287 tenders with incomplete cartels with a total value of more than 114 million CHF. In those tenders, cartel participants submitted 1'414 bids and external firms 650 bids. Finally, we observe 2'398 competitive tenders with a value of roughly 1'700 million CHF and 13'925 submitted bids.

\begin{figure}[!htp] \begin{center}
\includegraphics[height=8cm, width=14cm]{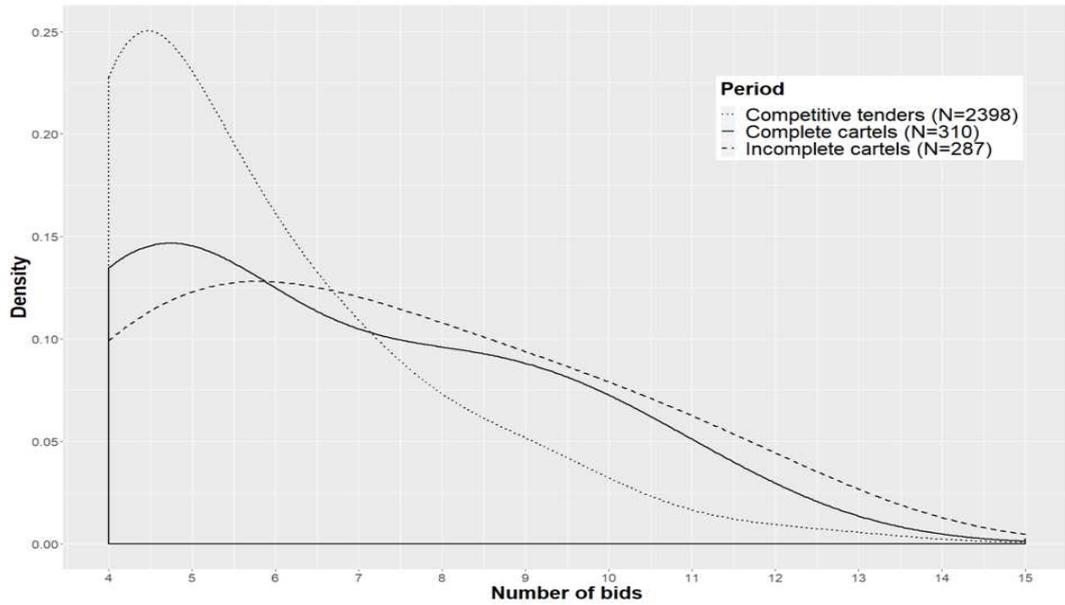}
\caption{Distribution of tenders for a predetermined number of bids in the Swiss data\label{densitySWI}}
\end{center}
\end{figure}

Figure \ref{densitySWI} visualizes the distribution of the number of bids per tender for complete cartels, incomplete cartels and competitive tenders, respectively. While tenders with four to seven bids dominate, there is also a sufficient number of tenders with eight or more bids (see also Table \ref{nbrbidSWI} in Appendix). Table \ref{empiridistSWI} depicts the empirical distribution of the bids for each type of tenders. The empirical distributions for tenders with complete cartels and with incomplete cartels are similar. This is, however, not the case for competitive tenders. That have many more contracts, varying from one thousand CHF to 148 million CHF. As for the data from the Ticino cartel, all empirical distributions of the bids are right-skewed (such that the mean is higher than the median), but stronger so for competitive tenders than for complete and incomplete cartels.

{\renewcommand{\arraystretch}{1.1}
\begin{table} [!htp]
\caption{Empirical distributions of bids in the Swiss data} \label{empiridistSWI}
\begin{center}
\begin{tabular}{lccc}\hline\hline
&Complete cartels&Incomplete cartels&Competitive tenders\\\hline
Mean&0.36&0.4&0.72\\
Std&0.36&0.47&3.81\\
Min&0.03&0.02&0.001\\
Lower Q.&0.16&0.12&0.13\\
Median&0.29&0.25&0.31\\
Upper Q.&0.44&0.50&0.66\\
Max&3.45&3.46&147.73\\
N&310&287&2398\\\hline
\hline
\end{tabular}
\end{center}
\par
{\footnotesize Note: “Mean”, “Std”, “Min”, “Lower Q.”, “Median”, “Upper Q.”, “Max”, and “N” denote the mean, standard deviation, minimum, lower quartile, median, upper quartile, maximum, and number of observations, respectively.}
\end{table}}

\section{Detection Methods}

This section outlines the suggested method to detect bid rigging. We first describe the concept of a random forest, the machine learning algorithm used for training and testing our predictive models for collusion \citep[see][]{Ho1995, Breiman2001}.
Second, we present in detail the screens that enter the algorithm as potential predictors. Third, we discuss five different predictive models applied to our data that differ in terms of included screens. Finally, we provide descriptive statistics for two important screens in each dataset.

\subsection{Random Forest}
We use the random forest as machine learning algorithm for predicting collusive and competitive tenders. In our data, the outcome takes the value 1 for collusive tenders, including both incomplete and complete bid-rigging cartels, and 0 for competitive tenders. Note that we intentionally do not distinguish between incomplete and complete cartels, as our aim is to construct a  reliable detection method for any form of bid rigging. Tenders are therefore either collusive or competitive.

Machine learning requires randomly splitting the data into the so-called training data, used for developing the predictive model, and the test data, used for evaluating the performance of the model. We randomly split the data such that the training and test data consist of 75\% and 25\% of the observations, respectively. The random forest is a so-called ensemble method that averages over multiple decision trees for predicting the outcome. Tree-based methods split the predictor space (\textit{i.e.} the values the screens might take) of the training data recursively into a number of non-overlapping regions. Each split aims at maximizing homogeneity of the dependent variable within the newly created regions according to a goodness of fit criterion like the Gini coefficient. The latter measures the average gain in purity (or homogeneity) of outcome values when splitting and is popular for binary variables like our collusion dummy. Splitting is continued until the decision tree reaches a specific stopping rule, \textit{e.g.} a minimum amount of observations in a region or maximum number of splits. Tree-based prediction of bid rigging (1) or competition (0) is based on whether collusive or competitive tenders dominate in that region which contains the values of the screens for which the outcome is to be predicted.

Importantly, there exists a bias-variance trade-off in out of (training) sample prediction when using such tree-based (and other machine learning) methods w.r.t. model generality. More splits reduce the bias and increase the flexibility of the model specification, however, at the cost of a greater variance in unseen data (not used for training), as the test sample, due to smaller regions. The issue of a too large variance can be mitigated by repeatedly drawing many subsamples from the initial training data and estimating the predictive model, \textit{i.e.} the tree (or splitting) structure, in each of the newly generated samples. In fact, a random forest consists of predicting the outcome by the majority rule across the individual trees, \textit{i.e.} based on whether the majority of the trees estimated in the various subsamples predict collusion or competition for specific values of the screens. A further feature of the random forest is that at each splitting step in a specific subsample, only a random subsample of possible predictors (\textit{i.e.} screens) is considered, which reduces the correlation of tree structures across subsamples and thus further reduces the variance of prediction. In our application, we use the randomForest package by \citet{Breiman2018} for the statistical software R with growing 1’000 trees to estimate the predictive models in the training data and assess their performance in the test data based on the correct classification rate.

We note that we repeat the random sample splitting into 75\% training and 25\% test data as well as the assessment of the predictive performance in the latter 100 times. Our reported correct classification rate corresponds to the average of the correct classification rates across the 100 repetitions. This procedure likely entails a smaller variance in the estimation of the correct classification rate than when relying on a single random data split.

\subsection{Predictors}
Screens are statistics that permit analyzing economic data with the aim to flag anomalous outcomes that suggest potential anticompetitive issues. The literature usually differentiates structural from behavioral screens in cartel detection \citep[see][]{Harrington2006a, OECD2014, Froeb2014}. Structural screens focus on the factors facilitating the emergence of collusive agreements and help to identify markets in which collusion is more likely. Among those factors, one distinguishes between market structure, demand-related and supply-related factors \citep{OECD2014}. In contrast, behavioral screens empirically measure the behavior of market participants and assess whether the observed behavior significantly departs from competition for flagging it as a potential issue worth scrutinizing deeper. Following \citet{Huberimhof2019} we propose using various descriptive statistics as screens and combining them with machine learning, however, not only for uncovering complete, but also incomplete bid-rigging cartels.\footnote{In contrast to the context of causal inference, causality goes from the dependent variable (collusion or competition) to the predictors (screens) rather than the other way around. The incidence of collusion as explanatory variable affects the distribution of the bids and thus the screens in causal terms. As in \citet{Huberimhof2019} our prediction problem consists of analyzing a reverse causality. By investigating the screens we infer the existence of their cause: Collusion.} We consider three classes of screens: Variance, asymmetry and uniformity.

As variance screens, we implement the coefficient of variation (CV) and the kurtosis statistic (KURTO), as suggested by \citet{Huberimhof2019} and \citet{Imhof2020}. In addition, we also implement the spread (SPD) of the distribution of the bids as screen.

The coefficient of variation is widely discussed in the literature \citep[see][]{Abrantes2006, Esposito2006, Jimenez2012, Abrantes2012, Imhof2020} and is defined as the standard deviation divided by the arithmetic mean of all bids submitted in a tender:

\begin{equation}\label{eqcvMLS}
CV_{t}=\frac{s_{t}}{\bar{b}_{t}},
\end{equation}
where $s_{t}$ is the standard deviation and $\bar{b}_{t}$ is the mean of the bids in some tender $t$.
As coordination and manipulation of bids by cartel participants might affect the convergence in the distribution of the bids, we also consider the following kurtosis statistic as screen:

\begin{equation} \label{kurtoMLS}
KURTO_{t}=\frac{n_{t}(n_{t}+1)}{(n_{t}-1)(n_{t}-2)(n_{t}-3)}\sum_{i=1}^{n_{t}}(\frac{b_{it}-{\bar{b}_{t}}}{s_{t}})^{4} - \frac{3(n_{t}-1)^3}{(n_{t}-2)(n_{t}-3)},
\end{equation}

where $b_{it}$ denotes the bid $i$ in tender $t$, $n_{t}$ the number of bids in tender t, $s_{t}$ the standard deviation of bids, and $\bar{b}_{t}$ the mean of bids in that tender. Note that we calculate the kurtosis statistic only for tenders with four bids or more. Furthermore, we estimate the spread using the following formula:

\begin{equation} \label{SPD}
SPD_{t}=\frac{b_{max,t}-b_{min,t}}{b_{min,t}},
\end{equation}

where $b_{max,t}$ denotes the maximum bid and $b_{min,t}$ the minimum bid in some tender $t$.

Bid rigging may produce asymmetries in the distribution of the bids and we for this reason implement the following cover-bidding screens as in \citet{Huberimhof2019}: The percentage difference (DIFFP), the skewness (SKEW), the relative distance (RD) and the normalized distance (RDNOR). We also add an alternative measure for calculating the relative difference, the alternative relative distance (RDALT).

It seems plausible that cartel participants manipulate the difference between the lowest and second lowest bids to secure contract allocation with the designated winner by the cartel. To analyze the difference between the two lowest bids, we use the following formula to calculate the percentage difference:

\begin{equation} \label{DiffPerMLS}
DIFFP_{t}=\frac{b_{2t}-b_{1t}}{b_{1t}},
\end{equation}
where $b_{1t}$ is the lowest bid and $b_{2t}$ the second lowest bid in some tender $t$. We also consider the absolute difference between the first and second lowest bids $ D_{t}=b_{2t}-b_{1t}$ in the empirical analysis.

The manipulation of the bids by cartel participants might simultaneously affect both the difference between the first and second lowest bids and the differences across the losing bids. Following Imhof et al. (2018), we calculate a relative distance (relative to a measure of dispersion) in a tender by dividing the difference between the first and second lowest bids by the standard deviation of the losing bids:

\begin{equation} \label{RDTMLS}
RD_{t}=\frac{b_{2t}-b_{1t}}{s_{losing bids,t}},
\end{equation}
where $b_{1t}$ denotes the lowest bid, $b_{2t}$ the second lowest bid, and $s_{t, losing bids}$ the standard deviation calculated among the losing bids in some tender $t$. In \citet{Huberimhof2019} the RD was in terms of predictive power outperformed by the difference between the second and first lowest bids divided (or normalized) by the average of the differences between all adjacent bids. We also consider this normalized distance in our study:

\begin{equation} \label{RDNORMTMLS}
RDNOR_{t}=\frac{b_{2t}-b_{1t}}{\frac{(\sum_{i=1,j=i+1}^{n_{t}-1}b_{jt}-b_{it})}{n_{t}-1}},
\end{equation}
where $b_{1t}$ is the lowest bid, $b_{2t}$ the second lowest bid, $n_{t}$ is the number of bids and $b_{it}$, $b_{jt}$ are adjacent bids (in terms of price) in tender $t$, with bids being ordered increasing order.

We consider a further alternative measure for the relative distance, initially suggested by \citet{Imhof2017a}:
\begin{equation} \label{ALTRDTMLS}
RDALT_{t}=\frac{b_{2t}-b_{1t}}{\frac{(\sum_{i=2,j=i+1}^{n_{t}-1}b_{jt}-b_{it})}{n_{t}-2}},
\end{equation}
where $b_{1t}$ is the lowest bid, $b_{2t}$ the second lowest bid, $n_{t}$ is the number of bids and $b_{it}$, $b_{jt}$ are adjacent losing bids in a tender $t$, with bids being ordered increasing order.
In contrast to the normalized distance, the mean of the differences in the denominator is calculated using only losing bids. Furthermore, bid manipulation might affect the symmetry of the distribution of bids. We therefore include the skewness as screen:

\begin{equation} \label{skewMLS}
SKEW_{t}=\frac{n_{t}}{(n_{t}-1)(n_{t}-2)}\sum_{i=1}^{n_{t}}(\frac{b_{it}-{\bar{b}_{t}}}{s_{t}})^{3},
\end{equation}
where $n_{t}$ denotes the number of the bids, $b_{it}$ the $i^{\textrm{th}}$ bid, $s_{t}$ the standard deviation of the bids, and $\bar{b}_{t}$ the mean of the bids in tender $t$.

Finally, we consider the nonparametric Kolmogorov-Smirnov statistic (KS) for verifying if the bids in a tender follow a uniform distribution:

\begin{equation} \label{kolmostat}
D_{t}^{+}=max_{i}(x_{it}-\frac{i_{t}}{n_{t}+1}), D_{t}^{-}=max_{i}(\frac{i_{t}}{n_{t}+1}-x_{it}),KS_{t}=max(D_{t}^{+},D_{t}^{-}),
\end{equation}

where $n_t$ is the number of bids in a tender, $i_t$ the rank of a bid and $x_it$ the standardized bid for the $i^{\textrm{th}}$ rank in tender $t$. The standardized bids $x_it$ are the bids $b_it$ divided by the standard deviation of bids in tender t to facilitate the comparison of tenders with different contract values. We would suspect that competitive bids are closer to a uniform distribution than collusive bids such that the statistic should generally differ across cartels and competitive periods.

In incomplete cartels, competitive bidders distort the statistical signals produced by bid rigging in the distribution of bids in a tender. Therefore, the screens can fail to recognize bid rigging if calculated for all bids. We circumvent that distortion by calculating the screens not (only) for all bids in a tender, but for all possible subgroups of three and four bids. Table \ref{examSUBgroup} gives the number of possible subgroups of three or four bids, respectively, when the total number of bids in a tender varies between four to ten bids.

{\renewcommand{\arraystretch}{1.1}
\begin{table} [!htp]
\caption{Example of possible subgroups for three and four bids in a tender} \label{examSUBgroup}
\begin{center}
\begin{tabular}{ccc}\hline\hline
Bids in a tender&Subgroups formed with three bids&Subgroups formed with four bids\\\hline
4&4&1\\
5&10&5\\
6&20&15\\
7&35&35\\
8&56&70\\
9&84&126\\
10&120&210\\\hline\hline
\end{tabular}
\end{center}
\end{table}}

For instance, in a tender with a total number of six bids, we calculate the same screen but for different 15 subgroups containing four bids and for 20 different subgroups containing three bids. In each tender, we then include summary statistics for each screen: the mean, the median, the minimum and the maximum of the respective screen across the various subgroups of three or four bids. These summary statistics are themselves used as predictive screens for flagging collusive and competitive tenders and permit comparing tenders with different numbers of bids. We subsequently exemplify the computation of such screens by means of the coefficient of variation for subgroups formed on four bids.

The mean of all coefficients of variation calculated for subgroups of four bids in each tender is:

\begin{equation}\label{eqmean}
MEAN4CV_{t}=\sum_{s=1}^{S_{t}}(\frac{s_{st}/\bar{b}_{st}}{S_{t}}),
\end{equation}

where $s$ and $t$ denote the indices for some sub-group $s$ and some tender $t$, respectively, $S_{t}$ is the number of all the possible subgroups of four bids in tender $t$ and $s_{st}$ and $\bar{b}_{t}$ are the standard deviation and the mean of the bids, respectively. Likewise, the minimum and maximum of coefficients of variation across subgroups in a tender correspond to, respectively:

\begin{equation}\label{eqmin}
MIN4CV_{t}=min_{s}{\frac{s_{st}}{\bar{b}_{st}}},
\end{equation}

\begin{equation}\label{eqmax}
MAX4CV_{t}=max_{s}{\frac{s_{st}}{\bar{b}_{st}}},
\end{equation}

In order to calculate the median for subgroups of four bids in each tender, define the coefficient of variation in subgroup $s$ and tender $t$ as $CV_{st}=\frac{s_{st}}{\bar{b}_{st}}$ and order the coefficients in so that $$CV_{1t} \leq CV_{2t}\leq ... \leq CV_{st} \leq ... \leq CV_{S_{t}t} $$. If the number of subgroups $S_{t}$ in a tender is uneven, the median of the coefficient of variation in tender $t$ is calculated as follows:

\begin{equation}\label{eqmed1}
MEDIAN4CV_{t}=CV_{(S_{t}+1)/2,t},
\end{equation}

If the number of subgroups is even, the median corresponds to:

\begin{equation}\label{eqmed2}
MEDIAN4CV_{t}=\frac{CV_{S_{t}/2,t}+CV_{S_{t}/2+1,t}}{2},
\end{equation}

These approaches are applied to all the screens discussed above across different tenders.
Note also that we do not calculate screens for subgroups of two bidders because of the impossibility to calculate some screens as the RD, RDALT, RDNOR, KURTO, or SKEW. Moreover, cartel participants were in general more than two in tenders characterized by incomplete cartels. We also renounce calculating screens for sub-groups of five bidders or more. It makes sense only for tenders with six bids and more to use screens calculated for subgroups of five bidders. Using tenders with six bids or more would have restricted too much our sample and limited the application of our suggested methods in other cases. Calculating screens for subgroups of more than five bids would have only increased such limitation.

\subsection{Model Specification}

In the empirical analyses, we consider five different predictive models that vary in terms of screens considered, as well as the method of \citet{Imhof2017a}. Model 1 only applies screens calculated for all bids in a tender (rather than subgroups) as in \citet{Huberimhof2019}, but extends the set of predictors compared to the latter study by including the relative measure for the alternative distance (RDALT), the spread (SPD) and the Kolmogorov-Smirnov statistic (KS). In total, we use nine predictors and exclude any screens based on the absolute bid value to consider only scale-invariant screens in model 1. The results in \citet{Huberimhof2019} suggest that scale-invariant screens work well for predicting collusive and competitive tenders and that the value added of statistics that are sensitive to the scale is limited. Furthermore, the transfer of a method to other cases (with possibly different contract values) is easier if screens are scale invariant. Model 1 therefore represents the baseline approach for detecting bid-rigging cartels based on focusing on scale-invariant screens at the tender level.

In contrast, model 2 exclusively includes the summary statistics (mean, minimum, maximum, median) of the screens, which are calculated for all possible subgroups of three bids in a tender, as predictors. In total, we consider 32 of these summary statistics applied to eight screens, namely all of model 1 but the kurtosis (KURTO) which requires at least four bids. Model 3 uses summary statistics of the nine screens for all possible subgroups of four bids in a tender, thus in all 36 predictors (now including the kurtosis). Model 4 considers all predictors included in models 1, 2 and 3, this results in 77 statistics in total. Finally, model 5 additionally includes three screens that are based on absolute bid values (and thus, not scale-invariant) as well as the number of bids in a tender (NBRBIDS),\footnote{The motivation for including the number of bids is that it might be easier to settle an agreement in a tender with few bidders than with many.} thus 81 predictors in total. The three value-based screens are the mean bid in a tender as proxy for the contract value (MEANBIDS), the standard deviation of the bids in a tender (STDBIDS) and the absolute difference between the first and the second lowest bids (D).

\subsection{Descriptive Statistics for Predictors}

In the appendix, we present the Tables of descriptive statistics for all the different samples used in the empirical analyses for both the Ticino simulation and the Swiss data. We review here the most important key information drawn from the descriptive statistics for the coefficient of variation (CV) and the normalized distance (RDNOR). Similar interpretation can be drawn with other screens.

For the Ticino cartel, the coefficient of variation exhibits a mean of 3.25 and a median of 2.97 with a low standard deviation of 1.18 (see Table \ref{cartelDESCti}). This contrasts with the post-cartel period in which the mean and the median of the coefficient of variation amount to 9.51 and 8.49, respectively, with a larger standard deviation of 5.38 compared to the collusive period (see Table \ref{postcartelDESCti}). If a great majority of observations in the cartel period is below 3.83, we only find few CVs below 5.65 in the post-cartel period considering the upper and lower quartile, respectively. For the Swiss data, we find similar values for the cartel period with a mean of 3.66, a median of 3.29 and a standard deviation of 2.09 (see Table \ref{cartelCOMPDESCswiss}). All of them contrast with the values found for the post-cartel period (competitive tenders) in the Swiss data with a mean of 10.12, a median of 8.45 and a standard deviation of 7.89 (see Table \ref{competDESCswiss}). Note that we select in the following empirical analyses only competitive tenders with an anonymised year superior or equal to 8. Since collusive tenders are absent in the anonymised years superior or equal to 8, we conclude that both bid-rigging cartels has collapsed in this period called post-cartel period. It ensures that a competitive tender in the post-cartel period is really a "competitive" one.

If we look at the coefficient of variation for the incomplete bid-rigging cartel in sample 1 of the Swiss data, the CV is affected by the presence of competitive bids with a mean of 7.79, a median of 6.79 and a standard deviation of 3.89 (see Table \ref{incompleteDESCswiss1}).
Looking more precisely at the minimum of all coefficients of variation calculated for subgroups of four bids in a tender (MIN4CV), we find a mean of 3.16, a median of 2.26 and a standard deviation of 2.97 for the incomplete bid-rigging cartels in sample 1 (see Table \ref{incompleteDESCswiss1}). However, the MIN4CV for the competitive tenders exhibits higher values with a mean of 6.24, a median of 4.49 and a standard deviation of 6.77 (see Table \ref{competDESCswiss}). Noteworthy, the differences are weaker for the maxima of all coefficients of variation calculated for subgroups of four bids (MAX4CV), between incomplete cartels in sample 1 (mean of 10.63, median of 9.43 and a standard deviation of 5.46 in Table \ref{incompleteDESCswiss1}) and competitive tenders (mean of 12.14, median of 10.13 and a standard deviation of 9.73 in Table \ref{competDESCswiss}). This example is crucial to understand the benefit delivered by summary statistics of the screens. Even if the maxima of the coefficient of variation is high in both cases of incomplete bid-rigging cartels and competition, the minima notably diverge and could be used to differentiate between competition and collusion.

The normalized distance (RDNOR) take higher values in collusive periods than in competitive periods. For example, the RDNOR exhibits a mean of 2.93 and a median of 2.72 with a standard deviation of 1.35 for the Ticino cartel (see Table \ref{cartelDESCti}). In the post-cartel period, the values of the RDNOR are lower with a mean of 1.02, a median of 0.74, and a standard deviation of 0.80 (see Table \ref{postcartelDESCti}). Although less notable, we find a divergence in the Swiss data between collusive tenders (with a mean of 1.38, a median of 1.24 and a standard deviation of 0.79 in Table \ref{cartelCOMPDESCswiss}) and competitive tenders (with a mean of 1.04, a median of 0.87 and a standard deviation of 0.82 in Table \ref{competDESCswiss}). We find similar values for the minima of the normalized distance (MIN4RDNOR) between incomplete bid-rigging cartels in sample 1 (mean of 0.37, median of 0.15,standard deviation of 0.54 in Table \ref{incompleteDESCswiss1}) and competitive tenders in the Swiss data (with a mean of 0.51, a median of 0.29 and a standard deviation of 0.56 in Table \ref{competDESCswiss}). The values are more divergent for the maxima (MAX4RDNOR) between both types of tenders. For the incomplete bid-rigging cartels in sample 1, we observe a mean of 2.18, a median of 2.37 and a standard deviation of 0.68 in Table \ref{incompleteDESCswiss1} contrasting with competitive periods, which exhibit a mean of 1.62, a median of 1.74 and a standard deviation of 0.81 in Table \ref{competDESCswiss}. The result indicates that the maxima of the RDNOR could be used to discriminate between incomplete bid-rigging cartels and competition.

\section{Flagging Incomplete Bid-rigging Cartels}

\subsection{The Ticino Simulation}
We use the data from the Ticino cartel to investigate how the predictive models presented above perform in detecting bid-rigging cartels when competitive bids distort the statistical pattern produced by bid rigging. Since the Ticino cartel was complete, we use the data from the competitive periods to simulate competitive bids and progressively add them to the collusive tenders, creating five additional datasets for the cartel period. The first dataset includes only one simulated competitive bid in each collusive tender, the second two, and the fifth dataset includes five simulated competitive bids. This stepwise approach permits investigating how different levels of partial collusion affect the performance of each model.

We generate simulated bids from competitive bids using the following formula:

\begin{equation}\label{eqnormbid}
b_{t,simulated}=\bar{b}_{t}(1+\frac{b_{i,drawcomp}-\bar{b}_{drawcomp}}{\bar{b}_{drawcomp}}),
\end{equation}

$i$ and $t$ are indices for bids and tenders, respectively, $\bar{b}_{t}$ is the mean bid of tender t (without the simulated bid), while $\ b_{i,drawcomp}$ and $\bar{b}_{drawcomp}$ are the bid the mean bid randomly drawn from competitive tenders, respectively. We simulate competitive bids to be added to collusive tenders in four steps. First, we generate the normalized mean differences for all bids in competitive tenders. Normalizing by the mean rather than the standard deviation avoids losing information about the dispersion of the bids. In a second step, we randomly draw normalized mean differences with replacement from their empirical distribution in the competitive period and randomly assign them a collusive tender. In a third step, we multiply the normalized mean difference with the mean bid in tender $t$. In a fourth step, we add the mean bid of the assigned collusive tender to the normalized mean difference of the competitive in order to simulate a competitive bid in a tender with bid rigging.

We end up with seven different datasets for the Ticino cartel: The dataset of the post-cartel period including only competitive bids, the dataset of the collusive period including only collusive bids and five different datasets including the collusive tenders with one to five competitive bids in each tender.

We verify if the simulated competitive bids are similar to the competitive bids of the post-cartel period. Since we generate five simulated competitive bids for each tender in the collusive period, we calculate the screens for those five simulated competitive bids only. We test if the screens based on the simulated competitive bids are statistically significantly different from the competitive bids of the post-cartel period. The results presented in the Appendix show that most statistical tests do not reject the null hypothesis of no distributional differences, such that our simulation process adequately generates competitive bids.

We first apply the screening methods of \citet{Imhof2017a}, using their benchmarks (rather than machine learning) for classifying conspicuous tenders,\footnote{Tenders with a CV below 6 and a RD above 1 are classified as conspicuous.} and find a correct classification rate of 84.8\% in the test data in the absence of competitive bids, see Table \ref{corrclassTI}. However, when adding one simulated competitive bid to the collusive tenders, the correct classification rate falls to 66.7\%. It continuously decreases in the number of competitive bids added such that with five competitive bids in the collusive tenders, the method suggested by \citet{Imhof2017a} does not perform better than throwing a coin. As expected, the collusive tenders exclusively drive the decrease in the overall correct classification rate. With the addition of only one competitive bid, the correct classification rate among collusive tenders decreases to 48.5\%. With five competitive bids, it decreases to 15.2\%. This suggests that while the screening method of \citet{Imhof2017a} might be well suited for detecting complete bid-rigging cartels, it seems inappropriate for flagging incomplete bid-rigging cartels. This is explained by too many false negative predictions. Note that concerning the competitive tenders, the correct classification rate of 84.8\% remains unaffected since the sample of the post-cartel period does not vary across the various simulations.

We turn now to the results for model 1, inspired by \citet{Huberimhof2019}, but using a larger set of screens and the random forest as machine learner. Applied to the Ticino simulation, we find that the correct classification rate of model 1 shrinks between 5.3 and 11.3 percentage points depending on the number of competitive bids added to the collusive tenders. Model 4 includes the screens calculated in all possible subgroups of three and of four bids in a tender, and with the change of the correct classification rate varying between -6.3 and +3.1 percentage points with the occurrence of competitive bids. Noticeably, the correct classification rate of model 4 outperforms that of model 1 in all simulated datasets. When competitive bids are absent, model 1 performs slightly better than model 4, while, model 4 outperforms model 1 by 10.1 and 10.3 percentage points when including four and five competitive bids, respectively. This result illustrates the gain of our approach based on calculating the screens for subgroups of bids in a tender in terms of statistical power. Considering the error rate, it amounts to 24.2\% for model 1 such that almost one tender out of four is incorrectly classified. In model 4, the  error rate is only 13.9\%. Such a decrease of 42.6\% in the error rate when compared to model 1 is substantial, especially when considering the legal consequences for firms being flagged as a potential cartel.

Models 2 and 3 include summary statistics applied to the screens calculated in subgroups of three and four bids, respectively. On average, model 2 performs slightly better than model 3, but the correct classification rates are very similar. The maximum difference in (overall) correct classification rates across model 4 and models 2 or 3 amounts to 2.2 percentage points in the sample with three competitive bids. Overall, the correct classification rates of models 2 and 3 thus hardly differ from model 4.

\begin{table} [!htp]
\caption{Correct classification rate for the Ticino simulation} \label{corrclassTI}
\begin{center}
\begin{tabular}{clccccc}\hline\hline
Comp.B &Tenders&Im.&M1&M2&M3&M4\\\hline
\multirow{3}{1cm}{\centering 0}&All&0.848&0.835&0.832&0.834&0.830\\
&Comp.&0.848&0.816&0.811&0.806&0.811\\
&Coll.&0.848&0.863&0.861&0.87&0.859\\\hline
\multirow{3}{1cm}{\centering 1}&All&0.667&0.722&0.766&0.756&0.767\\
&Comp.&0.848&0.701&0.781&0.739&0.777\\
&Coll.&0.485&0.758&0.765&0.788&0.769\\\hline
\multirow{3}{1cm}{\centering 2}&All&0.652&0.752&0.819&0.786&0.802\\
&Comp.&0.848&0.751&0.856&0.786&0.832\\
&Coll.&0.455&0.758&0.79&0.803&0.784\\\hline
\multirow{3}{1cm}{\centering 3}&Overall&0.576&0.782&0.826&0.795&0.817\\
&Comp&0.848&0.796&0.862&0.807&0.838\\
&Coll.&0.303&0.775&0.804&0.793&0.806\\\hline
\multirow{3}{1cm}{\centering 4}&All&0.561&0.727&0.808&0.832&0.828\\
&Comp.&0.848&0.713&0.843&0.83&0.837\\
&Coll.&0.273&0.747&0.777&0.832&0.819\\\hline
\multirow{3}{1cm}{\centering 5}&All&0.500&0.758&0.871&0.871&0.861\\
&Comp.&0.848&0.745&0.885&0.879&0.887\\
&Coll.&0.152&0.778&0.837&0.862&0.837\\
\hline\hline
\end{tabular}
\end{center}
\par
{\footnotesize Note: "Comp.B", "Tenders", "Im.", "M1", "M2", "M3" and "M4" denote the number of competitive bids in the collusive tenders, the type of tenders, the results produced by the screening methods of \citet{Imhof2017a}, model 1, model 2, model 3 and model 4, respectively. For the type of tenders, "All", "Comp." and "Coll." denote the prediction for all types of tenders, the prediction for the competitive tenders and the prediction for the collusive tenders, respectively.}
\end{table}

We examine variable importance in the random forest for predicting collusive and competitive tenders in each dataset according to the mean decrease in Gini index (hereafter: MDG) when omitting the respective predictor, which ranks variables according to their predictive power. It, however, does not allow for a direct comparison between models since the MDG depends on the number of predictors. As we use less variables in model 1 than in model 4, the MDG of the former model is therefore higher. For each dataset and models 1 to 4, we depict the five most important variables in Table \ref{ImpPredTI}.

We observe for model 1 almost the same important predictors in all six data sets (zero to five competitive bids), namely the Kolmogorov-Smirnov statistic (KS), the coefficient of variation (CV), the kurtosis statistic (KURTO) and the normalized distance (RDNOR). For zero or one competitive bid, also the spread (SPD) is important, which is replaced by the alternative relative distance (RDALT) having more than one competitive bid. The order of importance changes with the number of competitive bids. With zero or one competitive bid, the Kolmogorov-Smirnov statistic (KS), the coefficient of variation (CV) and the kurtosis statistic (KURTO) are the most relevant variables with a MDG larger than three. In the presence of two or more competitive bids, the normalized distance (RDNOR) and the alternative distance (RDALT) exhibit a greater predictive power alongside with the Kolmogorov-Smirnov (KS) and the kurtosis statistic (KURTO).

In models 2 to 4, the Kolmogorov-statistic (KS), the spread (SPD), the difference in percentage (DIFFP) and the coefficient of variation (CV) are the best predictors. Contrary to model 1, screens based on the relative distance (RDNOR, RD and RDALT) or the skewness (SKEW) are less important, while the difference between the first and second lowest bids in percentage (DIFFP) are more important. As in model 1, the order of the importance changes with the number of competitive bids. In models 2 to 4 for zero and one competitive bid, the Kolmogorov-statistic (KS) is the most important variable. Different, in the data sets with four and five competitive bids the difference in percentage (DIFFP) comes first.

For models 2 to 4, the set of important variables changes with the number of competitive bids. The median or mean of certain screens calculated in subgroups of three and four bids in a tender is most predictive in the presence of zero or one competitive bids, while the minima and the maxima of screens dominate under a larger number of competitive bids. Intuitively, the minimum or the maximum of some screen likely excludes competitive bids if the latter distort the distribution of collusive bids and should thus be relatively more predictive as the number of competitive bids increase. We also note that for three or more competitive bids, the random forest mainly selects screens calculated in subgroups of four bids in a tender (rather than three).

To sum up the results of the Ticino simulation, we find that our approach, based on screens for subgroups of three and four bids in a tender, is able to flag bid-rigging cartels even when we add competitive bids. When the number of competitive bids increases, the random forest puts more weight on minima or the maxima of screens across subgroups. For models 1 to 4, the algorithm selects a mix of screens from the three groups (variance, asymmetry and uniformity).

\begin{table} [!htp]
\caption{Important predictors for the Ticino simulation} \label{ImpPredTI}
\begin{center}
\begin{tabular}{clclclclc}\hline\hline
&\multicolumn{2}{c}{Model 1}&\multicolumn{2}{c}{Model 2}&\multicolumn{2}{c}{Model 3}&\multicolumn{2}{c}{Model 4}\\
Comp.B&IV&MDG &IV&MDG &IV&MDG &IV&MDG \\\hline
\multirow{5}{1cm}{\centering 0}
&KS&4.63&MEAN3KS&2.33&MEAN4KS&2.88&MEAN4KS&1.56\\
&CV&4.6&MEDIAN3SPD&1.88&MEDIAN4SPD&1.89&MEAN3KS&1.37\\
&KURTO&3.37&MEAN3SPD&1.85&MEAN4SPD&1.86&MEDIAN3SPD&0.99\\
&SPD&2.42&MEAN3DIFFP&1.78&MEAN4CV&1.83&MEAN3SPD&0.97\\
&RDNOR&2.36&MEAN3CV&1.74&MEDIAN4KS&1.82&MEAN3DIFFP&0.97\\
\hline
\multirow{5}{1cm}{\centering 1}
&KS&3.88&MEAN3KS&1.92&MEAN4KS&2.08&MEAN3KS&1.16\\
&KURTO&3.65&MEDIAN3SPD&1.18&MEDIAN4SPD&1.15&MEAN4KS&1.08\\
&CV&3.61&MEAN3SPD&1.17&MEAN4SPD&1.13&MEDIAN3SPD&0.65\\
&SPD&2.96&MEDIAN3KS&1.13&MEDIAN4CV&1.1&MEAN3SPD&0.65\\
&RDNOR&2.72&MEAN3CV&1.08&MIN4SPD&1.08&MEDIAN3KS&0.63\\
\hline
\multirow{5}{1cm}{\centering 2}
&RDNOR&4.2&MEAN3KS&1.45&MIN4DIFFP&1.45&MIN3DIFFP&0.82\\
&RDALT&3.37&MIN3DIFFP&1.4&MEAN4KS&1.4&MEAN3KS&0.81\\
&KS&3.1&MIN3SPD&1.14&MIN4CV&1.14&MIN4DIFFP&0.76\\
&KURTO&2.99&MAX3KS&1.1&MIN4SPD&1.1&MEAN4KS&0.74\\
&CV&2.92&MIN3CV&1.1&MAX4KS&1.1&RDNOR&0.72\\
\hline
\multirow{5}{1cm}{\centering 3}
&RDNOR&3.67&MEAN3KS&1.64&MIN4CV&1.76&MAX4KS&0.99\\
&KURTO&3.47&MAX3KS&1.46&MAX4KS&1.75&MIN4CV&0.99\\
&KS&3.45&MIN3CV&1.44&MIN4SPD&1.74&MIN4SPD&0.95\\
&RDALT&3.12&MIN3SPD&1.44&MIN4DIFFP&1.71&MEAN3KS&0.9\\
&CV&3.08&MIN3DIFFP&1.33&MEAN4KS&1.33&MIN4DIFFP&0.89\\
\hline
\multirow{5}{1cm}{\centering 4}
&KURTO&4.49&MIN3DIFFP&2.04&MIN4DIFFP&2.34&MIN4DIFFP&1.34\\
&RDNOR&3.72&MIN3SPD&1.69&MAX4KS&2.1&MIN4CV&1.27\\
&KS&3.16&MIN3CV&1.59&MIN4CV&2.09&MAX4KS&1.27\\
&RDALT&2.68&MAX3KS&1.57&MIN4SPD&2.02&MIN4SPD&1.23\\
&CV&2.52&MAX3RDNOR&1.47&MAX4RDALT&1.28&MIN3DIFFP&1.07\\
\hline
\multirow{5}{1cm}{\centering 5}
&KURTO&5.08&MIN3DIFFP&2.15&MIN4DIFFP&2.44&MIN4DIFFP&1.53\\
&RDNOR&3.87&MIN3SPD&1.62&MIN4SPD&2.31&MIN4SPD&1.51\\
&RDALT&2.77&MAX3KS&1.52&MAX4KS&2.05&MAX4KS&1.35\\
&KS&2.71&MIN3CV&1.5&MIN4CV&2.03&MIN4CV&1.34\\
&CV&2.15&MEAN3KS&1.5&MAX4RDALT&1.54&MIN3DIFFP&1.06\\
\hline\hline
\end{tabular}
\end{center}
\par
{\footnotesize Note: "Comp.B", "IV" and "MDG" denote the number of competitive bids in the collusive tenders, the important variables selected by the random forest and the mean decrease in Gini index.  "KS", "CV", "SPD", "RD",  "RDNOR", "RDALT", "SKEW", "DIFFP" and "KURTO" denote the Kolmogorov-Smirnov statistic, the coefficient of variation, the spread, the relative distance, the normalized distance, the alternative relative distance, the skewness statistic, the percentage difference and the kurtosis statistic, respectively.}
\end{table}

\subsection{Application to the Swiss Data}

A drawback of the simulation based on the Ticino case is that reactions of competitive and collusive bidders, aware of their reciprocal existence, are unconsidered. For instance, collusive bidders might adopt a more competitive behavior in the presence of competitive bidders, while the latter might try to benefit from the umbrella effect of a cartel by bidding higher than they would have in a fully competitive situation. We therefore apply our detection method to data of the cases See-Gaster and Strassenbau Graubünden, characterized by well-organized bid-rigging cartels that, however, faced competition from outsiders from time to time.

We investigate the performance of our various predictive models first considering complete cartels. As shown in Table \ref{corrclassSWI}, the correct classification rates do not differ importantly across machine learning-based models 1 to 5 and amount to 81.3\% to 83.3\%. For the screening method of \citet{Imhof2017a}, the correct classification rate of 61.7\% is clearly below that of models 1 to 5 and in addition strongly differs between competitive and collusive tenders, amounting to only 33.4\% in the latter case. Possible explanations for the poor performance are the reliance on only two screens, which are not necessarily the optimal predictors, as well as on benchmark values for these two screens from two previous investigations, that are not necessarily optimal in the data considered. In contrast, the machine learning approaches use a larger set of screens and weight their importance in a data-driven way.

However, if we adjust the benchmarks in \citet{Imhof2017a}, we can achieve a better prediction rate for complete cartels. In Appendix, we display a decision tree on figure \ref{tree} corresponding to the minimal cross-validation error. Our pruned tree, using as predictors only the RD and the CV as in \citet{Imhof2017a}, exhibits a correct classification rate of 81.6\% for complete cartels. Since \citet{Imhof2017a} have drawn their benchmarks from two previous investigations, one of them was the Ticino cartel, it is therefore coherent that the benchmarks produce better results for detecting complete bid-rigging cartels in Ticino than in the Swiss data. Such discrepancy illustrates the fundamental difference between a benchmark method and machine learning. Benchmarks are given exogenous whereas machine learning outperforms benchmarks since it chooses the best predictors in each case. If a benchmark can still be adapted to different cases, machine learning algorithms are far more precise. Nonetheless, a benchmark method has the advantage of requiring few information to be implemented and therefore remains a first step for flagging cartels.

Considering incomplete cartels, the correct classification rates vary between 61.2\% to 84.1\% depending on the sample and the model. When the proportion of competitive bidders increases, the correct predictions generally decrease, as depicted in Table \ref{corrclassSWI}. This result suggests that cartel participants anticipated competitive bids and decided not to collude in some peculiar tenders, as attested for example in the case of See-Gaster. The models with screens, calculated for subgroups outperform model 1. Among them, models 3 and 4 (different to the Ticino simulation) slightly outperform model 2, indicating that in our case, screens calculated for subgroups of four bids exhibit a higher predictive power than those calculated for subgroups of three bids. The fact that we have four cartel participants per tender in most cases likely explains this result. In contrast, screens calculated for subgroups of three bids possibly work better if we mostly observe three cartel participants per tender.

Model 5 (which is the only to also include the number of bidders or the contract value as predictors) outperforms all other models and has a 5 to 10 percentage points higher correct classification rate than model 1. The advantage of models 3 or 4 over model 1 is 3 to 5.7 percentage points. The gain of calculating screens for subgroups is therefore not quite as high as for the Ticino simulation (4.5 to 10.3 percentage points). The result again suggests that cartel participants anticipated competitive bids and adapted their behavior, which was absent in the simulation of competitive bids for the Ticino case. However, models 2 to 4 still outperform model 1 and therefore competition agencies should consider screens for subgroups in the presence of incomplete bid-rigging cartels.
Similar to the Ticino simulation, the method of \citet{Imhof2017a} performs poorly for flagging incomplete bid-rigging cartels and does not better than throwing a coin. Specifically for truly collusive tenders, the correct classification rate is only between 8.7\% and 14.7\%.

When looking at the variable importance as reported in Table \ref{ImpPredSWI}, we find for all models and samples that the Kolmogorov-Smirnov statistic (KS) is an important predictor. In many case, it is among the three most important variables. Even if both collusive and competitive tenders generally do not follow a uniform distribution, the collusive ones are generally less uniform such that the Kolmogorov-Smirnov statistic for deviations from the uniform distribution tends to exhibit higher values under bid rigging.

The random forest generally picks up a balanced set of screens for the variance and asymmetry along with the Kolmogorov-Smirnov statistic for all models in samples with incomplete cartels. However, for complete cartels and all models except model 1, the random forest selects screens for the variance, mainly the coefficient of variation (CV) and the spread (SPD), along with the Kolmogorov-Smirnov statistic (KS). In model 1, the random forest picks up screens for asymmetry as the percentage difference (DIFFP) and the normalized distance (RDNOR). Further, the results suggest that screens for asymmetry are less important in the case of incomplete cartels, especially if one uses the Kolmogorov-Smirnov statistic (KS).

For all the samples with incomplete cartels, the minima and the maxima of the screens calculated for subgroups are the most important variables, while under complete cartels it is the mean and the median. As for the Ticino case, the result suggests that a few competitive bids sufficiently disturb the statistical pattern produced by bid rigging so that it becomes difficult to detect collusion by screens based on the total of bids per tender. In contrast, the use of the minimum or the maximum of all possible subgroups mitigates the distortion by competitive bids.

\begin{table} [!htp]
\caption{Correct classification rate in the Swiss data} \label{corrclassSWI}
\begin{center}
\begin{tabular}{ccclccccccc}\hline\hline
Sample&Cart.F&Perc.Cart.F&Tenders&Im.&M1&M2&M3&M4&M5\\\hline
\multirow{3}{1cm}{\centering 0}&\multirow{3}{1cm}{$>1$}&\multirow{3}{1cm}{71\%}
&All&0.524&0.612&0.637&0.642&0.645&0.673\\
& & &Comp.&0.901&0.612&0.607&0.612&0.62&0.646\\
& & &Coll.&0.147&0.615&0.671&0.677&0.677&0.706\\
\hline
\multirow{3}{1cm}{\centering 2}& \multirow{3}{1cm}{$>2$}& \multirow{3}{1cm}{75\%}
&All&0.525&0.648&0.665&0.675&0.678&0.708\\
& & &Comp.&0.901&0.652&0.643&0.645&0.65&0.683\\
& & &Coll.&0.148&0.647&0.691&0.71&0.709&0.737\\
\hline
\multirow{3}{1cm}{\centering 3}& \multirow{3}{1cm}{$>3$}& \multirow{3}{1cm}{79\%}
&All&0.511&0.706&0.722&0.748&0.745&0.759\\
& & &Comp.&0.901&0.705&0.688&0.705&0.707&0.719\\
& & &Coll.&0.121&0.708&0.758&0.792&0.784&0.800\\
\hline
\multirow{3}{1cm}{\centering 4}& \multirow{3}{1cm}{$>4$}& \multirow{3}{1cm}{83\%}
&All&0.506&0.743&0.770&0.8&0.798&0.814\\
& & &Comp.&0.901&0.755&0.751&0.764&0.771&0.783\\
& & &Coll.&0.111&0.735&0.791&0.835&0.826&0.846\\
\hline
\multirow{3}{1cm}{\centering 5}& \multirow{3}{1cm}{$>5$}& \multirow{3}{1cm}{88\%}
&All&0.494&0.766&0.805&0.813&0.818&0.841\\
& & &Comp.&0.901&0.771&0.769&0.786&0.788&0.813\\
& & &Coll.&0.087&0.763&0.844&0.842&0.849&0.871\\
\hline
\multirow{3}{1cm}{\centering Compl. Cartel}& \multirow{3}{1cm}{All}& \multirow{3}{1cm}{100\%}
&All&0.617&0.826&0.813&0.82&0.823&0.833\\
& & &Comp.&0.900&0.83&0.818&0.819&0.827&0.833\\
& & &Coll.&0.334&0.823&0.808&0.823&0.82&0.834\\
\hline\hline
\end{tabular}
\end{center}
\par
{\footnotesize Note: "Sample", "Cartel.F", "Per.Cart.F", "Tenders", "Im.", "M1", "M2", "M3", "M4" and "M5" denote the sample, the number of cartel firms in the collusive tenders, the percentage of cartel firms in the collusive tenders, the type of tenders, the results produced by the screening methods of \citet{Imhof2017a}, model 1, model 2, model 3, model 4 and model 5, respectively. For the type of tenders, "All", "Comp." and "Coll." denote the prediction for all types of tenders, the prediction for the competitive tenders and the prediction for the collusive tenders, respectively.}
\end{table}

\begin{landscape}
\begin{table} [!htp]
\caption{Important predictors for the Swiss data} \label{ImpPredSWI}
\begin{center}
\begin{tabular}{clclclclclc}\hline\hline
&\multicolumn{2}{c}{Model 1}&\multicolumn{2}{c}{Model 2}&\multicolumn{2}{c}{Model 3}&\multicolumn{2}{c}{Model 4} &\multicolumn{2}{c}{Model 5}\\
Sample&IV&MDG &IV&MDG &IV&MDG &IV&MDG &IV&MDG\\\hline
\multirow{5}{1cm}{\centering 1}
&SKEW&22.48&MIN3CV&9.98&MIN4CV&9.87&MIN4CV&5.8&STDBIDS&5.94\\
&RDNOR&21.73&MAX3KS&9.93&MAX4KS&9.55&MAX4KS&5.56&MEANBIDS&5.67\\
&SPD&21.67&MIN3SPD&9.66&MIN4SPD&9.34&MIN4SPD&5.39&MIN4CV&5.27\\
&KS&21.31&MAX3DIFFP&6.56&MIN4SKEW&8.14&MAX4RD&4.66&MAX4KS&5.11\\
&DIFFP&20.75&MIN3DIFFP&6.47&MAX4RD&8.08&MIN4SKEW&4.61&MIN4SPD&4.92\\

\hline
\multirow{5}{1cm}{\centering 2}
&KS&19.64&MIN3CV&9.27&MIN4CV&9.26&MIN4CV&5.61&MIN4CV&5.23\\
&RDNOR&19.56&MAX3KS&9.23&MAX4KS&9.12&MAX4KS&5.54&STDBIDS&5.03\\
&SPD&19.51&MIN3SPD&8.98&MIN4SPD&8.87&MIN4SPD&5.32&MAX4KS&5.02\\
&SKEW&18.66&MIN3DIFFP&5.62&MAX4RD&7.71&MAX4RD&4.63&MEANBIDS&4.97\\
&CV&18.38&MAX3DIFFP&5.51&MIN4SKEW&7.67&MIN4SKEW&4.56&MIN4SPD&4.78\\

\hline
\multirow{5}{1cm}{\centering 3}
&RDNOR&16.31&MAX3KS&8.42&MIN4CV&8.32&MIN4CV&5.46&NBRBIDS&5.88\\
&SPD&15.04&MIN3CV&8.32&MAX4KS&8.08&MAX4KS&5.43&MIN4CV&4.9\\
&KURTO&14.62&MIN3SPD&8.2&MIN4SPD&7.52&MAX4RD&5.01&MAX4KS&4.87\\
&KS&14.35&MAX3RD&5.09&MAX4RD&7.5&MIN4SPD&4.96&MAX4RD&4.53\\
&RDALT&14.02&MAX3RDALT&5.09&MIN4SKEW&7.4&MIN4SKEW&4.94&MIN4SKEW&4.49\\

\hline
\multirow{5}{1cm}{\centering 4}
&RDNOR&13.28&MIN3SPD&8.15&MIN4CV&7.6&MIN4CV&5.15&NBRBIDS&7.49\\
&SPD&12.12&MAX3KS&8.13&MAX4KS&7.49&MAX4KS&5.09&MIN4CV&4.52\\
&KS&11.52&MIN3CV&8.09&MIN4SPD&7.46&MIN4SPD&4.99&MAX4KS&4.45\\
&RDALT&11.42&MAX3RD&4.51&MIN4SKEW&7.39&MIN4SKEW&4.96&MIN4SKEW&4.38\\
&CV&11.06&MIN3SKEW&4.49&MAX4RD&6.92&MAX4RD&4.66&MIN4SPD&4.33\\

\hline
\multirow{5}{1cm}{\centering 5}
&KS&10.35&MIN3SPD&6.43&MAX4KS&6.05&MIN4CV&4.08&NBRBIDS&7.27\\
&SPD&10.14&MIN3CV&5.98&MIN4CV&6.05&MAX4KS&3.99&MIN4CV&3.51\\
&CV&9.83&MAX3KS&5.94&MIN4SPD&6.02&MIN4SPD&3.96&MIN4SPD&3.4\\
&RDNOR&9.53&MAX3RD&3.75&MIN4SKEW&5.09&MAX4RDALT&3.25&MAX4KS&3.39\\
&RDALT&8.03&MAX3RDALT&3.74&MAX4RDALT&4.96&MAX4RDNOR&3.25&MAX4RDALT&2.86\\

\hline
\multirow{5}{1cm}{\centering Compl. Cartel}
&KS&54.65&MEDIAN3CV&20.04&MEDIAN4KS&20.23&MEDIAN4KS&11.4&MEDIAN4KS&10.86\\
&CV&50.9&MEDIAN3SPD&18.68&MEDIAN4SPD&20.04&MEDIAN4SPD&11.27&MEDIAN4CV&10.82\\
&SPD&32.96&MEAN3CV&18.49&MEDIAN4CV&19.56&MEDIAN4CV&11.04&MEDIAN4SPD&10.75\\
&DIFFP&18.37&MEDIAN3KS&18.46&MEAN4KS&16.91&MEDIAN3CV&10.29&MEDIAN3CV&9.96\\
&RDNOR&17.1&MEAN3SPD&17.64&MEAN4CV&16.67&MEDIAN3KS&9.37&MEDIAN3SPD&9.22\\

\hline\hline
\end{tabular}
\end{center}
\par
{\footnotesize Note: "Sample", "IV" and "MDG" denote the sample, the important variables selected by the random forest and the mean decrease in Gini index.  "KS", "CV", "SPD", "RD",  "RDNOR", "RDALT", "SKEW", "DIFFP" and "KURTO" denote the Kolmogorov-Smirnov statistic, the coefficient of variation, the spread, the relative distance, the normalized distance, the alternative relative distance, the skewness statistic, the percentage difference and the kurtosis statistic, respectively.}
\end{table}
\end{landscape}

\subsection{Robustness Analysis}

We investigate the robustness of our correct classification rates when discarding the most important predictors and applying the random forest to the remaining predictors. Since model 1 makes use of fewer predictors than the other ones, we leave out the three most important variables, while for models 2 to 5, we drop the five best predictors. Table \ref{rob1} reports the difference in percentage points in the correct classification rates when keeping vs. dropping the respective predictors.

\begin{table} [!htp]
\caption{Differences in percentage points in correct classification rates of original random forest minus the correct classification rates of the random forest with discarded variables} \label{rob1}
\begin{center}
\begin{tabular}{ccclccccc}\hline\hline
Sample&Cart.F&Perc.Cart.F&Tenders&M1&M2&M3&M4&M5\\\hline
\multirow{3}{1cm}{\centering 1}&\multirow{3}{1cm}{$>1$}&\multirow{3}{1cm}{71\%}
&All&3.4&1.1&-0.2&0&1.8\\
& & &Comp.&3.6&-0.9&-1.2&-0.3&1.8\\
& & &Coll.&3.2&3.1&0.9&0.3&1.8\\
\hline
\multirow{3}{1cm}{\centering 2}& \multirow{3}{1cm}{$>2$}& \multirow{3}{1cm}{75\%}
&All&1.8&1.3&-0.4&0&1.4\\
& & &Comp.&2.9&-0.4&-1.1&-0.3&1.4\\
& & &Coll.&0.6&3&0.3&0.3&1.3\\
\hline
\multirow{3}{1cm}{\centering 3}& \multirow{3}{1cm}{$>3$}& \multirow{3}{1cm}{79\%}
&All&4.8&0.1&0.1&0.1&0.5\\
& & &Comp.&6.3&-0.4&-0.2&-0.2&0.1\\
& & &Coll.&3.2&0.6&0.6&0.5&0.9\\
\hline
\multirow{3}{1cm}{\centering 4}& \multirow{3}{1cm}{$>4$}& \multirow{3}{1cm}{83\%}
&All&3.4&1.2&0.2&0.9&1.8\\
& & &Comp.&4.1&0.9&-0.2&0.5&0.9\\
& & &Coll.&2.6&1.6&0.6&1.3&2.7\\
\hline
\multirow{3}{1cm}{\centering 5}& \multirow{3}{1cm}{$>5$}& \multirow{3}{1cm}{88\%}
&All&0.6&2&0.2&0.2&1.7\\
& & &Comp.&1.3&0.9&0.5&0.2&1.5\\
& & &Coll.&-0.2&3.1&-0.4&0&1.7\\
\hline
\multirow{3}{1cm}{\centering Compl. Cartel}& \multirow{3}{1cm}{All}& \multirow{3}{1cm}{100\%}
&All&1&-0.2&0.4&0&0\\
& & &Comp.&0.7&0.5&0.6&0.2&-0.2\\
& & &Coll.&1.3&-0.8&0.3&-0.1&0.2\\
\hline\hline
\end{tabular}
\end{center}
\par
{\footnotesize Note: "Sample", "Cartel.F", "Per.Cart.F", "Tenders", "M1", "M2", "M3", "M4" and "M5" denote the sample, the number of cartel firms in the collusive tenders, the percentage of cartel firms in the collusive tenders, the type of tenders, model 1, model 2, model 3, model 4 and model 5, respectively. For the type of tenders, "All", "Comp." and "Coll." denote the prediction for all types of tenders, the prediction for the competitive tenders and the prediction for the collusive tenders, respectively.}
\end{table}

The overall correct classification rate of model 1 in samples 1, 3 and 4 keeping all variables dominates that when dropping the best three predictors by 3.4 to 4.8 percentage points. Considering the other models and samples, we observe more or less the same predictive power when discarding the most important variables. This suggests that the remaining predictors seem to be good substitutes for the discarded ones. Other variables become more important when the most important predictors are omitted, and the correct classification rate is hardly affected, which is in line with previous findings of \citet{Huberimhof2019}.

Furthermore, we investigate the robustness with respect to the type of contracts. For both the cartel and post-cartel periods, we subsequently only consider contracts for road construction and asphalting. We exclude contracts for civil engineering and mixed contracts, combining civil engineering and road construction or asphalting. The motivation is that some specific characteristics of contracts in civil engineering could possibly affect the screens and therefore the correct classification rate. Dropping mixed contracts and contracts for civil engineering permits verifying whether this importantly affects the correct classification rate among the remaining contracts for road construction and asphalting. Table \ref{rob2} reports the difference in percentage points in the correct classification rates when using all contracts vs. using contracts for road construction and asphalting only.

\begin{table} [!htp]
\caption{Differences in percentage points in correct classification rates of original random forest minus the correct classification rates of the random forest using only contracts for road construction and asphalting} \label{rob2}
\begin{center}
\begin{tabular}{ccclccccc}\hline\hline
Sample&Cart.F&Perc.Cart.F&Tenders&M1&M2&M3&M4&M5\\\hline
\multirow{3}{1cm}{\centering 1}&\multirow{3}{1cm}{$>1$}&\multirow{3}{1cm}{81\%}
&All&-6.2&-5.2&-6.7&-5.5&-3.6\\
& & &Comp.&-7.2&-4.6&-6.9&-5.7&-3.8\\
& & &Coll.&-5.2&-5.8&-6.2&-5.2&-3.3\\
\hline
\multirow{3}{1cm}{\centering 2}& \multirow{3}{1cm}{$>2$}& \multirow{3}{1cm}{82\%}
&All&-2.8&-3&-4.3&-3.6&-1.2\\
& & &Comp.&-3.6&-3.1&-4.7&-3.9&-1.4\\
& & &Coll.&-1.9&-2.9&-3.8&-3.2&-1\\
\hline
\multirow{3}{1cm}{\centering 3}& \multirow{3}{1cm}{$>3$}& \multirow{3}{1cm}{84\%}
&All&0.1&-0.6&-0.9&-0.6&0.2\\
& & &Comp.&-1.9&-1.5&-2.7&-2.1&-1.9\\
& & &Coll.&2&0&0.5&0.5&1.9\\
\hline
\multirow{3}{1cm}{\centering 4}& \multirow{3}{1cm}{$>4$}& \multirow{3}{1cm}{86\%}
&All&1.5&2.2&2&2.4&3.4\\
& & &Comp.&2.2&2.7&1.9&2.5&3.1\\
& & &Coll.&1&1.9&1.9&2.2&3.6\\
\hline
\multirow{3}{1cm}{\centering 5}& \multirow{3}{1cm}{$>5$}& \multirow{3}{1cm}{88\%}
&All&2.6&3.3&2&2.1&2.7\\
& & &Comp.&3.1&3.3&2.2&1.8&3\\
& & &Coll.&1.9&3.2&1.9&2.4&2.3\\
\hline
\multirow{3}{1cm}{\centering Compl. Cartel}& \multirow{3}{1cm}{All}& \multirow{3}{1cm}{100\%}
&All&0.7&0.3&0&0.3&0.3\\
& & &Comp.&0.9&-0.3&-0.4&0&-0.2\\
& & &Coll.&0.4&0.9&0.4&0.5&0.8\\
\hline\hline
\end{tabular}
\end{center}
\par
{\footnotesize Note: "Sample", "Cartel.F", "Per.Cart.F", "Tenders", "M1", "M2", "M3", "M4" and "M5" denote the sample, the number of cartel firms in the collusive tenders, the percentage of cartel firms in the collusive tenders, the type of tenders, model 1, model 2, model 3, model 4 and model 5, respectively. For the outcome classification, "All", "Comp." and "Coll." denote the prediction for all types of tenders, the prediction for the competitive tenders and the prediction for the collusive tenders, respectively.}
\end{table}

In samples 1 and 2, we find the correct classification rates of the random forest for road construction and asphalting contracts to be superior to the classification rate of the random forest with all types of contracts. For example, the difference of the (overall) classification rate of model 1 in samples 1 and 2 accounts for 6.2 and 2.8 percentage points, respectively. A possible explanation could be that when we keep only the road construction and asphalting contracts, we implicitly suppress some competitors. For example in sample 1, the relative amount of collusive bidders of 80.9\% is considerable higher as in the situation with all type of contracts (71.1\%, see Table \ref{corrclassSWI}). The cartel percent is therefore higher for this restricted sample of exclusively road construction and asphalting contracts and explains the higher performance in samples 1 and 2. In sample 3, the situation begins to change for both types and the correct classification rates are quasi identical. Noticeably for all models in samples 3 and 4, the differences rise again (not as strong as before), but in the opposite direction. That means for an almost identical relative amount of cartel participants, the correct classification rates of the random forest for all types of contracts are slightly superior to the ones for road construction and asphalting.

To investigate the robustness of the correct classification rate across different classes of machine learning, we also assess the performance of lasso regression and an ensemble method (including bagged trees, random forests and neural networks) as applied and outlined in \citet{Huberimhof2019} for all models and samples, which are explained in more detail in the appendix. Table \ref{rob3} reports the difference in the correct classification rates of the random forest minus the correct classification rates of the lasso and the ensemble method.

\begin{table} [!htp]
\caption{Differences in percentage points in correct classification rates of original random forest minus the correct classification rates of the lasso and the ensemble method} \label{rob3}
\begin{center}
\begin{tabular}{clcccccccccccc}\hline\hline
&&\multicolumn{2}{c}{Sample 1}&\multicolumn{2}{c}{Sample 2}&\multicolumn{2}{c}{Sample 3}&\multicolumn{2}{c}{Sample 4} &\multicolumn{2}{c}{Sample 5}&\multicolumn{2}{c}{Compl. Cart.}\\
&Tenders&lasso&ens.&lasso&ens.&lasso&ens.&lasso&ens.&lasso&ens.&lasso&ens.\\\hline
\multirow{3}{0.5cm}{\centering M1}
&All&-1.5&-2&-1.6&-2.3&-3.7&-2.7&-5.6&-6.7&-6&-4.3&0.9&0.4\\
&Comp.&-3.5&-1.4&-3.1&-0.7&-4.3&0.4&-5.9&0.9&-4&2.2&4.7&0.6\\
&Coll.&0.1&-2.9&-0.4&-3.9&-3.1&-5.9&-5.4&-14.1&-8.3&-11&-3&0.1\\

\hline
\multirow{3}{0.5cm}{\centering M2}
&All&-1.3&-1.1&-0.8&-0.7&-1.8&-1.3&-1.6&-1.8&-0.8&0.4&0.2&-0.8\\
&Comp.&-13&-9.9&4.8&1.5&2.2&1.4&2.7&1.3&2&2.4&4&-0.8\\
&Coll.&10.3&7.5&-6.5&-2.9&-5.9&-4&-5.7&-4.7&-3.7&-2&-3.6&-0.9\\

\hline
\multirow{3}{0.5cm}{\centering M3}
&All&-1.4&-0.9&-1.5&-1.2&-1.5&-1&-0.9&0.2&-0.3&0.6&0.1&0.1\\
&Comp.&0.9&-0.8&0.5&-0.6&1.9&0.7&0.3&1.9&2.6&-6.1&4.6&1.2\\
&Coll.&-3.8&-1.1&-3.5&-1.8&-4.9&-2.6&-2.2&-1.7&-3.4&7.5&-4.4&-1\\

\hline
\multirow{3}{0.5cm}{\centering M4}
&All&-0.7&-0.7&-0.7&-0.6&-1.4&-0.8&-0.9&-0.5&-0.3&1&0.2&-0.1\\
&Comp.&2.7&0.3&2.9&0.2&2.2&0.9&1.5&1.2&2.6&2.8&4.7&0.9\\
&Coll.&-4.3&-1.8&-4.4&-1.5&-5&-2.4&-3.2&-2&-3.4&-0.9&-4.3&-1.1\\

\hline
\multirow{3}{0.5cm}{\centering M5}
&All&-1.5&-2.9&-1.3&-1.4&-2.5&-1.1&-1.6&-1.8&1&-2.4&0.6&-0.1\\
&Comp.&-3.4&-3.4&-2.4&-2.2&-2.2&-1.4&-3&-0.1&2.4&5.9&4.4&0.6\\
&Coll.&0.3&-2.2&-0.1&-0.6&-2.7&-0.8&-0.3&-3.3&-3.5&-13.8&-3.2&-0.9\\

\hline\hline
\end{tabular}
\end{center}
\par
{\footnotesize Note: "Tenders", "lasso", "ens.", "M1", "M2", "M3", "M4" and "M5" denote the type of tenders, the lasso, the ensemble of method, model 1, model 2, model 3, model 4 and model 5, respectively. For the type of tenders, "All", "Comp." and "Coll." denote the prediction for all types of tenders, the prediction for the competitive tenders and the prediction for the collusive tenders, respectively.}
\end{table}

Considering samples 1 and 2 in Table \ref{rob3}, we find that the lasso and ensemble method slightly outperform the random forest. The maximum difference in (overall) correct classification rates across models and samples amounts to 2.9 percentage points. While the slightly lower rates speak against the random forest, it shows a more uniform performance and therefore less divergence across competitive and collusive periods, which may be an important aspect for practitioners. For samples 3, 4 and 5 the lasso and ensemble method in general slightly outperform the random forest, too, in two cases even more profoundly with a 4.3 to 6.7 percentage points higher correct classification rate for model 1 in samples 4 and 5. This implies that in samples 4 and 5 (with a high amount of collusive bidders), considering screens for subgroups does not significantly improve the predictive power of lasso and ensemble method, in contrast to the random forest. On the other hand and as for samples 1 and 2, the random forest shows a more uniform performance (\textit{i.e.} correct classification rates are not too different for competitive and collusive tenders). In complete cartels, we find a similar performance of (overall) correct classification rates between the random forest and the ensemble method. However, the random forest slightly dominates the lasso regression. Considering the deviation across prediction of competitive and collusive periods, the random forest and the ensemble method show a less divergent performance than the lasso. To conclude, the random forest shows in Table \ref{rob3} a somewhat lower correct classification rate than the lasso and the ensemble of method, but exhibits a more homogeneous correct classification rate across competitive and collusive tenders.

\section{Conclusion}

In this paper, we suggested a method for flagging bid rigging in tenders that is likely more powerful for detecting incomplete cartels than previously suggested methods. Our approach combined screens, \textit{i.e.} statistics derived from the distribution of bids in a tender, with machine learning to predict the probability of collusion. As a methodological innovation, we calculated the screens for all possible subgroups of three or four bids within a tender and considered summary statistics as the mean, median, maximum, and minimum for each screen as predictors in the machine learning algorithm. By this approach, we tackled the issue that competitive bids in incomplete cartels distort the statistical signals produced by bid rigging.

We first applied the methods to the Ticino bid-rigging cartel and found the method to attain a correct out-of-sample classification rate of of 77\% to 86\% even in the presence of simulated competitive bids. Our approach increasingly outperformed other methods based on conventional screens using all bids in a tender as the number of competitive bids per tender increased. In this simulation, there was by design no strategic reaction or interaction of competitive and collusive bidders. To allow for such reactions, we also applied our method to rather unique data from the investigations involving partial cartels in the regions See-Gaster and Graubünden in Switzerland. The out of sample performance of machine learning using summary statistics of screens (calculated for all possible subgroups of three and four bids) as predictors outperformed other screening methods. However, the performance of all machine learning-based methods decreased in the number of competitive bids, indicating that cartel participants partially adapt their bids in the presence of competitive bidders.

Compared to screens calculated with all bids in a tender, summary statistics of screens calculated for subgroups increased the correct classification rate by 5 to 10 and 3 to 7.5 percentage points for incomplete cartels in the Ticino simulation and the Swiss data from See-Gaster and Graubünden, respectively. This implies a substantial decrease in error rate (which is one minus the correct classification rate) of 42.6\% and 22.2\% for the Ticino simulation and the Swiss data, respectively. As screening by competition agencies can trigger investigations with legal consequences for potential cartel candidates, such decreases in the error rate appears more than desirable. Our results demonstrate the usefulness of combining machine learning with an improved set of statistical screens reducing distortions of competitive bids in partial cartels. The method appears promising for detecting collusion in other industries or countries as an agenda for future research.
\newpage

\section*{Appendix}

\subsection*{Appendix 1: Descriptive Statistics for the Ticino Cartel}

{\renewcommand{\arraystretch}{1.1}
\begin{table} [!htp]
\caption{Numbers of bids in a tender for the cartel and post-cartel periods in Ticino} \label{nbrbidTI}
\begin{center}
\begin{tabular}{lcccccccc}\hline\hline
Number of bids in a tender&4&5&6&7&8&9&10&10+\\\hline
Cartel Period&32&24&23&28&15&12&7&8\\
Post-cartel Period&8&2&8&3&5&4&2&1\\\hline\hline
\end{tabular}
\end{center}
\end{table}}
\newpage

{\renewcommand{\arraystretch}{0.75}
\tablecaption{Descriptive statistics for the collusive tenders in Ticino (without simulated bids)\label{cartelDESCti}}
\tablehead{\hline Predictors&Mean&Std&Min&Lower Q.&Median&Upper Q.&Max&N\\  \hline}
\tabletail{\hline \multicolumn{4}{r}{See next page}\\}
\tablelasttail{\hline}
\begin{center}
\begin{supertabular}{>{$}l<{$} >{$}c<{$} >{$}c<{$} >{$}c<{$}>{$}c<{$}>{$}c<{$}>{$}c<{$}>{$}c<{$}>{$}c<{$}}
NBRBIDS&6.54&2.16&4&5&6&8&13&149\\
MEANBIDS&1134.58&1048.71&23.16&379.12&836.59&1547.76&4967.50&149\\
STDBIDS&34.08&30.13&0.82&12.77&25.84&44.68&136.90&149\\
CV&3.25&1.18&1.52&2.45&2.97&3.83&10.2&149\\
KURTO&2.71&2.12&-3.08&1.26&2.84&3.8&8.14&149\\
SKEW&-1.13&0.96&-2.76&-1.85&-1.34&-0.6&2.21&149\\
SPD&0.1&0.04&0.04&0.07&0.09&0.12&0.37&149\\
D&49.15&43.01&1.64&15.90&39.10&70.17&272.45&149\\
RD&4.09&3.37&0.31&1.77&3.14&5.13&23.02&149\\
RDNOR&2.93&1.35&0.53&1.95&2.72&3.6&6.95&149\\
RDALT&7.06&6.4&0.43&2.73&5.47&8.22&40.02&149\\
DIFFP&5.09&1.98&1.03&4.19&5.13&5.55&21.74&149\\
KS&34.27&9.93&9.77&26.5&34.26&41.17&66.05&149\\\hline
MIN3CV&0.7&0.8&0&0.13&0.41&0.99&4.36&149\\
MAX3CV&5.03&1.87&1.99&3.71&4.72&5.87&16.64&149\\
MEAN3CV&2.89&1.12&1.24&2.11&2.63&3.39&8.61&149\\
MEDIAN3CV&2.87&1.29&0.4&1.93&2.94&3.49&6.58&149\\\hline
MIN3SKEW&-1.67&0.17&-1.73&-1.73&-1.73&-1.71&-0.29&149\\
MAX3SKEW&1.33&0.72&-1.38&1.31&1.7&1.73&1.73&149\\
MEAN3SKEW&-0.44&0.48&-1.56&-0.76&-0.39&-0.12&0.94&149\\
MEDIAN3SKEW&-0.68&0.69&-1.69&-1.27&-0.74&-0.27&1.26&149\\\hline
MIN3D&7.67&18.44&0.00&0.21&1.17&5.89&106.92&149\\
MAX3D&74.23&60.37&1.73&25.78&60.09&101.12&319.13&149\\
MEAN3D&37.58&36.04&1.03&11.60&28.73&45.99&186.08&149\\
MEDIAN3D&36.60&41.83&1.10&7.51&21.92&51.49&272.45&149\\\hline
MIN3RD&0.43&0.75&0&0.02&0.1&0.46&4.86&149\\
MAX3RD&3177.54&31051.89&1.72&22.81&73.49&181.85&376725.28&149\\
MEAN3RD&48.87&279.28&0.75&5.09&8.67&13.66&3060.42&149\\
MEDIAN3RD&4.47&14.02&0.49&1.71&2.6&4.39&171.76&149\\\hline
MIN3RDNOR&0.32&0.38&0&0.03&0.14&0.49&1.55&149\\
MAX3RDNOR&1.89&0.15&1.1&1.88&1.96&1.98&2&149\\
MEAN3RDNOR&1.2&0.22&0.61&1.05&1.17&1.35&1.74&149\\
MEDIAN3RDNOR&1.27&0.28&0.5&1.09&1.27&1.49&1.85&149\\\hline
MIN3RDALT&0.3&0.53&0&0.02&0.07&0.33&3.44&149\\
MAX3RDALT&2246.86&21957&1.22&16.13&51.97&128.59&266385&149\\
MEAN3RDALT&34.56&197.48&0.53&3.6&6.13&9.66&2164.04&149\\
MEDIAN3RDALT&3.16&9.92&0.34&1.21&1.84&3.1&121.45&149\\\hline
MIN3DIFFP&0.51&0.76&0&0.05&0.16&0.63&4.03&149\\
MAX3DIFFP&8&3.34&1.87&6.08&7.18&9.49&34.62&149\\
MEAN3DIFFP&3.68&1.58&1.19&2.56&3.35&4.55&15.43&149\\
MEDIAN3DIFFP&3.42&2.12&0.35&1.35&3.23&5.03&10.15&149\\\hline
MIN3SPD&0.01&0.02&0&0&0.01&0.02&0.09&149\\
MAX3SPD&0.1&0.04&0.04&0.07&0.09&0.12&0.37&149\\
MEAN3SPD&0.06&0.02&0.02&0.04&0.05&0.07&0.19&149\\
MEDIAN3SPD&0.06&0.03&0.01&0.04&0.06&0.07&0.14&149\\\hline
MIN3KS&22.12&6.72&6.04&17.08&21.28&26.94&50.12&149\\
MAX3KS&851.31&3306.73&23.21&101.26&244.9&742.41&39320&149\\
MEAN3KS&88.19&88.53&17.01&40.86&64.31&100.72&863.76&149\\
MEDIAN3KS&49.42&38.86&15.83&29.28&34.73&52.31&247.55&149\\\hline
MIN4CV&1.35&1.31&0.04&0.28&0.86&2.29&6.49&149\\
MAX4CV&4.21&1.6&1.69&3.07&3.92&4.94&13.92&149\\
MEAN4CV&3.06&1.16&1.35&2.27&2.81&3.55&9.2&149\\
MEDIAN4CV&3.29&1.46&0.54&2.55&2.99&4.06&12.88&149\\\hline
MIN4SKEW&-1.67&0.58&-2&-1.99&-1.95&-1.62&1.47&149\\
MAX4SKEW&0.82&1.29&-1.92&-0.29&1.47&1.95&2&149\\
MEAN4SKEW&-0.65&0.66&-1.92&-1.12&-0.57&-0.2&1.47&149\\
MEDIAN4SKEW&-0.82&0.76&-1.96&-1.48&-0.9&-0.35&1.47&149\\\hline
MIN4D&18.48&35.07&0.00&0.28&1.54&15.88&171.35&149\\
MAX4D&61.85&49.83&1.70&23.35&52.45&89.22&307.84&149\\
MEAN4D&40.84&38.74&1.34&12.48&29.47&52.38&229.44&149\\
MEDIAN4D&43.70&43.33&0.98&9.98&31.25&58.21&272.45&149\\\hline
MIN4RD&1.15&2.28&0&0.04&0.19&1.14&13.56&149\\
MAX4RD&46.81&160.33&0.4&5.27&16.33&42.7&1865.04&149\\
MEAN4RD&5.2&4.81&0.4&2.47&3.79&6.54&34.88&149\\
MEDIAN4RD&3.27&2.83&0.32&1.44&2.34&4.24&19.6&149\\\hline
MIN4RDNOR&0.65&0.75&0&0.06&0.28&1.11&2.62&149\\
MAX4RDNOR&2.47&0.53&0.53&2.22&2.69&2.87&3&149\\
MEAN4RDNOR&1.59&0.41&0.53&1.29&1.55&1.84&2.62&149\\
MEDIAN4RDNOR&1.66&0.5&0.44&1.29&1.65&2.05&2.73&149\\\hline
MIN4RDALT&1.2&2.33&0&0.04&0.2&1.18&13.86&149\\
MAX4RDALT&49.69&170.21&0.43&5.68&17.11&43.97&1966.14&149\\
MEAN4RDALT&5.48&5.21&0.43&2.57&4.02&6.8&40.06&149\\
MEDIAN4RDALT&3.41&2.88&0.34&1.51&2.53&4.31&20.01&149\\\hline
MIN4DIFFP&1.28&1.87&0&0.06&0.25&1.59&6.99&149\\
MAX4DIFFP&6.84&3.12&1.03&5.6&6.26&7.84&34.1&149\\
MEAN4DIFFP&4.05&1.79&1.03&2.88&3.58&5.02&18.1&149\\
MEDIAN4DIFFP&4.4&2.46&0.35&3.02&4.6&5.42&21.74&149\\\hline
MIN4KURTO&-2.38&3.14&-6&-5.32&-3.19&0.35&3.76&149\\
MAX4KURTO&3.17&1.37&-3.08&2.93&3.83&3.98&4&149\\
MEAN4KURTO&1.47&1.18&-3.08&0.86&1.55&2.31&3.76&149\\
MEDIAN4KURTO&1.93&1.27&-3.08&1.45&2.09&2.77&3.87&149\\\hline
MIN4SPD&0.03&0.03&0&0.01&0.02&0.06&0.17&149\\
MAX4SPD&0.1&0.04&0.04&0.07&0.09&0.12&0.37&149\\
MEAN4SPD&0.07&0.03&0.03&0.05&0.07&0.08&0.24&149\\
MEDIAN4SPD&0.08&0.04&0.01&0.06&0.07&0.1&0.34&149\\\hline
MIN4KS&26.58&8.13&7.22&20.48&25.64&32.53&59.33&149\\
MAX4KS&267.41&363.93&15.73&43.86&116.15&355.69&2643.96&149\\
MEAN4KS&55.64&36.47&15.73&31.62&43.12&64.75&195.25&149\\
MEDIAN4KS&38.72&25.53&7.82&24.73&33.39&39.56&184.51&149\\
\hline
\end{supertabular}
\end{center}
\begin{spacing}{1}
{\footnotesize  Note: “Mean”, “Std”, “Min”, “Lower Q.”, “Median”, “Upper Q.”, “Max”, and “N” denote the mean, standard deviation, minimum, lower quartile, median, upper quartile, maximum, and number of observations, respectively. The value for "MEANBIDS", "STDBIDS", "D", "MIN3D", "MAX3D", "MEAN3D", "MEDIAN3D", "MIN4D", "MAX4D", "MEAN4D" and "MEDIAN4D" are expressed in thousand CHF. "KS", "CV", "SPD", "RD",  "RDNOR", "RDALT", "SKEW", "DIFFP", "KURTO", "D", "STDBIDS", "MEANBIDS" and "NBRBIDS" denote the Kolmogorov-Smirnov Statistic, the coefficient of variation, the spread, the relative distance, the normalized distance, the alternative relative distance, the skewness statistic, the percentage difference, the kurtosis statistic, the difference in absolute between the first and second lowest bids, the standard deviation of the bids in a tender, the mean of the bids in a tender and the number of the bids in a tender, respectively.}
\end{spacing}
\newpage

{\renewcommand{\arraystretch}{0.75}
\tablecaption{Descriptive statistics for the Ticino cartel in the post-cartel period\label{postcartelDESCti}}
\tablehead{\hline Predictors&Mean&Std&Min&Lower Q.&Median&Upper Q.&Max&N\\  \hline}
\tabletail{\hline \multicolumn{4}{r}{See next page}\\}
\tablelasttail{\hline}
\begin{center}
\begin{supertabular}{>{$}l<{$} >{$}c<{$} >{$}c<{$} >{$}c<{$}>{$}c<{$}>{$}c<{$}>{$}c<{$}>{$}c<{$}>{$}c<{$}}
NBRBIDS&6.73&2.34&4&5&6&8&13&33\\
MEANBIDS&756.79&785.43&43.97&279.90&482.05&695.68&3191.78&33\\
STDBIDS&54.83&49.50&2.79&27.78&39.64&61.86&209.64&33\\
CV&9.51&5.38&1.71&5.65&8.49&12.7&21.12&33\\
KURTO&-0.08&1.78&-2.83&-1.31&-0.16&1.19&6.06&33\\
SKEW&0.24&0.85&-1.46&-0.31&0.31&0.93&2.36&33\\
SPD&0.31&0.2&0.04&0.16&0.26&0.47&0.84&33\\
D&29.67&36.08&0.68&7.87&17.42&31.26&149.28&33\\
RD&0.77&0.89&0.02&0.24&0.41&0.95&4.03&33\\
RDNOR&1.02&0.8&0.06&0.45&0.74&1.33&3.67&33\\
RDALT&1.22&1.23&0.05&0.42&0.72&1.46&4.84&33\\
DIFFP&5.16&5.02&0.23&2.21&3.73&5.07&20.6&33\\
KS&16.52&12.23&5.4&8.86&12.4&18.49&58.56&33\\\hline
MIN3CV&2.58&2.8&0.07&1.03&1.73&3.11&12.23&33\\
MAX3CV&14.45&8.63&2.15&7.75&13.09&21.45&36.56&33\\
MEAN3CV&8.72&4.76&1.65&5.13&7.87&11.43&19.63&33\\
MEDIAN3CV&8.67&4.73&1.68&5.21&7.97&10.7&20.71&33\\\hline
MIN3SKEW&-1.48&0.55&-1.73&-1.72&-1.7&-1.53&0.7&33\\
MAX3SKEW&1.48&0.47&-0.27&1.6&1.69&1.73&1.73&33\\
MEAN3SKEW&0.1&0.51&-1.05&-0.2&0.21&0.42&1.2&33\\
MEDIAN3SKEW&0.14&0.79&-1.26&-0.5&0.29&0.74&1.2&33\\\hline
MIN3D&14.25&32.20&0.16&1.29&2.58&8.54&149.28&33\\
MAX3D&107.42&100.76&2.88&48.31&76.20&133.87&471.50&33\\
MEAN3D&47.30&47.20&1.75&23.50&31.57&48.04&221.08&33\\
MEDIAN3D&41.00&38.93&1.44&19.67&28.88&39.00&183.76&33\\\hline
MIN3RD&0.26&0.38&0.01&0.03&0.12&0.23&1.7&33\\
MAX3RD&39.4&59.34&0.86&6.69&17.33&39.98&274.21&33\\
MEAN3RD&4.37&4.62&0.51&1.85&3.4&4.4&25.42&33\\
MEDIAN3RD&1.65&1.17&0.53&0.83&1.16&2.09&4.24&33\\\hline
MIN3RDNOR&0.25&0.28&0.01&0.04&0.16&0.28&1.09&33\\
MAX3RDNOR&1.74&0.29&0.76&1.65&1.85&1.93&1.99&33\\
MEAN3RDNOR&0.96&0.22&0.49&0.83&0.91&1.09&1.43&33\\
MEDIAN3RDNOR&0.96&0.29&0.53&0.74&0.9&1.17&1.49&33\\\hline
MIN3RDALT&0.19&0.27&0.01&0.02&0.08&0.16&1.2&33\\
MAX3RDALT&27.86&41.96&0.61&4.73&12.25&28.27&193.89&33\\
MEAN3RDALT&3.09&3.27&0.36&1.31&2.4&3.11&17.97&33\\
MEDIAN3RDALT&1.17&0.83&0.37&0.59&0.82&1.48&3&33\\\hline
MIN3DIFFP&1.53&2.45&0.05&0.34&0.58&1.75&13.25&33\\
MAX3DIFFP&21.91&14.6&2.6&10.75&17.43&30.98&62.88&33\\
MEAN3DIFFP&8.32&4.73&1.43&4.44&7.63&10.3&20.24&33\\
MEDIAN3DIFFP&7.1&4.23&1.3&3.98&6.12&8.64&20.6&33\\\hline
MIN3SPD&0.05&0.06&0&0.02&0.03&0.06&0.28&33\\
MAX3SPD&0.31&0.2&0.04&0.16&0.26&0.47&0.84&33\\
MEAN3SPD&0.19&0.11&0.03&0.11&0.16&0.25&0.47&33\\
MEDIAN3SPD&0.19&0.11&0.03&0.11&0.17&0.23&0.49&33\\\hline
MIN3KS&11.84&10.28&3.14&5.06&8.04&13.19&46.56&33\\
MAX3KS&113.2&230.7&8.41&32.52&58.13&97.09&1360.3&33\\
MEAN3KS&22.18&16.84&5.85&11.68&15.88&25.57&74.31&33\\
MEDIAN3KS&16.75&11.89&5.34&9.69&12.79&19.4&60.05&33\\\hline
MIN4CV&4.4&4.32&0.5&1.85&2.93&6.33&20.66&33\\
MAX4CV&12.55&7.62&1.91&6.57&11.2&18.1&32.31&33\\
MEAN4CV&9.09&5&1.7&5.36&8.27&11.97&20.66&33\\
MEDIAN4CV&9.01&4.89&1.78&5.25&8.14&11.76&20.66&33\\\hline
MIN4SKEW&-1.16&0.91&-2&-1.81&-1.46&-0.79&1.32&33\\
MAX4SKEW&1.14&0.99&-1.46&0.81&1.57&1.89&2&33\\
MEAN4SKEW&0.11&0.7&-1.46&-0.35&0.22&0.58&1.32&33\\
MEDIAN4SKEW&0.19&0.73&-1.46&-0.24&0.22&0.77&1.34&33\\\hline
MIN4D&20.50&38.09&0.16&1.62&3.62&15.10&149.28&33\\
MAX4D&76.56&77.30&1.26&36.48&62.49&92.32&435.92&33\\
MEAN4D&37.81&35.59&1.26&17.18&25.70&42.78&149.28&33\\
MEDIAN4D&34.53&35.39&1.26&15.94&23.39&35.73&149.28&33\\\hline
MIN4RD&0.49&0.92&0.01&0.04&0.17&0.48&4.03&33\\
MAX4RD&10.2&22.25&0.17&1.9&4.15&8.5&128.38&33\\
MEAN4RD&1.55&1.44&0.17&0.83&1.03&1.71&7.36&33\\
MEDIAN4RD&1.03&0.94&0.1&0.48&0.67&1.32&4.03&33\\\hline
MIN4RDNOR&0.43&0.52&0.02&0.06&0.23&0.59&2.01&33\\
MAX4RDNOR&1.95&0.68&0.23&1.48&2.07&2.46&2.95&33\\
MEAN4RDNOR&0.97&0.41&0.23&0.71&0.87&1.18&2.01&33\\
MEDIAN4RDNOR&0.91&0.49&0.15&0.59&0.77&1.26&2.01&33\\\hline
MIN4RDALT&0.51&0.92&0.01&0.04&0.17&0.49&4.03&33\\
MAX4RDALT&10.47&22.63&0.17&1.94&4.48&9.13&130.64&33\\
MEAN4RDALT&1.62&1.48&0.17&0.87&1.06&1.75&7.68&33\\
MEDIAN4RDALT&1.07&0.96&0.11&0.49&0.68&1.45&4.03&33\\\hline
MIN4DIFFP&2.43&4.4&0.16&0.4&0.84&2.75&20.6&33\\
MAX4DIFFP&16.11&11.32&0.54&7.23&15.77&20.11&45.9&33\\
MEAN4DIFFP&6.78&4.45&0.54&3.91&6.06&9.05&20.6&33\\
MEDIAN4DIFFP&5.97&4.5&0.54&3.28&5.07&7.1&20.6&33\\\hline
MIN4KURTO&-3.77&2.63&-6&-5.62&-4.78&-2.83&2.2&33\\
MAX4KURTO&2.65&1.73&-2.83&2.2&3.29&3.83&4&33\\
MEAN4KURTO&-0.08&1.17&-2.83&-0.8&-0.08&0.48&2.2&33\\
MEDIAN4KURTO&0.17&1.52&-3.68&-0.53&0.69&1.18&2.2&33\\\hline
MIN4SPD&0.11&0.13&0.01&0.04&0.07&0.13&0.63&33\\
MAX4SPD&0.31&0.2&0.04&0.16&0.26&0.47&0.84&33\\
MEAN4SPD&0.24&0.15&0.04&0.13&0.2&0.32&0.63&33\\
MEDIAN4SPD&0.23&0.14&0.04&0.13&0.2&0.31&0.63&33\\\hline
MIN4KS&13.88&12.22&3.79&6.05&9.27&15.47&52.64&33\\
MAX4KS&43.49&37.89&5.4&16.37&34.46&54.38&201.12&33\\
MEAN4KS&17.88&12.29&5.4&9.86&13.17&22.37&60.1&33\\
MEDIAN4KS&16.68&12.04&5.4&9.1&12.71&19.31&56.12&33\\
\hline
\end{supertabular}
\end{center}
\par
\begin{spacing}{1}
{\footnotesize Note: “Mean”, “Std”, “Min”, “Lower Q.”, “Median”, “Upper Q.”, “Max”, and “N” denote the mean, standard deviation, minimum, lower quartile, median, upper quartile, maximum, and number of observations, respectively. The value for "MEANBIDS", "STDBIDS", "D", "MIN3D", "MAX3D", "MEAN3D", "MEDIAN3D", "MIN4D", "MAX4D", "MEAN4D" and "MEDIAN4D" are expressed in thousand CHF. "KS", "CV", "SPD", "RD",  "RDNOR", "RDALT", "SKEW", "DIFFP", "KURTO", "D", "STDBIDS", "MEANBIDS" and "NBRBIDS" denote the Kolmogorov-Smirnov statistic, the coefficient of variation, the spread, the relative distance, the normalized distance, the alternative relative distance, the skewness statistic, the percentage difference, the kurtosis statistic, the difference in absolute between the first and second lowest bids, the standard deviation of the bids in a tender, the mean of the bids in a tender and the number of the bids in a tender, respectively.}
\end{spacing}
\newpage

{\renewcommand{\arraystretch}{0.75}
\tablecaption{Descriptive statistics for collusive tenders of the Ticino cartel with five competitive bids\label{cartelsimulDESCti}}
\tablehead{\hline Predictors&Mean&Std&Min&Lower Q.&Median&Upper Q.&Max&N\\  \hline}
\tabletail{\hline \multicolumn{4}{r}{See next page}\\}
\tablelasttail{\hline}
\begin{center}
\begin{supertabular}{>{$}l<{$} >{$}c<{$} >{$}c<{$} >{$}c<{$}>{$}c<{$}>{$}c<{$}>{$}c<{$}>{$}c<{$}>{$}c<{$}}
NBRBIDS&10.79&2.49&6&9&11&12&18&184\\
MEANBIDS&1417.60&1328.38&19.00&383.19&897.16&1962.42&6080.35&184\\
STDBIDS&112.94&140.47&1.83&21.68&60.74&143.96&859.50&184\\
CV&7.38&3.23&2.77&5.16&6.89&8.95&23.83&184\\
KURTO&1.78&2.37&-1.91&-0.11&1.24&3.27&9.87&184\\
SKEW&0.11&1.24&-2.6&-0.87&-0.08&1.01&3.07&184\\
SPD&0.29&0.14&0.08&0.18&0.26&0.36&0.89&184\\
D&633.31&93.76&0.11&6.86&25.49&69.55&585.65&184\\
RD&0.93&1.04&0&0.23&0.64&1.13&5.77&184\\
RDNOR&1.84&1.55&0&0.66&1.52&2.57&8.19&184\\
RDALT&2.51&2.94&0&0.64&1.62&3.17&18.31&184\\
DIFFP&5.27&4.81&0.01&1.79&4.23&7.66&27.56&184\\
KS&16.68&5.95&5.44&12.41&15.44&19.79&37.34&184\\\hline
MIN3CV&0.49&0.52&0&0.12&0.29&0.7&2.9&184\\
MAX3CV&13.83&6.03&4.57&9.3&12.54&17.06&37.62&184\\
MEAN3CV&6.2&2.41&2.4&4.54&5.9&7.49&19.05&184\\
MEDIAN3CV&5.6&2.23&1.48&4.04&5.08&6.52&13.65&184\\\hline
MIN3SKEW&-1.72&0.03&-1.73&-1.73&-1.73&-1.73&-1.51&184\\
MAX3SKEW&1.71&0.06&1.22&1.73&1.73&1.73&1.73&184\\
MEAN3SKEW&-0.06&0.4&-1.19&-0.33&-0.12&0.18&1.09&184\\
MEDIAN3SKEW&-0.1&0.77&-1.56&-0.67&-0.25&0.44&1.73&184\\\hline
MIN3D&4.61&10.12&0.00&0.16&0.66&2.96&70.32&184\\
MAX3D&239.79&245.60&4.90&62.47&146.83&354.79&1327.34&184\\
MEAN3D&82.36&91.89&1.86&20.25&45.74&109.63&391.88&184\\
MEDIAN3D&71.06&82.74&1.55&14.69&41.10&92.43&394.87&184\\\hline
MIN3RD&0.05&0.08&0&0&0.01&0.05&0.52&184\\
MAX3RD&5367.51&51859.16&6.4&53.19&129.19&604.61&693080.61&184\\
MEAN3RD&43.17&245.44&0.95&4.22&7.71&16.32&2710.21&184\\
MEDIAN3RD&1.82&1.01&0.25&1.11&1.69&2.28&7.49&184\\\hline
MIN3RDNOR&0.06&0.09&0&0.01&0.02&0.06&0.54&184\\
MAX3RDNOR&1.96&0.06&1.64&1.95&1.98&2&2&184\\
MEAN3RDNOR&1.04&0.17&0.6&0.94&1.06&1.16&1.56&184\\
MEDIAN3RDNOR&1.05&0.26&0.3&0.88&1.08&1.23&1.68&184\\\hline
MIN3RDALT&0.03&0.06&0&0&0.01&0.03&0.37&184\\
MAX3RDALT&3795.41&36669.96&4.53&37.61&91.35&427.52&490082&184\\
MEAN3RDALT&30.53&173.55&0.67&2.98&5.45&11.54&1916.41&184\\
MEDIAN3RDALT&1.29&0.72&0.18&0.79&1.19&1.61&5.29&184\\\hline
MIN3DIFFP&0.23&0.34&0&0.02&0.11&0.29&2.01&184\\
MAX3DIFFP&20.05&8.11&7.05&13.94&18.21&24.27&53.73&184\\
MEAN3DIFFP&6.33&2.43&2.48&4.4&5.8&7.74&13.45&184\\
MEDIAN3DIFFP&5.21&2.33&0.87&3.55&5&6.2&14.71&184\\\hline
MIN3SPD&0.01&0.01&0&0&0.01&0.01&0.06&184\\
MAX3SPD&0.29&0.14&0.08&0.18&0.26&0.36&0.89&184\\
MEAN3SPD&0.13&0.05&0.05&0.09&0.12&0.15&0.45&184\\
MEDIAN3SPD&0.11&0.05&0.03&0.08&0.1&0.14&0.3&184\\\hline
MIN3KS&8.83&3.51&3.06&6.22&8.1&11&22.3&184\\
MAX3KS&938.54&3130.28&34.68&144.12&339.85&830.93&39320&184\\
MEAN3KS&39.88&25.59&8.37&23.92&32.91&46.72&192.41&184\\
MEDIAN3KS&20.92&8.4&7.62&15.44&19.97&25.03&67.92&184\\\hline
MIN4CV&0.94&0.95&0.04&0.28&0.73&1.27&7.67&184\\
MAX4CV&12.03&5.14&4.23&8.2&11.03&14.48&33.57&184\\
MEAN4CV&6.57&2.64&2.53&4.7&6.26&7.89&20.74&184\\
MEDIAN4CV&6.27&2.82&1.9&4.49&5.74&7.44&26&184\\\hline
MIN4SKEW&-1.92&0.19&-2&-2&-1.99&-1.93&-0.67&184\\
MAX4SKEW&1.87&0.29&0.22&1.89&1.99&2&2&184\\
MEAN4SKEW&-0.08&0.53&-1.53&-0.48&-0.17&0.26&1.2&184\\
MEDIAN4SKEW&-0.06&0.7&-1.54&-0.59&-0.14&0.33&1.74&184\\\hline
MIN4D&5.32&11.79&0.00&0.16&0.68&3.10&70.32&184\\
MAX4D&196.40&202.67&3.82&56.66&112.66&278.55&903.60&184\\
MEAN4D&75.31&87.29&1.35&18.73&41.48&96.57&449.88&184\\
MEDIAN4D&70.54&87.18&0.82&14.65&37.46&87.74&449.27&184\\\hline
MIN4RD&0.06&0.11&0&0&0.02&0.06&0.74&184\\
MAX4RD&117.38&392.24&1.49&14.96&33.85&95.96&4710.41&184\\
MEAN4RD&3.59&4.07&0.37&1.59&2.58&4.27&44.88&184\\
MEDIAN4RD&1.31&0.78&0.12&0.72&1.21&1.76&4.63&184\\\hline
MIN4RDNOR&0.09&0.14&0&0.01&0.03&0.1&0.81&184\\
MAX4RDNOR&2.74&0.29&1.32&2.65&2.84&2.94&3&184\\
MEAN4RDNOR&1.18&0.31&0.4&0.98&1.23&1.4&2.21&184\\
MEDIAN4RDNOR&1.13&0.41&0.17&0.83&1.18&1.43&2.12&184\\\hline
MIN4RDALT&0.07&0.12&0&0.01&0.02&0.07&0.74&184\\
MAX4RDALT&124.1&415.85&1.56&15.04&34.79&99.1&4965.75&184\\
MEAN4RDALT&3.73&4.24&0.39&1.69&2.66&4.48&47.93&184\\
MEDIAN4RDALT&1.37&0.81&0.12&0.77&1.28&1.81&4.82&184\\\hline
MIN4DIFFP&0.27&0.41&0&0.03&0.11&0.3&2.5&184\\
MAX4DIFFP&16.67&6.63&6&11.57&15.6&20.15&40.38&184\\
MEAN4DIFFP&5.99&2.63&1.71&3.99&5.49&7.6&14.87&184\\
MEDIAN4DIFFP&5.32&2.7&0.58&3.57&4.86&6.48&19.21&184\\\hline
MIN4KURTO&-5.57&0.76&-6&-5.96&-5.84&-5.55&-0.04&184\\
MAX4KURTO&3.9&0.2&2.88&3.91&3.98&4&4&184\\
MEAN4KURTO&0.64&0.77&-1.74&0.15&0.72&1.11&2.39&184\\
MEDIAN4KURTO&1.32&0.98&-2.78&0.88&1.51&1.86&3.24&184\\\hline
MIN4SPD&0.02&0.02&0&0.01&0.02&0.03&0.18&184\\
MAX4SPD&0.29&0.14&0.08&0.18&0.26&0.36&0.89&184\\
MEAN4SPD&0.16&0.07&0.06&0.11&0.15&0.19&0.57&184\\
MEDIAN4SPD&0.15&0.08&0.05&0.11&0.13&0.18&0.75&184\\\hline
MIN4KS&10.09&3.78&3.67&7.33&9.25&12.34&23.85&184\\
MAX4KS&288.31&349.99&13.07&79.29&137.96&363.99&2643.96&184\\
MEAN4KS&26.2&11.64&6.8&17.79&23.37&32.35&72.53&184\\
MEDIAN4KS&19.03&7.35&4.49&13.87&17.82&22.57&53.22&184\\
\hline
\end{supertabular}
\end{center}
\par
\begin{spacing}{1}
{\footnotesize Note: “Mean”, “Std”, “Min”, “Lower Q.”, “Median”, “Upper Q.”, “Max”, and “N” denote the mean, standard deviation, minimum, lower quartile, median, upper quartile, maximum, and number of observations, respectively. The value for "MEANBIDS", "STDBIDS", "D", "MIN3D", "MAX3D", "MEAN3D", "MEDIAN3D", "MIN4D", "MAX4D", "MEAN4D" and "MEDIAN4D" are expressed in thousand CHF. "KS", "CV", "SPD", "RD",  "RDNOR", "RDALT", "SKEW", "DIFFP", "KURTO", "D", "STDBIDS", "MEANBIDS" and "NBRBIDS" denote the Kolmogorov-Smirnov statistic, the coefficient of variation, the spread, the relative distance, the normalized distance, the alternative relative distance, the skewness statistic, the percentage difference, the kurtosis statistic, the difference in absolute between the first and second lowest bids, the standard deviation of the bids in a tender, the mean of the bids in a tender and the number of the bids in a tender, respectively.}
\end{spacing}
\newpage

\subsection*{Appendix 2: Descriptive Statistics for the Swiss Data}

{\renewcommand{\arraystretch}{1.1}
\begin{table} [!htp]
\caption{Numbers of bids in a tender in the Swiss data} \label{nbrbidSWI}
\begin{center}
\begin{tabular}{lcccccccc}\hline\hline
Number of bids in a tender&4&5&6&7&8&9&10&10+\\\hline
Tenders with complete cartels&94&50&29&24&33&33&23&24\\
Tenders with incomplete cartels&56&36&38&40&27&28&24&38\\
Competitive tenders&786&559&365&257&158&129&74&70\\\hline\hline
\end{tabular}
\end{center}
\end{table}}

\newpage

{\renewcommand{\arraystretch}{0.75}
\tablecaption{Descriptive statistics for the collusive tenders including only cartel participants in the Swiss data\label{cartelCOMPDESCswiss}}
\tablehead{\hline Predictors&Mean&Std&Min&Lower Q.&Median&Upper Q.&Max&N\\  \hline}
\tabletail{\hline \multicolumn{4}{r}{See next page}\\}
\tablelasttail{\hline}
\begin{center}
\begin{supertabular}{>{$}l<{$} >{$}c<{$} >{$}c<{$} >{$}c<{$}>{$}c<{$}>{$}c<{$}>{$}c<{$}>{$}c<{$}>{$}c<{$}}
NBRBIDS&6.57&2.44&4&4&6&9&13&308\\
MEANBIDS&379.88&376.09&34.42&172.13&305.81&460.98&3509.71&308\\
STDBIDS&13.15&13.67&0.49&5.59&9.75&15.88&109.94&308\\
CV&3.66&2.09&0.6&2.22&3.29&4.51&15.73&308\\
KURTO&0.16&1.65&-5.4&-0.99&0.21&1.33&4.37&308\\
SKEW&0.08&0.81&-1.94&-0.42&0.07&0.7&1.78&308\\
SPD&0.11&0.07&0.01&0.06&0.09&0.13&0.5&308\\
D&9.20&13.32&0.14&2.93&6.02&10.11&121.30&308\\
RD&1.16&1.36&0.01&0.46&0.75&1.31&13.66&308\\
RDNOR&1.38&0.79&0.02&0.78&1.24&1.72&5.03&308\\
RDALT&1.83&1.73&0.01&0.74&1.33&2.2&13.89&308\\
DIFFP&2.76&2.91&0.06&1.3&1.94&3.24&34.11&308\\
KS&36.54&20.15&6.59&22.69&31.2&45.55&167.93&308\\\hline
MIN3CV&1.08&1.01&0.04&0.48&0.81&1.26&8.46&308\\
MAX3CV&5.49&3.41&0.73&3.14&4.72&6.88&21.17&308\\
MEAN3CV&3.41&1.94&0.57&2.1&3.04&4.18&15.32&308\\
MEDIAN3CV&3.46&2&0.63&2.1&3&4.25&16.87&308\\\hline
MIN3SKEW&-1.48&0.44&-1.73&-1.73&-1.7&-1.42&0.69&308\\
MAX3SKEW&1.41&0.63&-1.49&1.43&1.68&1.73&1.73&308\\
MEAN3SKEW&0&0.49&-1.61&-0.26&0.05&0.33&1.16&308\\
MEDIAN3SKEW&0.03&0.69&-1.7&-0.41&0.07&0.55&1.53&308\\\hline
MIN3D&2.67&6.13&0.00&0.38&1.03&2.99&73.33&308\\
MAX3D&26.81&27.14&0.59&10.58&20.08&32.20&200.26&308\\
MEAN3D&11.62&13.40&0.43&4.63&8.44&13.24&118.06&308\\
MEDIAN3D&10.88&13.28&0.42&3.89&7.87&12.01&121.30&308\\\hline
MIN3RD&0.38&0.73&0&0.04&0.14&0.38&6.13&308\\
MAX3RD&117.7&665.15&0.87&5.26&16.67&54.56&9157.71&308\\
MEAN3RD&7.53&23.86&0.55&1.92&3.04&5.7&291.4&308\\
MEDIAN3RD&1.95&2.79&0.29&0.97&1.37&1.98&39.01&308\\\hline
MIN3RDNOR&0.3&0.33&0&0.06&0.17&0.42&1.63&308\\
MAX3RDNOR&1.73&0.28&0.76&1.58&1.85&1.95&2&308\\
MEAN3RDNOR&1&0.22&0.51&0.86&0.98&1.11&1.8&308\\
MEDIAN3RDNOR&0.99&0.26&0.33&0.81&0.98&1.14&1.85&308\\\hline
MIN3RDALT&0.27&0.52&0&0.03&0.1&0.27&4.34&308\\
MAX3RDALT&83.23&470.33&0.62&3.72&11.79&38.58&6475.48&308\\
MEAN3RDALT&5.32&16.87&0.39&1.36&2.15&4.03&206.05&308\\
MEDIAN3RDALT&1.38&1.97&0.2&0.68&0.97&1.4&27.58&308\\\hline
MIN3DIFFP&0.69&0.85&0&0.15&0.44&0.91&8.62&308\\
MAX3DIFFP&8.1&6.07&1.13&4.26&6.76&10.05&47.45&308\\
MEAN3DIFFP&3.37&2.32&0.77&1.99&2.77&4.08&22.76&308\\
MEDIAN3DIFFP&3.13&2.31&0.69&1.79&2.51&3.95&23.91&308\\\hline
MIN3SPD&0.02&0.02&0&0.01&0.02&0.03&0.18&308\\
MAX3SPD&0.11&0.07&0.01&0.06&0.09&0.13&0.5&308\\
MEAN3SPD&0.07&0.04&0.01&0.04&0.06&0.08&0.37&308\\
MEDIAN3SPD&0.07&0.04&0.01&0.04&0.06&0.08&0.41&308\\\hline
MIN3KS&26.02&17.06&4.76&14.83&21.48&31.93&137.18&308\\
MAX3KS&181.77&213.13&12.09&79.31&124.09&206.54&2751.23&308\\
MEAN3KS&48.26&25.64&7.44&30.53&42.03&60.32&199.62&308\\
MEDIAN3KS&38.19&20.28&6.19&24.04&33.58&47.94&158.95&308\\\hline
MIN4CV&1.93&1.66&0.22&0.91&1.5&2.37&15.73&308\\
MAX4CV&4.74&2.95&0.6&2.75&4.07&5.83&19.16&308\\
MEAN4CV&3.54&2.01&0.6&2.18&3.13&4.35&15.73&308\\
MEDIAN4CV&3.6&2.07&0.6&2.22&3.21&4.39&15.73&308\\\hline
MIN4SKEW&-1.07&0.98&-2&-1.89&-1.5&-0.41&1.78&308\\
MAX4SKEW&1.07&1.02&-1.93&0.46&1.48&1.87&2&308\\
MEAN4SKEW&0.04&0.68&-1.93&-0.35&0.07&0.5&1.78&308\\
MEDIAN4SKEW&0.05&0.71&-1.93&-0.3&0.03&0.54&1.78&308\\\hline
MIN4D&4.47&11.64&0.00&0.50&1.36&3.86&121.30&308\\
MAX4D&20.38&20.98&0.36&7.07&14.66&25.33&121.30&308\\
MEAN4D&10.11&12.76&0.36&3.81&6.99&11.59&121.30&308\\
MEDIAN4D&9.73&13.27&0.17&3.48&6.40&10.73&121.30&308\\\hline
MIN4RD&0.68&1.38&0&0.06&0.23&0.67&13.66&308\\
MAX4RD&9.14&19.83&0.15&1.51&4.31&10.73&298.7&308\\
MEAN4RD&1.66&1.53&0.15&0.87&1.26&1.94&13.66&308\\
MEDIAN4RD&1.26&1.37&0.07&0.61&0.87&1.35&13.66&308\\\hline
MIN4RDNOR&0.53&0.58&0&0.1&0.31&0.77&2.62&308\\
MAX4RDNOR&1.9&0.72&0.22&1.31&2.09&2.55&2.98&308\\
MEAN4RDNOR&1.07&0.4&0.22&0.8&1.02&1.27&2.62&308\\
MEDIAN4RDNOR&1.02&0.44&0.1&0.72&0.93&1.23&2.62&308\\\hline
MIN4RDALT&0.71&1.44&0&0.07&0.23&0.69&13.89&308\\
MAX4RDALT&9.67&22.08&0.15&1.55&4.57&11.2&340.47&308\\
MEAN4RDALT&1.74&1.62&0.15&0.91&1.33&2.02&13.89&308\\
MEDIAN4RDALT&1.31&1.42&0.07&0.64&0.9&1.38&13.89&308\\\hline
MIN4DIFFP&1.19&1.93&0&0.19&0.67&1.55&23.91&308\\
MAX4DIFFP&6.2&5.28&0.64&2.59&4.98&7.89&39.3&308\\
MEAN4DIFFP&2.95&2.42&0.61&1.66&2.29&3.56&24.48&308\\
MEDIAN4DIFFP&2.82&2.74&0.2&1.49&2.08&3.28&34.11&308\\\hline
MIN4KURTO&-2.96&2.98&-6&-5.63&-4.23&-0.28&3.75&308\\
MAX4KURTO&2.44&1.83&-5.4&1.66&3.14&3.77&4&308\\
MEAN4KURTO&0.14&1.36&-5.4&-0.58&0.09&0.95&3.75&308\\
MEDIAN4KURTO&0.34&1.54&-5.4&-0.28&0.61&1.31&3.75&308\\\hline
MIN4SPD&0.05&0.04&0.01&0.02&0.03&0.05&0.47&308\\
MAX4SPD&0.11&0.07&0.01&0.06&0.09&0.13&0.5&308\\
MEAN4SPD&0.08&0.05&0.01&0.05&0.07&0.1&0.47&308\\
MEDIAN4SPD&0.09&0.05&0.01&0.05&0.08&0.1&0.47&308\\\hline
MIN4KS&30.68&20.89&5.42&17.65&24.8&36.76&167.93&308\\
MAX4KS&88.22&69.07&6.59&42.47&67.21&110.66&458.02&308\\
MEAN4KS&39.97&20.41&6.59&26.34&35.08&50.1&167.93&308\\
MEDIAN4KS&36.66&19.93&6.59&23.01&31.38&45.53&167.93&308\\
\hline
\end{supertabular}
\end{center}
\par
\begin{spacing}{1}
{\footnotesize Note: “Mean”, “Std”, “Min”, “Lower Q.”, “Median”, “Upper Q.”, “Max”, and “N” denote the mean, standard deviation, minimum, lower quartile, median, upper quartile, maximum, and number of observations, respectively. The value for "MEANBIDS", "STDBIDS", "D", "MIN3D", "MAX3D", "MEAN3D", "MEDIAN3D", "MIN4D", "MAX4D", "MEAN4D" and "MEDIAN4D" are expressed in thousand CHF. "KS", "CV", "SPD", "RD",  "RDNOR", "RDALT", "SKEW", "DIFFP", "KURTO", "D", "STDBIDS", "MEANBIDS" and "NBRBIDS" denote the Kolmogorov-Smirnov statistic, the coefficient of variation, the spread, the relative distance, the normalized distance, the alternative relative distance, the skewness statistic, the percentage difference, the kurtosis statistic, the difference in absolute between the first and second lowest bids, the standard deviation of the bids in a tender, the mean of the bids in a tender and the number of the bids in a tender, respectively.}
\end{spacing}
\newpage

{\renewcommand{\arraystretch}{0.75}
\tablecaption{Descriptive statistics for the competitive tenders in the Swiss data\label{competDESCswiss}}
\tablehead{\hline Predictors&Mean&Std&Min&Lower Q.&Median&Upper Q.&Max&N\\  \hline}
\tabletail{\hline \multicolumn{4}{r}{See next page}\\}
\tablelasttail{\hline}
\begin{center}
\begin{supertabular}{>{$}l<{$} >{$}c<{$} >{$}c<{$} >{$}c<{$}>{$}c<{$}>{$}c<{$}>{$}c<{$}>{$}c<{$}>{$}c<{$}}
NBRBIDS&5.73&1.86&4&4&5&7&13&1082\\
MEANBIDS&828.06&1803.84&13.63&203.87&423.97&858.67&37786.87&1082\\
STDBIDS&87.35&216.52&0.41&13.96&32.74&81.27&3996.24&1082\\
CV&10.12&7.89&0.76&5.91&8.45&11.64&128&1082\\
KURTO&0.25&2.27&-6&-1.27&0.13&1.75&8.03&1082\\
SKEW&0.26&0.97&-2.68&-0.37&0.28&0.92&2.47&1082\\
SPD&2.5&29.79&0.02&0.16&0.24&0.35&730.71&1082\\
D&54.94&223.18&0.00&5.18&14.19&39.12&4656.85&1082\\
RD&1.16&2.45&0&0.23&0.57&1.13&41.26&1082\\
RDNOR&1.04&0.82&0&0.43&0.87&1.44&6.95&1082\\
RDALT&1.61&3.19&0&0.36&0.84&1.66&47.49&1082\\
DIFFP&176.79&2246.36&0&1.84&4.36&8.43&50228.95&1082\\
KS&15.07&10.98&1.48&9.12&12.24&17.44&132.33&1082\\\hline
MIN3CV&3.37&3.13&0.02&1.16&2.33&4.68&24.05&1082\\
MAX3CV&14.22&11.46&0.93&8.19&11.77&16.52&122.06&1082\\
MEAN3CV&9.37&6.9&0.73&5.56&7.92&10.9&91.8&1082\\
MEDIAN3CV&9.7&7.83&0.6&5.61&8.19&11.41&121.69&1082\\\hline
MIN3SKEW&-1.38&0.62&-1.73&-1.73&-1.67&-1.34&1.68&1082\\
MAX3SKEW&1.47&0.53&-1.61&1.48&1.69&1.73&1.73&1082\\
MEAN3SKEW&0.13&0.58&-1.66&-0.22&0.16&0.51&1.71&1082\\
MEDIAN3SKEW&0.22&0.83&-1.73&-0.35&0.26&0.83&1.73&1082\\\hline
MIN3D&13.03&30.15&0.00&1.10&3.79&11.64&364.58&1082\\
MAX3D&154.93&433.88&0.72&22.68&52.94&132.40&7506.55&1082\\
MEAN3D&69.05&178.70&0.44&11.05&25.66&60.66&2536.05&1082\\
MEDIAN3D&64.40&171.21&0.36&10.29&23.09&56.37&2640.97&1082\\\hline
MIN3RD&0.32&0.68&0&0.03&0.12&0.34&8.83&1082\\
MAX3RD&69.28&265.38&0.13&4.44&12.27&38.7&5315.77&1082\\
MEAN3RD&8.2&25.53&0.08&1.55&2.78&6.05&421.57&1082\\
MEDIAN3RD&2.22&4.95&0.01&0.78&1.21&1.98&72.77&1082\\\hline
MIN3RDNOR&0.27&0.3&0&0.05&0.16&0.39&1.72&1082\\
MAX3RDNOR&1.68&0.34&0.17&1.52&1.8&1.93&2&1082\\
MEAN3RDNOR&0.94&0.27&0.11&0.77&0.93&1.11&1.85&1082\\
MEDIAN3RDNOR&0.92&0.34&0.01&0.7&0.91&1.12&1.94&1082\\\hline
MIN3RDALT&0.22&0.48&0&0.02&0.09&0.24&6.24&1082\\
MAX3RDALT&48.99&187.65&0.1&3.14&8.67&27.36&3758.82&1082\\
MEAN3RDALT&5.8&18.05&0.06&1.1&1.96&4.27&298.09&1082\\
MEDIAN3RDALT&1.57&3.5&0&0.55&0.85&1.4&51.46&1082\\\hline
MIN3DIFFP&2.07&2.77&0&0.36&0.98&2.56&22.67&1082\\
MAX3DIFFP&226.26&2740.18&0.89&9.94&15.42&23.4&63802.49&1082\\
MEAN3DIFFP&92.86&1118.61&0.64&4.6&7.07&10.73&23734.93&1082\\
MEDIAN3DIFFP&60.05&1011.71&0.54&4.08&6.52&10.25&25112.51&1082\\\hline
MIN3SPD&0.07&0.07&0&0.02&0.05&0.09&0.58&1082\\
MAX3SPD&2.5&29.79&0.02&0.16&0.24&0.35&730.71&1082\\
MEAN3SPD&1.14&12.52&0.01&0.11&0.16&0.23&279.23&1082\\
MEDIAN3SPD&0.76&10.69&0.01&0.11&0.17&0.25&264.32&1082\\\hline
MIN3KS&10.96&8.72&1.1&6.33&8.72&12.45&107.96&1082\\
MAX3KS&82.65&179.47&4.21&21.54&43&86.79&4045.44&1082\\
MEAN3KS&21.45&21.35&2.62&11.62&16.51&24.77&463.06&1082\\
MEDIAN3KS&15.78&13.43&1.2&9.07&12.57&18.27&167.49&1082\\\hline
MIN4CV&6.24&6.77&0.07&2.29&4.49&8.14&128&1082\\
MAX4CV&12.14&9.73&0.76&7.07&10.13&14.17&128&1082\\
MEAN4CVB&9.82&7.64&0.76&5.74&8.29&11.45&128&1082\\
MEDIAN4CV&10.24&8.21&0.76&6&8.55&11.74&128&1082\\\hline
MIN4SKEW&-0.75&1.09&-2&-1.72&-1.02&0&2&1082\\
MAX4SKEW&1.02&1&-1.99&0.34&1.4&1.85&2&1082\\
MEAN4SKEW&0.19&0.8&-1.99&-0.31&0.22&0.74&2&1082\\
MEDIAN4SKEW&0.22&0.86&-1.99&-0.27&0.19&0.79&2&1082\\\hline
MIN4D&22.95&76.73&0.00&1.76&5.64&18.51&1226.06&1082\\
MAX4D&111.58&328.01&0.02&13.13&34.32&86.30&5764.43&1082\\
MEAN4D&59.42&181.05&0.02&8.14&19.23&46.70&3035.79&1082\\
MEDIAN4D&59.94&224.26&0.00&6.98&18.35&45.41&4656.85&1082\\\hline
MIN4RD&0.71&2.05&0&0.06&0.2&0.62&41.26&1082\\
MAX4RD&6.73&17.51&0&0.88&2.55&6.3&266.72&1082\\
MEAN4RD&1.69&3.23&0&0.56&1&1.72&67.61&1082\\
MEDIAN4RD&1.24&2.28&0&0.39&0.74&1.25&41.26&1082\\\hline
MIN4RDNOR&0.51&0.56&0&0.09&0.29&0.76&2.87&1082\\
MAX4RDNOR&1.62&0.81&0&0.96&1.74&2.3&2.98&1082\\
MEAN4RDNOR&0.95&0.51&0&0.61&0.9&1.21&2.87&1082\\
MEDIAN4RDNOR&0.9&0.55&0&0.51&0.84&1.19&2.87&1082\\\hline
MIN4RDALT&0.75&2.18&0&0.06&0.21&0.67&43.5&1082\\
MAX4RDALT&7.07&18.5&0&0.93&2.69&6.57&267.55&1082\\
MEAN4RDALT&1.77&3.36&0&0.59&1.05&1.8&68.45&1082\\
MEDIAN4RDALT&1.3&2.4&0&0.41&0.78&1.31&43.5&1082\\\hline
MIN4DIFFP&20.35&549.37&0&0.51&1.54&4.44&18073.22&1082\\
MAX4DIFFP&206.69&2603.79&0.03&5.44&9.88&16.43&62175.31&1082\\
MEAN4DIFFP&116.56&1450.99&0.03&3.24&5.46&9.08&30433.18&1082\\
MEDIAN4DIFFP&140.96&2011.38&0&2.73&5.24&8.95&50228.95&1082\\\hline
MIN4KURTO&-2.61&3.02&-6&-5.39&-3.37&0.07&4&1082\\
MAX4KURTO&2.07&2.28&-6&1.43&2.94&3.72&4&1082\\
MEAN4KURTO&0.12&1.83&-6&-0.83&0.1&1.31&4&1082\\
MEDIAN4KURTO&0.29&2.06&-6&-0.75&0.6&1.62&4&1082\\\hline
MIN4SPD&0.34&6.03&0&0.05&0.1&0.2&198.43&1082\\
MAX4SPD&2.5&29.79&0.02&0.16&0.24&0.35&730.71&1082\\
MEAN4SPD&1.54&17.19&0.02&0.14&0.2&0.29&388.02&1082\\
MEDIAN4SPD&1.78&23.58&0.02&0.14&0.21&0.3&621.75&1082\\\hline
MIN4KS&12.89&10.47&1.38&7.42&10.24&14.5&132.33&1082\\
MAX4KS&36.64&66.05&1.48&12.61&22.53&44.04&1433.11&1082\\
MEAN4KS&16.65&11.9&1.48&9.73&13.55&19.78&132.33&1082\\
MEDIAN4KS&14.87&11.3&1.48&8.9&12.14&16.99&132.33&1082\\
\hline
\end{supertabular}
\end{center}
\par
\begin{spacing}{1}
{\footnotesize Note: “Mean”, “Std”, “Min”, “Lower Q.”, “Median”, “Upper Q.”, “Max”, and “N” denote the mean, standard deviation, minimum, lower quartile, median, upper quartile, maximum, and number of observations, respectively. The value for "MEANBIDS", "STDBIDS", "D", "MIN3D", "MAX3D", "MEAN3D", "MEDIAN3D", "MIN4D", "MAX4D", "MEAN4D" and "MEDIAN4D" are expressed in thousand CHF. "KS", "CV", "SPD", "RD",  "RDNOR", "RDALT", "SKEW", "DIFFP", "KURTO", "D", "STDBIDS", "MEANBIDS" and "NBRBIDS" denote the Kolmogorov-Smirnov statistic, the coefficient of variation, the spread, the relative distance, the normalized distance, the alternative relative distance, the skewness statistic, the percentage difference, the kurtosis statistic, the difference in absolute between the first and second lowest bids, the standard deviation of the bids in a tender, the mean of the bids in a tender and the number of the bids in a tender, respectively.}
\end{spacing}
\newpage

{\renewcommand{\arraystretch}{0.75}
\tablecaption{Descriptive statistics for incomplete bid-rigging cartels in sample 1\label{incompleteDESCswiss1}}
\tablehead{\hline Predictors&Mean&Std&Min&Lower Q.&Median&Upper Q.&Max&N\\  \hline}
\tabletail{\hline \multicolumn{4}{r}{See next page}\\}
\tablelasttail{\hline}
\begin{center}
\begin{supertabular}{>{$}l<{$} >{$}c<{$} >{$}c<{$} >{$}c<{$}>{$}c<{$}>{$}c<{$}>{$}c<{$}>{$}c<{$}>{$}c<{$}}
NBRBIDS&7.49&2.54&4&5&7&9&13&252\\
MEANBIDS&435.90&476.67&18.01&146.59&296.62&538.73&3460.91&252\\
STDBIDS&35.12&50.17&1.54&9.60&19.62&33.72&362.86&252\\
CV&7.79&3.89&1.77&5.14&6.79&9.6&23.92&252\\
KURTO&0.41&2.12&-5.9&-0.94&0.04&1.54&6.97&252\\
SKEW&-0.07&0.99&-2.59&-0.67&-0.06&0.5&2.57&252\\
SPD&0.26&0.16&0.05&0.16&0.21&0.32&0.89&252\\
D&21.07&39.03&0.08&4.33&9.10&20.87&351.83&252\\
RD&1.39&2.67&0.01&0.26&0.58&1.42&28.37&252\\
RDNOR&1.41&1.11&0.01&0.59&1.14&2&5.48&252\\
RDALT&2.32&3.52&0.01&0.55&1.17&2.65&28.54&252\\
DIFFP&6.34&8.85&0.03&1.8&3.69&7.02&73.53&252\\
KS&16.62&8.04&4.15&10.99&15.24&19.97&57.54&252\\\hline
MIN3CV&1.68&1.73&0&0.65&1.24&2.14&14.77&252\\
MAX3CV&12.26&6.39&2.9&7.87&10.66&15.21&38.48&252\\
MEAN3CV&7.12&3.48&1.55&4.7&6.32&8.98&22.96&252\\
MEDIAN3CV&7.06&3.81&1.01&4.59&6.28&8.51&29.82&252\\\hline
MIN3SKEW&-1.6&0.36&-1.73&-1.73&-1.72&-1.65&0.97&252\\
MAX3SKEW&1.53&0.48&-1.1&1.64&1.72&1.73&1.73&252\\
MEAN3SKEW&-0.04&0.5&-1.43&-0.3&-0.03&0.25&1.47&252\\
MEDIAN3SKEW&-0.05&0.77&-1.71&-0.59&-0.05&0.51&1.73&252\\\hline
MIN3D&3.64&8.58&0.00&0.43&1.31&3.23&80.15&252\\
MAX3D&76.68&107.56&1.58&18.87&41.15&80.60&842.70&252\\
MEAN3D&29.56&44.27&0.62&8.00&16.57&28.64&437.36&252\\
MEDIAN3D&26.54&40.57&0.41&7.70&14.60&27.21&424.30&252\\\hline
MIN3RD&0.23&0.45&0&0.02&0.07&0.19&3.35&252\\
MAX3RD&1203.97&9993.9&0.68&10.98&31.05&96.88&122393.82&252\\
MEAN3RD&27.54&191.86&0.31&2.5&4.5&9.16&2742.11&252\\
MEDIAN3RD&2.09&2.42&0.21&1.02&1.49&2.21&27.08&252\\\hline
MIN3RDNOR&0.21&0.28&0&0.03&0.09&0.24&1.41&252\\
MAX3RDNOR&1.83&0.23&0.65&1.77&1.92&1.97&2&252\\
MEAN3RDNOR&1.02&0.22&0.33&0.89&1.02&1.12&1.73&252\\
MEDIAN3RDNOR&1.02&0.3&0.26&0.83&1.02&1.2&1.89&252\\\hline
MIN3RDALT&0.16&0.32&0&0.02&0.05&0.14&2.37&252\\
MAX3RDALT&851.33&7066.76&0.48&7.76&21.95&68.51&86545.5&252\\
MEAN3RDALT&19.47&135.66&0.22&1.77&3.18&6.48&1938.96&252\\
MEDIAN3RDALT&1.48&1.71&0.15&0.72&1.06&1.56&19.15&252\\\hline
MIN3DIFFP&0.97&1.58&0&0.17&0.47&1.15&12.6&252\\
MAX3DIFFP&19.61&12.81&2.29&11.28&16.77&25.1&87.65&252\\
MEAN3DIFFP&7.76&5.62&1.39&4.79&6.4&8.94&49.6&252\\
MEDIAN3DIFFP&7.11&6.59&1.14&3.9&5.72&8.21&73.53&252\\\hline
MIN3SPD&0.03&0.04&0&0.01&0.02&0.04&0.32&252\\
MAX3SPD&0.26&0.16&0.05&0.16&0.21&0.32&0.89&252\\
MEAN3SPD&0.15&0.08&0.03&0.09&0.13&0.19&0.62&252\\
MEDIAN3SPD&0.15&0.09&0.02&0.09&0.13&0.18&0.83&252\\\hline
MIN3KS&10.96&6.18&3&6.88&9.62&12.9&34.52&252\\
MAX3KS&357.52&3429.85&6.79&47.14&80.58&155&54476.99&252\\
MEAN3KS&27.25&34.52&6.03&15.6&21.55&29.85&495.16&252\\
MEDIAN3KS&18.68&11.16&3.36&12.03&16.1&22.21&99.02&252\\\hline
MIN4CV&3.16&2.97&0.12&1.31&2.26&4.2&23.92&252\\
MAX4CV&10.63&5.46&2.38&6.97&9.43&13.25&33.94&252\\
MEAN4CV&7.43&3.65&1.64&4.89&6.54&9.33&23.92&252\\
MEDIAN4CV&7.65&3.91&1.69&5.02&6.76&9.66&25.38&252\\\hline
MIN4SKEW&-1.42&0.79&-2&-1.95&-1.78&-1.22&1.87&252\\
MAX4SKEW&1.28&0.97&-1.98&1.09&1.79&1.94&2&252\\
MEAN4SKEW&-0.07&0.69&-1.98&-0.43&-0.05&0.38&1.87&252\\
MEDIAN4SKEW&-0.07&0.79&-1.98&-0.47&-0.03&0.32&1.87&252\\\hline
MIN4D&4.98&11.53&0.00&0.48&1.63&4.68&102.18&252\\
MAX4D&60.49&87.82&0.18&13.66&33.62&66.93&771.65&252\\
MEAN4D&25.08&38.97&0.18&6.81&14.32&25.80&410.17&252\\
MEDIAN4D&24.93&42.34&0.14&6.02&13.17&25.38&424.30&252\\\hline
MIN4RD&0.59&2.26&0&0.03&0.09&0.31&28.37&252\\
MAX4RD&25.25&166.8&0.04&3.1&7.14&16.87&2627.72&252\\
MEAN4RD&2.25&2.96&0.04&0.91&1.52&2.37&28.37&252\\
MEDIAN4RD&1.53&2.43&0.04&0.53&1.01&1.66&28.37&252\\\hline
MIN4RDNOR&0.37&0.54&0&0.05&0.15&0.42&2.8&252\\
MAX4RDNOR&2.18&0.68&0.07&1.83&2.37&2.7&3&252\\
MEAN4RDNOR&1.11&0.46&0.07&0.8&1.08&1.33&2.8&252\\
MEDIAN4RDNOR&1.06&0.54&0.07&0.66&1.05&1.38&2.8&252\\\hline
MIN4RDALT&0.61&2.29&0&0.03&0.1&0.32&28.54&252\\
MAX4RDALT&27.16&186.06&0.04&3.13&7.47&18.11&2933.56&252\\
MEAN4RDALT&2.35&3.08&0.04&0.96&1.62&2.48&28.54&252\\
MEDIAN4RDALT&1.59&2.46&0.04&0.56&1.08&1.7&28.54&252\\\hline
MIN4DIFFP&1.83&4.55&0&0.19&0.58&1.58&57.15&252\\
MAX4DIFFP&16.09&11.96&0.3&8.81&13.53&20.12&77.43&252\\
MEAN4DIFFP&7.05&6.61&0.3&3.75&5.57&7.76&60.05&252\\
MEDIAN4DIFFP&6.8&7.49&0.24&3.19&5.09&7.32&73.53&252\\\hline
MIN4KURTO&-4&2.64&-6&-5.86&-5.28&-3.15&3.94&252\\
MAX4KURTO&2.99&1.68&-5.9&2.79&3.68&3.91&4&252\\
MEAN4KURTO&0.16&1.38&-5.9&-0.48&0.04&0.84&3.94&252\\
MEDIAN4KURTO&0.51&1.67&-5.9&-0.08&0.64&1.49&3.94&252\\\hline
MIN4SPRD&0.08&0.08&0&0.03&0.05&0.09&0.81&252\\
MAX4SPD&0.26&0.16&0.05&0.16&0.21&0.32&0.89&252\\
MEAN4SPD&0.19&0.11&0.04&0.12&0.16&0.23&0.81&252\\
MEDIAN4SPD&0.19&0.12&0.04&0.12&0.16&0.24&0.89&252\\\hline
MIN4KS&12.65&7.28&3.64&7.85&10.91&14.56&42.15&252\\
MAX4KS&64.97&79.07&4.15&24.21&44.41&76.85&862.88&252\\
MEAN4KS&19.96&11.22&4.15&12.92&17.51&23.75&97.43&252\\
MEDIAN4KS&16.85&8.23&3.87&11.12&15.26&20.29&59.16&252\\
\hline
\end{supertabular}
\end{center}
\par
\begin{spacing}{1}
{\footnotesize Note: “Mean”, “Std”, “Min”, “Lower Q.”, “Median”, “Upper Q.”, “Max”, and “N” denote the mean, standard deviation, minimum, lower quartile, median, upper quartile, maximum, and number of observations, respectively. The value for "MEANBIDS", "STDBIDS", "D", "MIN3D", "MAX3D", "MEAN3D", "MEDIAN3D", "MIN4D", "MAX4D", "MEAN4D" and "MEDIAN4D" are expressed in thousand CHF. "KS", "CV", "SPD", "RD",  "RDNOR", "RDALT", "SKEW", "DIFFP", "KURTO", "D", "STDBIDS", "MEANBIDS" and "NBRBIDS" denote the Kolmogorov-Smirnov statistic, the coefficient of variation, the spread, the relative distance, the normalized distance, the alternative relative distance, the skewness statistic, the percentage difference, the kurtosis statistic, the difference in absolute between the first and second lowest bids, the standard deviation of the bids in a tender, the mean of the bids in a tender and the number of the bids in a tender, respectively.}
\end{spacing}
\newpage

{\renewcommand{\arraystretch}{0.75}
\tablecaption{Descriptive statistics for incomplete bid-rigging cartels in sample 2\label{incompleteDESCswiss2}}
\tablehead{\hline Predictors&Mean&Std&Min&Lower Q.&Median&Upper Q.&Max&N\\  \hline}
\tabletail{\hline \multicolumn{4}{r}{See next page}\\}
\tablelasttail{\hline}
\begin{center}
\begin{supertabular}{>{$}l<{$} >{$}c<{$} >{$}c<{$} >{$}c<{$}>{$}c<{$}>{$}c<{$}>{$}c<{$}>{$}c<{$}>{$}c<{$}}
NBRBIDS&7.77&2.47&4&6&8&10&13&223\\
MEANBIDS&405.24&372.35&27.84&149.78&302.21&535.40&3002.37&223\\
STDBIDS&30.96&37.64&1.63&9.65&19.51&33.19&270.82&223\\
CV&7.6&3.77&1.77&5.14&6.66&9.07&23.92&223\\
KURTO&0.44&2.08&-5.75&-0.94&0.04&1.47&6.97&223\\
SKEW&-0.07&0.97&-2.59&-0.62&-0.08&0.46&2.57&223\\
SPD&0.26&0.16&0.05&0.16&0.21&0.31&0.89&223\\
D&20.13&36.04&0.08&4.13&8.90&21.08&351.83&223\\
RD&1.29&2.12&0.01&0.27&0.58&1.36&18.04&223\\
RDNOR&1.45&1.14&0.01&0.61&1.14&2.06&5.48&223\\
RDALT&2.3&3.24&0.01&0.58&1.17&2.83&20.74&223\\
DIFFP&6.3&9.13&0.03&1.88&3.69&6.78&73.53&223\\
KS&16.85&7.91&4.15&11.89&15.54&19.99&57.54&223\\\hline
MIN3CV&1.47&1.22&0&0.62&1.15&1.93&7.05&223\\
MAX3CV&12.24&6.43&2.93&8&10.57&14.98&38.48&223\\
MEAN3CV&6.92&3.33&1.55&4.69&6.09&8.33&22.96&223\\
MEDIAN3CV&6.75&3.56&1.01&4.58&6&8.06&29.82&223\\\hline
MIN3SKEW&-1.63&0.29&-1.73&-1.73&-1.72&-1.68&0.62&223\\
MAX3SKEW&1.58&0.38&-1.1&1.68&1.72&1.73&1.73&223\\
MEAN3SKEW&-0.04&0.45&-1.33&-0.28&-0.04&0.25&1.31&223\\
MEDIAN3SKEW&-0.05&0.74&-1.69&-0.58&-0.06&0.45&1.73&223\\\hline
MIN3D&2.76&5.42&0.00&0.38&1.17&2.87&56.23&223\\
MAX3D&69.79&84.97&2.38&20.13&42.35&79.53&670.57&223\\
MEAN3D&26.13&30.47&1.42&8.28&16.72&28.29&235.67&223\\
MEDIAN3D&23.11&25.43&1.52&7.92&14.63&26.60&171.87&223\\\hline
MIN3RD&0.18&0.37&0&0.02&0.06&0.15&3.35&223\\
MAX3RD&1355.45&10617.14&0.92&14.74&37.36&128.17&122393.82&223\\
MEAN3RD&30.39&203.82&0.43&2.65&4.66&9.35&2742.11&223\\
MEDIAN3RD&1.92&1.72&0.26&1.06&1.5&2.14&18.63&223\\\hline
MIN3RDNOR&0.18&0.25&0&0.03&0.08&0.19&1.41&223\\
MAX3RDNOR&1.85&0.2&0.79&1.82&1.93&1.98&2&223\\
MEAN3RDNOR&1.02&0.2&0.41&0.9&1.02&1.12&1.66&223\\
MEDIAN3RDNOR&1.02&0.28&0.3&0.85&1.02&1.2&1.85&223\\\hline
MIN3RDALT&0.13&0.26&0&0.01&0.04&0.11&2.37&223\\
MAX3RDALT&958.45&7507.45&0.65&10.42&26.42&90.63&86545.5&223\\
MEAN3RDALT&21.49&144.12&0.3&1.88&3.29&6.61&1938.96&223\\
MEDIAN3RDALT&1.36&1.21&0.18&0.75&1.06&1.51&13.18&223\\\hline
MIN3DIFFP&0.81&1.25&0&0.14&0.43&0.99&10.47&223\\
MAX3DIFFP&19.73&12.93&2.29&11.81&16.81&24.42&87.65&223\\
MEAN3DIFFP&7.59&5.58&1.39&4.82&6.35&8.75&49.6&223\\
MEDIAN3DIFFP&6.92&6.67&1.14&3.91&5.71&8.1&73.53&223\\\hline
MIN3SPD&0.03&0.02&0&0.01&0.02&0.04&0.15&223\\
MAX3SPD&0.26&0.16&0.05&0.16&0.21&0.31&0.89&223\\
MEAN3SPD&0.15&0.08&0.03&0.09&0.13&0.18&0.62&223\\
MEDIAN3SPD&0.14&0.09&0.02&0.09&0.12&0.17&0.83&223\\\hline
MIN3KS&10.9&6.01&3&6.98&9.7&12.52&34.52&223\\
MAX3KS&394.81&3645.21&14.27&52.14&87.43&161.06&54476.99&223\\
MEAN3KS&27.97&36.27&6.25&16&21.99&30.38&495.16&223\\
MEDIAN3KS&19.22&11.35&3.36&12.64&16.97&22.22&99.02&223\\\hline
MIN4CV&2.8&2.48&0.12&1.2&2.17&3.87&23.92&223\\
MAX4CV&10.63&5.48&2.48&7.05&9.36&13.08&33.94&223\\
MEAN4CV&7.22&3.5&1.64&4.88&6.32&8.58&23.92&223\\
MEDIAN4CV&7.41&3.75&1.69&5.01&6.51&8.9&25.38&223\\\hline
MIN4SKEW&-1.51&0.71&-2&-1.95&-1.82&-1.42&1.84&223\\
MAX4SKEW&1.38&0.85&-1.96&1.23&1.82&1.94&2&223\\
MEAN4SKEW&-0.06&0.64&-1.96&-0.4&-0.07&0.35&1.84&223\\
MEDIAN4SKEW&-0.07&0.74&-1.96&-0.46&-0.04&0.29&1.84&223\\\hline
MIN4D&3.82&8.45&0.00&0.43&1.43&3.77&96.23&223\\
MAX4D&58.39&72.08&0.18&16.06&37.52&68.97&535.26&223\\
MEAN4D&23.35&29.62&0.18&7.26&14.45&25.83&274.73&223\\
MEDIAN4D&23.01&32.93&0.14&6.07&13.49&25.40&351.83&223\\\hline
MIN4RD&0.41&1.4&0&0.03&0.09&0.25&18.04&223\\
MAX4RD&27.52&177.13&0.04&3.91&7.86&17.75&2627.72&223\\
MEAN4RD&2.19&2.53&0.04&0.96&1.62&2.37&24.7&223\\
MEDIAN4RD&1.42&1.77&0.04&0.57&1.02&1.52&18.04&223\\\hline
MIN4RDNOR&0.32&0.46&0&0.04&0.14&0.36&2.7&223\\
MAX4RDNOR&2.25&0.61&0.07&2&2.42&2.71&3&223\\
MEAN4RDNOR&1.11&0.42&0.07&0.82&1.09&1.3&2.7&223\\
MEDIAN4RDNOR&1.06&0.51&0.07&0.69&1.06&1.31&2.7&223\\\hline
MIN4RDALT&0.42&1.42&0&0.03&0.09&0.27&18.05&223\\
MAX4RDALT&29.63&197.6&0.04&4&8.35&18.57&2933.56&223\\
MEAN4RDALT&2.29&2.67&0.04&1.02&1.72&2.49&27.4&223\\
MEDIAN4RDALT&1.48&1.8&0.04&0.6&1.1&1.55&18.05&223\\\hline
MIN4DIFFP&1.48&4.34&0&0.17&0.51&1.23&57.15&223\\
MAX4DIFFP&16.59&11.97&0.3&9.6&13.87&20.28&77.43&223\\
MEAN4DIFFP&7&6.68&0.3&3.97&5.51&7.55&60.05&223\\
MEDIAN4DIFFP&6.74&7.66&0.24&3.24&5.08&7.23&73.53&223\\\hline
MIN4KURTO&-4.32&2.38&-6&-5.88&-5.48&-3.92&3.85&223\\
MAX4KURTO&3.16&1.43&-5.75&3.04&3.74&3.91&4&223\\
MEAN4KURTO&0.15&1.23&-5.75&-0.48&0.04&0.79&3.85&223\\
MEDIAN4KURTO&0.52&1.56&-5.75&-0.05&0.63&1.42&3.85&223\\\hline
MIN4SPD&0.07&0.07&0&0.03&0.05&0.09&0.81&223\\
MAX4SPD&0.26&0.16&0.05&0.16&0.21&0.31&0.89&223\\
MEAN4SPD&0.18&0.11&0.04&0.12&0.15&0.22&0.81&223\\
MEDIAN4SPD&0.19&0.11&0.04&0.12&0.16&0.22&0.89&223\\\hline
MIN4KS&12.55&7.02&3.64&7.95&11.06&14.33&40.78&223\\
MAX4KS&69.75&82.58&4.15&26.17&46.65&84.1&862.88&223\\
MEAN4KS&20.49&11.41&4.15&13.41&18.45&23.97&97.43&223\\
MEDIAN4KS&17.14&8.1&3.87&11.75&15.89&20.31&59.16&223\\
\hline
\end{supertabular}
\end{center}
\par
\begin{spacing}{1}
{\footnotesize Note: “Mean”, “Std”, “Min”, “Lower Q.”, “Median”, “Upper Q.”, “Max”, and “N” denote the mean, standard deviation, minimum, lower quartile, median, upper quartile, maximum, and number of observations, respectively. The value for "MEANBIDS", "STDBIDS", "D", "MIN3D", "MAX3D", "MEAN3D", "MEDIAN3D", "MIN4D", "MAX4D", "MEAN4D" and "MEDIAN4D" are expressed in thousand CHF. "KS", "CV", "SPD", "RD",  "RDNOR", "RDALT", "SKEW", "DIFFP", "KURTO", "D", "STDBIDS", "MEANBIDS" and "NBRBIDS" denote the Kolmogorov-Smirnov statistic, the coefficient of variation, the spread, the relative distance, the normalized distance, the alternative relative distance, the skewness statistic, the percentage difference, the kurtosis statistic, the difference in absolute between the first and second lowest bids, the standard deviation of the bids in a tender, the mean of the bids in a tender and the number of the bids in a tender, respectively.}
\end{spacing}
\newpage

{\renewcommand{\arraystretch}{0.75}
\tablecaption{Descriptive statistics for incomplete bid-rigging cartels in sample 3\label{incompleteDESCswiss3}}
\tablehead{\hline Predictors&Mean&Std&Min&Lower Q.&Median&Upper Q.&Max&N\\  \hline}
\tabletail{\hline \multicolumn{4}{r}{See next page}\\}
\tablelasttail{\hline}
\begin{center}
\begin{supertabular}{>{$}l<{$} >{$}c<{$} >{$}c<{$} >{$}c<{$}>{$}c<{$}>{$}c<{$}>{$}c<{$}>{$}c<{$}>{$}c<{$}}
NBRBIDS&8.5&2.15&5&7&8&10&13&173\\
MEANBIDS&439.36&388.89&30.24&182.64&329.61&572.01&3002.37&173\\
STDBIDS&33.50&40.38&2.01&11.89&20.48&33.44&270.82&173\\
CV&7.54&3.26&1.77&5.24&6.76&9.12&21.7&173\\
KURTO&0.46&1.97&-2.69&-0.88&-0.09&1.23&6.97&173\\
SKEW&-0.07&0.95&-2.59&-0.56&-0.08&0.43&2.57&173\\
SPD&0.26&0.14&0.06&0.17&0.22&0.31&0.88&173\\
D&21.95&39.01&0.08&4.53&9.46&22.94&351.83&173\\
RD&1.14&1.76&0.01&0.29&0.58&1.19&12.47&173\\
RDNOR&1.54&1.17&0.02&0.66&1.17&2.17&5.48&173\\
RDALT&2.33&3.2&0.02&0.63&1.19&2.81&20.74&173\\
DIFFP&5.91&7.48&0.1&2.13&3.7&6.74&45.42&173\\
KS&16.55&7.64&6.19&11.84&15.48&19.69&57.54&173\\\hline
MIN3CV&1.2&0.9&0&0.52&1.03&1.56&3.95&173\\
MAX3CV&12.6&5.86&2.93&8.78&10.99&15.19&38.48&173\\
MEAN3CV&6.82&2.83&1.55&4.79&6.11&8.34&16.5&173\\
MEDIAN3CV&6.51&2.79&1.01&4.67&5.92&8.02&14.2&173\\\hline
MIN3SKEW&-1.69&0.12&-1.73&-1.73&-1.73&-1.71&-0.88&173\\
MAX3SKEW&1.67&0.15&0.73&1.69&1.73&1.73&1.73&173\\
MEAN3SKEW&-0.03&0.39&-1.02&-0.22&-0.04&0.23&0.88&173\\
MEDIAN3SKEW&-0.03&0.69&-1.48&-0.47&-0.06&0.41&1.73&173\\\hline
MIN3D&2.35&4.05&0.00&0.34&1.03&2.40&30.19&173\\
MAX3D&78.60&90.43&3.70&24.82&51.54&89.92&670.57&173\\
MEAN3D&28.43&32.44&1.58&9.40&17.96&32.97&235.67&173\\
MEDIAN3D&24.31&26.46&1.52&8.53&15.79&26.36&171.87&173\\\hline
MIN3RD&0.11&0.19&0&0.02&0.04&0.12&1.76&173\\
MAX3RD&1737.96&12034.74&2.68&20.39&48.9&163.12&122393.82&173\\
MEAN3RD&37.63&231.03&0.92&2.91&5.15&9.8&2742.11&173\\
MEDIAN3RD&1.72&1.05&0.32&1.07&1.48&1.95&6.05&173\\\hline
MIN3RDNOR&0.12&0.16&0&0.02&0.06&0.16&1&173\\
MAX3RDNOR&1.9&0.13&1.31&1.87&1.94&1.98&2&173\\
MEAN3RDNOR&1.02&0.18&0.61&0.91&1.03&1.11&1.46&173\\
MEDIAN3RDNOR&1.02&0.25&0.37&0.86&1.02&1.16&1.62&173\\\hline
MIN3RDALT&0.08&0.14&0&0.01&0.03&0.09&1.25&173\\
MAX3RDALT&1228.92&8509.85&1.9&14.42&34.57&115.34&86545.5&173\\
MEAN3RDALT&26.61&163.36&0.65&2.06&3.64&6.93&1938.96&173\\
MEDIAN3RDALT&1.22&0.74&0.22&0.76&1.05&1.38&4.28&173\\\hline
MIN3DIFFP&0.58&0.71&0&0.12&0.37&0.81&4.49&173\\
MAX3DIFFP&20.62&11.43&3.84&13.06&17.99&25.01&72.01&173\\
MEAN3DIFFP&7.33&3.89&1.55&5&6.42&8.68&22.8&173\\
MEDIAN3DIFFP&6.18&3.26&1.33&3.91&5.67&7.45&21.76&173\\\hline
MIN3SPD&0.02&0.02&0&0.01&0.02&0.03&0.08&173\\
MAX3SPD&0.26&0.14&0.06&0.17&0.22&0.31&0.88&173\\
MEAN3SPD&0.14&0.07&0.03&0.1&0.13&0.18&0.38&173\\
MEDIAN3SPD&0.13&0.06&0.02&0.09&0.12&0.17&0.3&173\\\hline
MIN3KS&10.03&5.06&3&6.91&9.18&11.63&34.52&173\\
MAX3KS&489.07&4136.34&25.56&63.95&97.45&192.63&54476.99&173\\
MEAN3KS&29.3&40.7&9.3&16.34&21.99&30.15&495.16&173\\
MEDIAN3KS&19.58&12.05&7.28&12.7&17.13&21.76&99.02&173\\\hline
MIN4CV&2.14&1.45&0.12&1.03&1.76&2.94&8.25&173\\
MAX4CV&11.02&5.04&2.48&7.92&9.6&13.24&33.94&173\\
MEAN4CV&7.12&2.98&1.64&4.99&6.32&8.58&18.01&173\\
MEDIAN4CV&7.27&3.09&1.69&5.15&6.57&8.91&17.17&173\\\hline
MIN4SKEW&-1.69&0.44&-2&-1.96&-1.88&-1.64&-0.04&173\\
MAX4SKEW&1.62&0.52&-0.4&1.49&1.84&1.95&2&173\\
MEAN4SKEW&-0.05&0.55&-1.42&-0.34&-0.07&0.31&1.32&173\\
MEDIAN4SKEW&-0.06&0.67&-1.86&-0.42&-0.05&0.27&1.72&173\\\hline
MIN4D&2.86&4.84&0.00&0.34&1.13&3.02&30.28&173\\
MAX4D&66.68&76.42&1.56&20.02&43.87&78.43&535.26&173\\
MEAN4D&25.49&31.63&1.22&8.73&15.85&27.51&274.73&173\\
MEDIAN4D&24.96&35.20&0.96&7.98&15.12&27.02&351.83&173\\\hline
MIN4RD&0.16&0.26&0&0.02&0.06&0.17&1.93&173\\
MAX4RD&33.76&200.71&1.09&5.32&10.34&20.54&2627.72&173\\
MEAN4RD&2.22&2.44&0.35&1.04&1.68&2.38&24.7&173\\
MEDIAN4RD&1.31&1.24&0.16&0.69&1.05&1.39&9.81&173\\\hline
MIN4RDNOR&0.2&0.25&0&0.03&0.1&0.25&1.51&173\\
MAX4RDNOR&2.42&0.44&1.07&2.2&2.53&2.75&3&173\\
MEAN4RDNOR&1.11&0.35&0.43&0.85&1.1&1.28&2.06&173\\
MEDIAN4RDNOR&1.07&0.44&0.22&0.79&1.07&1.25&2.5&173\\\hline
MIN4RDALT&0.16&0.26&0&0.02&0.07&0.18&2.03&173\\
MAX4RDALT&36.41&223.95&1.1&5.54&10.76&21.56&2933.56&173\\
MEAN4RDALT&2.31&2.61&0.37&1.09&1.75&2.51&27.4&173\\
MEDIAN4RDALT&1.37&1.27&0.16&0.72&1.11&1.43&9.91&173\\\hline
MIN4DIFFP&0.71&0.9&0&0.13&0.41&0.98&6.52&173\\
MAX4DIFFP&17.69&10.69&3.03&11.06&15.55&21.32&70.09&173\\
MEAN4DIFFP&6.67&4.5&1.18&4.22&5.62&7.44&26.42&173\\
MEDIAN4DIFFP&6.33&5.07&0.98&3.44&5.23&7.15&34.35&173\\\hline
MIN4KURTO&-4.9&1.6&-6&-5.92&-5.55&-4.6&1.44&173\\
MAX4KURTO&3.56&0.61&1&3.49&3.82&3.93&4&173\\
MEAN4KURTO&0.17&0.89&-1.78&-0.46&-0.05&0.63&3.19&173\\
MEDIAN4KURTO&0.63&1.19&-4.42&0&0.63&1.34&3.57&173\\\hline
MIN4SPD&0.05&0.03&0&0.02&0.04&0.07&0.2&173\\
MAX4SPD&0.26&0.14&0.06&0.17&0.22&0.31&0.88&173\\
MEAN4SPD&0.18&0.08&0.04&0.12&0.15&0.22&0.48&173\\
MEDIAN4SPD&0.18&0.09&0.04&0.12&0.16&0.22&0.48&173\\\hline
MIN4KS&11.44&5.86&3.64&7.8&10.66&12.95&40.78&173\\
MAX4KS&81.53&89.76&12.72&34.28&57.17&97.8&862.88&173\\
MEAN4KS&20.95&12.1&8.31&13.75&18.47&23.97&97.43&173\\
MEDIAN4KS&16.96&7.85&5.9&11.68&15.89&19.94&59.16&173\\
\hline
\end{supertabular}
\end{center}
\par
\begin{spacing}{1}
{\footnotesize Note: “Mean”, “Std”, “Min”, “Lower Q.”, “Median”, “Upper Q.”, “Max”, and “N” denote the mean, standard deviation, minimum, lower quartile, median, upper quartile, maximum, and number of observations, respectively. The value for "MEANBIDS", "STDBIDS", "D", "MIN3D", "MAX3D", "MEAN3D", "MEDIAN3D", "MIN4D", "MAX4D", "MEAN4D" and "MEDIAN4D" are expressed in thousand CHF. "KS", "CV", "SPD", "RD",  "RDNOR", "RDALT", "SKEW", "DIFFP", "KURTO", "D", "STDBIDS", "MEANBIDS" and "NBRBIDS" denote the Kolmogorov-Smirnov statistic, the coefficient of variation, the spread, the relative distance, the normalized distance, the alternative relative distance, the skewness statistic, the percentage difference, the kurtosis statistic, the difference in absolute between the first and second lowest bids, the standard deviation of the bids in a tender, the mean of the bids in a tender and the number of the bids in a tender, respectively.}
\end{spacing}
\newpage

{\renewcommand{\arraystretch}{0.75}
\tablecaption{Descriptive statistics for incomplete bid-rigging cartels in sample 4\label{incompleteDESCswiss4}}
\tablehead{\hline Predictors&Mean&Std&Min&Lower Q.&Median&Upper Q.&Max&N\\  \hline}
\tabletail{\hline \multicolumn{4}{r}{See next page}\\}
\tablelasttail{\hline}
\begin{center}
\begin{supertabular}{>{$}l<{$} >{$}c<{$} >{$}c<{$} >{$}c<{$}>{$}c<{$}>{$}c<{$}>{$}c<{$}>{$}c<{$}>{$}c<{$}}
NBRBIDS&9.08&1.87&6&8&9&11&13&135\\
MEANBIDS&448.31&384.31&41.46&219.13&337.02&567.20&3002.37&135\\
STDBIDS&31.74&34.92&2.26&13.09&21.49&32.73&270.82&135\\
CV&7.19&2.93&1.77&5.22&6.3&8.62&18.9&135\\
KURTO&0.47&1.97&-2.26&-0.88&-0.18&1.18&6.97&135\\
SKEW&-0.09&0.94&-2.59&-0.54&-0.12&0.44&2.57&135\\
SPD&0.26&0.13&0.06&0.17&0.21&0.31&0.76&135\\
D&22.85&42.70&0.26&4.76&9.34&22.35&351.83&135\\
RD&1.05&1.5&0.01&0.3&0.57&1.12&10.58&135\\
RDNOR&1.6&1.18&0.04&0.7&1.22&2.28&5.48&135\\
RDALT&2.33&3.02&0.04&0.67&1.27&2.86&20.74&135\\
DIFFP&5.82&7.82&0.1&2.03&3.36&6.58&45.42&135\\
KS&16.91&7.21&6.62&12.48&16.29&19.83&57.54&135\\\hline
MIN3CV&0.94&0.62&0&0.43&0.82&1.28&3.04&135\\
MAX3CV&12.33&5.32&3.36&8.87&10.64&15.02&32.54&135\\
MEAN3CV&6.47&2.56&1.55&4.69&5.85&7.72&15.38&135\\
MEDIAN3CV&6.07&2.49&1.01&4.52&5.67&7.29&13.93&135\\\hline
MIN3SKEW&-1.71&0.07&-1.73&-1.73&-1.73&-1.72&-1.13&135\\
MAX3SKEW&1.69&0.1&0.93&1.71&1.73&1.73&1.73&135\\
MEAN3SKEW&-0.04&0.37&-1.02&-0.24&-0.04&0.21&0.87&135\\
MEDIAN3SKEW&-0.03&0.66&-1.45&-0.47&-0.07&0.4&1.73&135\\\hline
MIN3D&1.96&3.70&0.00&0.30&0.87&1.99&30.19&135\\
MAX3D&81.04&93.45&5.67&30.62&55.61&88.77&670.57&135\\
MEAN3D&28.50&33.46&2.14&11.22&18.00&30.40&235.67&135\\
MEDIAN3D&23.43&25.63&2.24&9.00&16.39&25.58&171.87&135\\\hline
MIN3RD&0.09&0.18&0&0.02&0.04&0.1&1.76&135\\
MAX3RD&2203.8&13597.91&3.46&27.6&62.53&192.17&122393.82&135\\
MEAN3RD&45.46&261.04&0.92&3.13&5.48&10.57&2742.11&135\\
MEDIAN3RD&1.69&0.96&0.45&1.08&1.5&1.95&5.61&135\\\hline
MIN3RDNOR&0.1&0.14&0&0.02&0.06&0.13&1&135\\
MAX3RDNOR&1.92&0.1&1.42&1.91&1.96&1.99&2&135\\
MEAN3RDNOR&1.02&0.17&0.61&0.92&1.04&1.11&1.46&135\\
MEDIAN3RDNOR&1.02&0.24&0.49&0.87&1.02&1.16&1.6&135\\\hline
MIN3RDALT&0.06&0.13&0&0.01&0.03&0.07&1.25&135\\
MAX3RDALT&1558.32&9615.17&2.44&19.52&44.22&135.89&86545.5&135\\
MEAN3RDALT&32.15&184.58&0.65&2.22&3.88&7.48&1938.96&135\\
MEDIAN3RDALT&1.2&0.68&0.32&0.76&1.06&1.38&3.97&135\\\hline
MIN3DIFFP&0.44&0.47&0&0.11&0.27&0.56&2.32&135\\
MAX3DIFFP&20.71&11.13&3.84&13.33&17.81&25.2&63.38&135\\
MEAN3DIFFP&7.11&3.99&1.55&4.82&6.24&8.15&22.8&135\\
MEDIAN3DIFFP&5.68&2.93&1.33&3.81&5.29&6.99&21.76&135\\\hline
MIN3SPD&0.02&0.01&0&0.01&0.02&0.02&0.06&135\\
MAX3SPD&0.26&0.13&0.06&0.17&0.21&0.31&0.76&135\\
MEAN3SPD&0.14&0.06&0.03&0.1&0.12&0.16&0.35&135\\
MEDIAN3SPD&0.12&0.05&0.02&0.09&0.11&0.15&0.3&135\\\hline
MIN3KS&9.9&4.37&3.47&7&9.69&11.51&30.19&135\\
MAX3KS&603.24&4679.33&32.93&77.85&122.26&231.34&54477&135\\
MEAN3KS&30.89&43.76&10.24&18.79&23.63&31.72&495.16&135\\
MEDIAN3KS&20.34&11.39&7.28&14.06&17.92&22.59&99.02&135\\\hline
MIN4CV&1.7&1.03&0.12&0.91&1.5&2.18&5.53&135\\
MAX4CV&10.83&4.53&2.8&8&9.4&13.15&28.01&135\\
MEAN4CV&6.76&2.69&1.64&4.93&6.08&8&16.54&135\\
MEDIAN4CV&6.88&2.88&1.69&5.01&6.17&8.11&17.17&135\\\hline
MIN4SKEW&-1.81&0.29&-2&-1.97&-1.92&-1.76&-0.28&135\\
MAX4SKEW&1.72&0.4&-0.26&1.7&1.86&1.96&2&135\\
MEAN4SKEW&-0.06&0.52&-1.37&-0.33&-0.1&0.28&1.18&135\\
MEDIAN4SKEW&-0.07&0.65&-1.69&-0.38&-0.07&0.26&1.72&135\\\hline
MIN4D&2.38&4.46&0.00&0.33&1.04&2.48&30.28&135\\
MAX4D&69.99&79.50&5.47&25.08&47.56&78.76&535.26&135\\
MEAN4D&25.93&33.56&1.39&9.45&16.57&26.24&274.73&135\\
MEDIAN4D&25.13&37.53&0.96&7.99&16.21&25.40&351.83&135\\\hline
MIN4RD&0.12&0.19&0&0.02&0.05&0.14&1.47&135\\
MAX4RD&41.1&226.79&1.09&6.95&13.21&23.81&2627.72&135\\
MEAN4RD&2.27&2.46&0.35&1.11&1.72&2.51&24.7&135\\
MEDIAN4RD&1.25&0.95&0.16&0.69&1.06&1.36&5.95&135\\\hline
MIN4RDNOR&0.16&0.19&0&0.03&0.09&0.21&1.02&135\\
MAX4RDNOR&2.53&0.35&1.07&2.34&2.61&2.79&3&135\\
MEAN4RDNOR&1.11&0.33&0.43&0.87&1.11&1.28&2.05&135\\
MEDIAN4RDNOR&1.07&0.42&0.22&0.8&1.07&1.25&2.27&135\\\hline
MIN4RDALT&0.12&0.17&0&0.02&0.06&0.15&1.04&135\\
MAX4RDALT&44.43&253.08&1.1&7.07&13.34&26.04&2933.56&135\\
MEAN4RDALT&2.37&2.65&0.37&1.16&1.79&2.62&27.4&135\\
MEDIAN4RDALT&1.31&1&0.16&0.73&1.11&1.43&6.3&135\\\hline
MIN4DIFFP&0.53&0.59&0&0.12&0.37&0.81&3.21&135\\
MAX4DIFFP&18.02&10.4&3.03&11.48&15.55&22.07&56.74&135\\
MEAN4DIFFP&6.55&4.74&1.18&4.08&5.51&7.01&26.42&135\\
MEDIAN4DIFFP&6.13&5.32&0.98&3.26&4.82&6.87&34.35&135\\\hline
MIN4KURTO&-5.26&1.06&-6&-5.93&-5.69&-5.08&-0.16&135\\
MAX4KURTO&3.71&0.39&2.18&3.62&3.86&3.95&4&135\\
MEAN4KURTO&0.16&0.8&-1.21&-0.4&-0.07&0.62&2.41&135\\
MEDIAN4KURTO&0.72&0.98&-3.63&0.06&0.68&1.38&3.11&135\\\hline
MIN4SPD&0.04&0.02&0&0.02&0.03&0.05&0.13&135\\
MAX4SPD&0.26&0.13&0.06&0.17&0.21&0.31&0.76&135\\
MEAN4SPD&0.17&0.08&0.04&0.12&0.15&0.2&0.45&135\\
MEDIAN4SPD&0.17&0.08&0.04&0.12&0.15&0.2&0.48&135\\\hline
MIN4KS&11.21&5.03&4.25&7.95&10.99&12.79&36.12&135\\
MAX4KS&91.93&91.08&18.25&46.29&67.17&110.97&862.88&135\\
MEAN4KS&21.62&10.72&8.43&15.07&19.14&24.63&97.28&135\\
MEDIAN4KS&17.62&7.69&5.9&13.03&16.55&20.31&59.16&135\\
\hline
\end{supertabular}
\end{center}
\par
\begin{spacing}{1}
{\footnotesize Note: “Mean”, “Std”, “Min”, “Lower Q.”, “Median”, “Upper Q.”, “Max”, and “N” denote the mean, standard deviation, minimum, lower quartile, median, upper quartile, maximum, and number of observations, respectively. The value for "MEANBIDS", "STDBIDS", "D", "MIN3D", "MAX3D", "MEAN3D", "MEDIAN3D", "MIN4D", "MAX4D", "MEAN4D" and "MEDIAN4D" are expressed in thousand CHF. "KS", "CV", "SPD", "RD",  "RDNOR", "RDALT", "SKEW", "DIFFP", "KURTO", "D", "STDBIDS", "MEANBIDS" and "NBRBIDS" denote the Kolmogorov-Smirnov statistic, the coefficient of variation, the spread, the relative distance, the normalized distance, the alternative relative distance, the skewness statistic, the percentage difference, the kurtosis statistic, the difference in absolute between the first and second lowest bids, the standard deviation of the bids in a tender, the mean of the bids in a tender and the number of the bids in a tender, respectively.}
\end{spacing}
\newpage

{\renewcommand{\arraystretch}{0.75}
\tablecaption{Descriptive statistics for incomplete bid-rigging cartels in sample 5\label{incompleteDESCswiss5}}
\tablehead{\hline Predictors&Mean&Std&Min&Lower Q.&Median&Upper Q.&Max&N\\  \hline}
\tabletail{\hline \multicolumn{4}{r}{See next page}\\}
\tablelasttail{\hline}
\begin{center}
\begin{supertabular}{>{$}l<{$} >{$}c<{$} >{$}c<{$} >{$}c<{$}>{$}c<{$}>{$}c<{$}>{$}c<{$}>{$}c<{$}>{$}c<{$}}
NBRBIDS&9.42&1.81&7&8&9&11&13&104\\
MEANBIDS&434.09&290.66&86.12&232.84&345.62&534.23&1559.96&104\\
STDBIDS&29.66&26.44&4.58&15.43&21.89&32.70&191.92&104\\
CV&6.87&2.53&1.77&5.19&6.22&8.05&14.4&104\\
KURTO&0.52&2.06&-2.26&-0.89&-0.15&1.17&6.97&104\\
SKEW&-0.1&0.94&-2.59&-0.62&-0.16&0.43&2.57&104\\
SPD&0.25&0.12&0.06&0.18&0.21&0.28&0.66&104\\
D&23.61&46.99&0.26&5.33&8.88&21.49&351.83&104\\
RD&1.03&1.56&0.01&0.3&0.57&1.09&10.58&104\\
RDNOR&1.65&1.24&0.04&0.72&1.25&2.39&5.48&104\\
RDALT&2.4&3.23&0.04&0.69&1.29&3.01&20.74&104\\
DIFFP&5.91&8.38&0.1&1.77&3.22&6.14&45.42&104\\
KS&17.25&7.04&6.93&13.12&16.66&19.81&57.54&104\\\hline
MIN3CV&0.86&0.6&0&0.38&0.75&1.21&3.04&104\\
MAX3CV&11.97&4.77&3.36&8.96&10.54&13.64&26.24&104\\
MEAN3CV&6.19&2.19&1.55&4.7&5.74&7.34&12.47&104\\
MEDIAN3CV&5.71&2.04&1.55&4.39&5.55&6.88&11.38&104\\\hline
MIN3SKEW&-1.71&0.08&-1.73&-1.73&-1.73&-1.72&-1.13&104\\
MAX3SKEW&1.71&0.05&1.48&1.71&1.73&1.73&1.73&104\\
MEAN3SKEW&-0.04&0.35&-1.02&-0.25&-0.05&0.16&0.87&104\\
MEDIAN3SKEW&-0.05&0.62&-1.45&-0.47&-0.07&0.36&1.32&104\\\hline
MIN3D&1.47&2.09&0.00&0.31&0.81&1.65&13.14&104\\
MAX3D&76.95&75.24&9.36&35.44&56.91&86.41&530.84&104\\
MEAN3D&27.35&30.06&3.76&12.85&18.95&28.64&235.67&104\\
MEDIAN3D&21.88&21.69&2.24&10.65&16.45&24.90&144.14&104\\\hline
MIN3RD&0.06&0.07&0&0.01&0.04&0.09&0.34&104\\
MAX3RD&2673.49&15390.37&3.46&35.67&75.98&246.78&122393.82&104\\
MEAN3RD&53.44&294.68&0.92&3.42&5.6&10.88&2742.11&104\\
MEDIAN3RD&1.66&0.88&0.45&1.11&1.48&1.95&5.61&104\\\hline
MIN3RDNOR&0.08&0.08&0&0.02&0.06&0.11&0.39&104\\
MAX3RDNOR&1.93&0.09&1.42&1.92&1.96&1.99&2&104\\
MEAN3RDNOR&1.02&0.16&0.61&0.92&1.04&1.11&1.46&104\\
MEDIAN3RDNOR&1.02&0.23&0.49&0.88&1.02&1.16&1.6&104\\\hline
MIN3RDALT&0.04&0.05&0&0.01&0.03&0.06&0.24&104\\
MAX3RDALT&1890.44&10882.63&2.44&25.22&53.73&174.5&86545.5&104\\
MEAN3RDALT&37.79&208.37&0.65&2.42&3.96&7.69&1938.96&104\\
MEDIAN3RDALT&1.17&0.62&0.32&0.78&1.05&1.38&3.97&104\\\hline
MIN3DIFFP&0.37&0.4&0&0.09&0.24&0.5&2.32&104\\
MAX3DIFFP&20.4&11.4&3.84&13.56&17.28&24.31&63.38&104\\
MEAN3DIFFP&6.93&4.04&1.55&4.74&5.94&7.75&22.8&104\\
MEDIAN3DIFFP&5.4&2.68&1.33&3.54&5&6.68&14.62&104\\\hline
MIN3SPD&0.02&0.01&0&0.01&0.01&0.02&0.06&104\\
MAX3SPD&0.25&0.12&0.06&0.18&0.21&0.28&0.66&104\\
MEAN3SPD&0.13&0.05&0.03&0.1&0.12&0.15&0.3&104\\
MEDIAN3SPD&0.12&0.04&0.03&0.09&0.11&0.14&0.25&104\\\hline
MIN3KS&9.91&4.06&3.91&7.59&9.79&11.48&30.19&104\\
MAX3KS&748.1&5328.42&32.93&82.68&133.82&260.68&54476.99&104\\
MEAN3KS&32.57&48.85&11.03&20.32&24.65&32.78&495.16&104\\
MEDIAN3KS&20.44&8.91&9.05&14.76&18.19&22.87&65.09&104\\\hline
MIN4CV&1.54&0.97&0.12&0.85&1.34&2.04&5.53&104\\
MAX4CV&10.53&4.01&2.8&8.13&9.12&12.21&21.92&104\\
MEAN4CV&6.46&2.3&1.64&4.9&5.96&7.61&13.1&104\\
MEDIAN4CV&6.52&2.52&1.69&4.99&6.07&7.47&17.17&104\\\hline
MIN4SKEW&-1.82&0.31&-2&-1.98&-1.94&-1.82&-0.28&104\\
MAX4SKEW&1.78&0.31&0.11&1.75&1.88&1.96&2&104\\
MEAN4SKEW&-0.06&0.5&-1.37&-0.34&-0.09&0.24&1.18&104\\
MEDIAN4SKEW&-0.07&0.62&-1.69&-0.38&-0.07&0.24&1.72&104\\\hline
MIN4D&1.95&3.57&0.00&0.31&0.93&2.26&30.28&104\\
MAX4D&68.42&69.32&8.37&31.49&50.56&780.10&495.98&104\\
MEAN4D&25.58&33.77&2.31&10.61&17.47&25.20&274.73&104\\
MEDIAN4D&24.81&39.62&0.96&8.97&15.70&23.58&351.83&104\\\hline
MIN4RD&0.09&0.1&0&0.02&0.05&0.11&0.45&104\\
MAX4RD&49.4&257.96&1.09&8.05&15.98&26.88&2627.72&104\\
MEAN4RD&2.33&2.67&0.35&1.15&1.76&2.51&24.7&104\\
MEDIAN4RD&1.24&0.92&0.16&0.74&1.07&1.35&5.95&104\\\hline
MIN4RDNOR&0.13&0.14&0&0.03&0.09&0.18&0.58&104\\
MAX4RDNOR&2.56&0.37&1.07&2.44&2.67&2.81&3&104\\
MEAN4RDNOR&1.11&0.31&0.43&0.92&1.12&1.28&2.05&104\\
MEDIAN4RDNOR&1.07&0.4&0.22&0.84&1.08&1.25&2.27&104\\\hline
MIN4RDALT&0.1&0.11&0&0.02&0.06&0.13&0.48&104\\
MAX4RDALT&53.65&287.9&1.1&8.68&16.19&29.49&2933.56&104\\
MEAN4RDALT&2.46&2.92&0.37&1.19&1.81&2.61&27.4&104\\
MEDIAN4RDALT&1.3&0.98&0.16&0.78&1.12&1.43&6.3&104\\\hline
MIN4DIFFP&0.46&0.52&0&0.1&0.31&0.6&2.5&104\\
MAX4DIFFP&18.1&10.77&3.03&11.85&15.64&20.51&56.74&104\\
MEAN4DIFFP&6.47&4.92&1.18&3.87&5.34&6.87&26.42&104\\
MEDIAN4DIFFP&5.98&5.32&0.98&3.21&4.61&6.49&34.35&104\\\hline
MIN4KURTO&-5.39&0.94&-6&-5.94&-5.77&-5.33&-1.12&104\\
MAX4KURTO&3.75&0.33&2.27&3.66&3.88&3.95&4&104\\
MEAN4KURTO&0.17&0.79&-1.07&-0.4&-0.06&0.56&2.41&104\\
MEDIAN4KURTO&0.71&0.98&-3.63&0.11&0.62&1.3&3.11&104\\\hline
MIN4SPD&0.03&0.02&0&0.02&0.03&0.05&0.13&104\\
MAX4SPD&0.25&0.12&0.06&0.18&0.21&0.28&0.66&104\\
MEAN4SPD&0.16&0.07&0.04&0.12&0.14&0.19&0.39&104\\
MEDIAN4SPD&0.16&0.07&0.04&0.12&0.14&0.17&0.48&104\\\hline
MIN4KS&11.18&4.58&4.66&8.45&11.24&12.71&36.12&104\\
MAX4KS&101.52&99.91&18.25&49.53&75.09&118.27&862.88&104\\
MEAN4KS&22.21&10.63&9.43&16.22&19.85&25.26&97.28&104\\
MEDIAN4KS&18.1&7.29&5.9&13.93&17.06&20.66&59.16&104\\
\hline
\end{supertabular}
\end{center}
\par
\begin{spacing}{1}
{\footnotesize Note: “Mean”, “Std”, “Min”, “Lower Q.”, “Median”, “Upper Q.”, “Max”, and “N” denote the mean, standard deviation, minimum, lower quartile, median, upper quartile, maximum, and number of observations, respectively. The value for "MEANBIDS", "STDBIDS", "D", "MIN3D", "MAX3D", "MEAN3D", "MEDIAN3D", "MIN4D", "MAX4D", "MEAN4D" and "MEDIAN4D" are expressed in thousand CHF. "KS", "CV", "SPD", "RD",  "RDNOR", "RDALT", "SKEW", "DIFFP", "KURTO", "D", "STDBIDS", "MEANBIDS" and "NBRBIDS" denote the Kolmogorov-Smirnov statistic, the coefficient of variation, the spread, the relative distance, the normalized distance, the alternative relative distance, the skewness statistic, the percentage difference, the kurtosis statistic, the difference in absolute between the first and second lowest bids, the standard deviation of the bids in a tender, the mean of the bids in a tender and the number of the bids in a tender, respectively.}
\end{spacing}
\newpage

\subsection*{Appendix 3: Classification tree adjusting the benchmarking rule of \citet{Imhof2017a} in the Swiss data in sample only with complete cartels}

\begin{figure}[!htp] \begin{center}
\includegraphics[height=8cm, width=10cm]{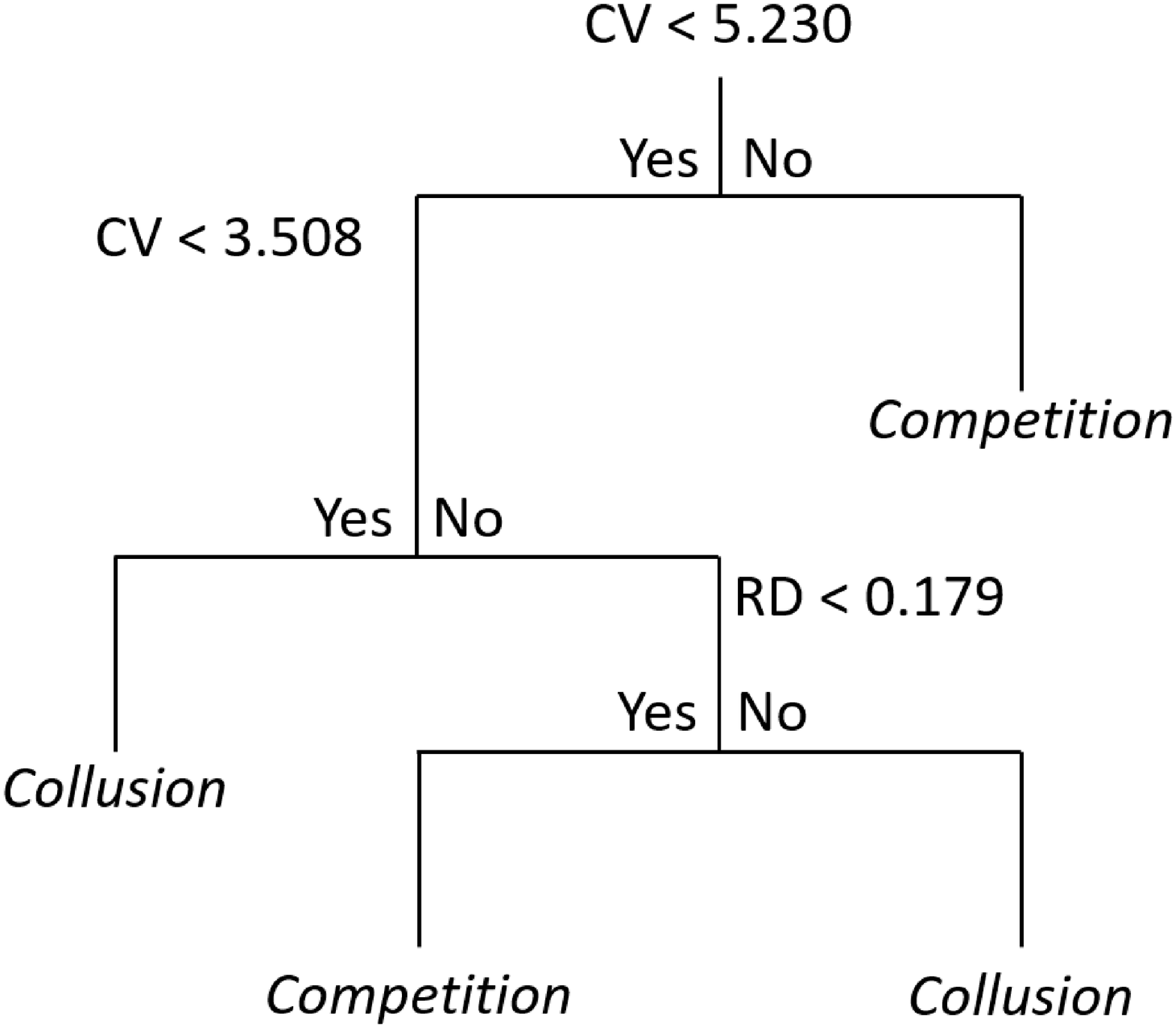}
\caption{Adjusted classification tree\label{tree}}
\end{center}
\end{figure}

\subsection*{Appendix 4: Details about lasso regression and the ensemble method}
We subsequently discuss in more detail the machine learning approaches of the lasso regression and ensemble method applied by \citet{Huberimhof2019}. Similar to the random forest, the lasso regression and ensemble method require randomly splitting the data into training (used for estimating the model parameters) and test data (used for out of sample prediction and performance evaluation). Again, our training and test samples contain 75\% and 25\% of the observations, respectively. Lasso regression corresponds to a penalized logit regression, where the penalty term restricts the sum of absolute coefficients on the regressors. Coefficients of less predictive variables shrink towards or even exactly to zero depending on the penalty term. Therefore, lasso regression may perform predictor selection. The estimation of the lasso logit coefficients is based on the following optimization problem:

\begin{equation}\label{maxlaslogitMLS}
\max_{\beta_{0},\bm{\beta} }\left\lbrace \sum_{i=1}^{n} \left[y_{i}\left(\beta_{0}+\sum_{j=1}^{p}\beta_{j}x_{ij}\right)-log\left(1+e^{\beta_{0}+\sum_{j=1}^{p}\beta_{j}x_{ij}}\right) \right] -\lambda\sum_{j=1}^{p}\vert\beta_{j}\vert\right\rbrace.
\end{equation}

where $\beta_{0}$ denotes intercept and slope coefficients on the predictors, $\beta$ the slope coefficients on the predictors, $x$ the vector of predictors, $i$ indexes an observation in our data set (with n being the number of observations), $j$ indexes a predictor (with p being the number of predictors), and $\lambda$ a penalty term larger than zero. We use the same predictors as described in the main text for the different models. In our application, we use the hdm package by \citet{Chernov2016} for the statistical software R. We apply 15-fold cross-validation to select the penalty term $\lambda$ based on minimizes the mean squared error of prediction.

For the ensemble method, we apply as in \citet{Huberimhof2019} the “SuperLearner” package for “R” by \citet{Laan2008} with default values for bagged regression tree, random forest, and neural network algorithms in the “ipredbagg”, “cforest”, and “nnet” packages, respectively. Ensemble method also relies on training data for estimating the model parameters and test data for prediction and performance evaluation. However, any estimation step now consists of a weighted average of bagged classification trees, random forest and neural networks. Bagged trees consist of estimating single trees (rather than random forests) repeatedly using the outcome residuals of the respective previous tree as outcome. Rather than splitting the predictor space, neural networks aim at fitting a system of nonlinear functions that models the influence of the predictors on collusion in a flexible way. To do so, we model the association between the predictors and the outcomes via a network of non-linear intermediate functions, so-called hidden notes. Several layers of hidden nodes allow modelling associations as well as interactions between the predictors in a flexible way, with more nodes and layers increasing the variance but reducing the bias.

\subsection*{Appendix 5: Results for the statistical tests between the simulated bids and the competitive bids in the Ticino case}

In the following, we test if the simulated competitive bids are similar to the competitive bids of the post-cartel period. We calculate the screens for the five simulated competitive bids for each collusive tender. We test if the screens differ from the screens calculated with the competitive bids of the post-cartel period. Since the screens are not normally distributed, we apply non-parametric tests to our data, which do not assume any particular distribution in the test procedures \citep[see also][]{Imhof2017a, Imhof2020}. First, we apply the Mann-Whitney test (also called the Wilcoxon rank sum test).\footnote{See \citet{Rice2007} chapter 11, page 435 ff.; \citet{Hollander2014} chapter 4, page 115 ff.} To insure the robustness of the results, we second use the Kolmogorov-Smirnov test, a more general test examining for any kind of differences between samples.\footnote{See \citet{Hollander2014} chapter 5, page 190 ff.}

Table \ref{stattestscreen} indicates the test results. We find no rejection for all the tests (at the 5\% significance level), except for the Kolmogorov-Smirnov test applied to the percentage difference (DIFFP). To sum up, the screens calculated with the simulated competitive bids do not significantly differ with the screens calculated with the "real" competitive bids of the post-cartel period. Therefore, the simulated competitive bids exhibit more or less the same statistical pattern as the "real" competitive bids. This result indicates that our simulation process adequately generates competitive bids for the purpose of our analyses.

{\renewcommand{\arraystretch}{1.1}
\begin{table} [!htp]
\caption{Statistical tests for the screens calculated with the simulated competitive bids against the competitive bids of the post-cartel period} \label{stattestscreen}
\begin{center}
\begin{tabular}{lcccc}\hline\hline
Screens&z-statistic&p-value MW&KSa&p-value KS\\\hline
CV&-1.14&0.2525&1.24&0.0934\\
KURTO&-0.93&0.3545&1.04&0.2311\\
SPD&-0.45&0.6541&1.12&0.1623\\
DIFFP&-1.64&0.1014&1.90&0.0015\\
SKEW&-0.06&0.9524&0.87&0.4377\\
RD&-0.10&0.9215&0.78&0.5820\\
RDNOR&0.1290&0.9874&0.83&0.4913\\
RDALT&0.1179&0.9061&0.77&0.5901\\
KS&1.31&0.1890&1.21&0.1084\\
\hline\hline
\end{tabular}
\end{center}
\par
{\footnotesize Note: "Screens", "z-statistic", "p-value MW" denote the screens tested, the z-statistic of the Mann-Whitney test and the p-value of the Mann-Whitney test,  respectively. "KSa" and "p-value KS" denote the asymptotic Kolmogorov-Smirnov statistic and the p-value of the Kolmogorov-Smirnov test, respectively.}
\end{table}}

\newpage

\begin{spacing}{1.0}
\bibliographystyle{agu}
\bibliography {bibliothesis}
\end{spacing}

\end{document}